\documentclass[hyper(*)]{JHEP3}

\usepackage{epsfig,multicol,bbm}
\usepackage{enumerate}
\usepackage{bibentry}
\usepackage{amsmath}
\usepackage{amsthm}
\usepackage{array}
\newtheorem{mydef}{Example}
\newtheorem{mydef1}{Conjecture}
\newtheorem{mydef2}{Statement}
\newtheorem{mydef3}{Step}
\newtheorem{mydef4}{Definition}
\usepackage{amsmath, amssymb, graphics, setspace}

\newcommand{\mathsym}[1]{{}}
\newcommand{\unicode}[1]{{}}

\title{BPS spectrum, wall crossing and quantum dilogarithm identity}

\author{Dan Xie

\\ School of Natural Sciences, Institute for Advanced Study \\
Princeton, NJ 08540, USA}

\abstract{ BPS spectrum with finite number of states are found for higher rank four dimensional $\mathcal{N}=2$ theory engineered from six dimensional $A_{N-1}$ $(2,0)$ 
 theory on a Riemann surface with various kinds of defects. The wall crossing formula is interpreted as 
the quantum dilogarithm identity. Various methods  including quiver
representation theory, maximal green mutation,  and cluster algebra are used extensively. The spectral generator and its refined version for 
the higher rank theory are written down using the explicit spectrum information. The finite chamber has an interesting $N^3$ behavior in the large $N$ limit.  }

\begin{document}
\maketitle  
\section{Introduction}
Understanding the BPS spectrum of the quantum field theory with extended supersymmetry always
provides important information about the dynamics of the theory. Such objects gave the key insights to the discovery of 
the electric-magnetic duality of 4d $\mathcal{N}=4$ 
super Yang-Mills theory \cite{Montonen:1977sn}, and the exact solution of the Coulomb branch of 4d $\mathcal{N}=2$ theory \cite{Seiberg:1994rs, Seiberg:1994aj}, etc. 

The BPS spectrum for four dimensional $\mathcal{N}=2$ theory has the interesting wall crossing 
behavior, i.e. the spectrum is not smooth in crossing some marginal stability walls on the Coulomb branch.  Such wall crossing behavior is very important for 
the consistency of the solution \cite{Seiberg:1994rs, Seiberg:1994aj}.  The Seiberg-Witten solution gives us the mass formula for the
BPS particle, but it does not teach us explicitly the  BPS spectrum at a given  point 
on the Coulomb branch. Therefore, the BPS spectrum is only found for very few examples
in the early days, say  $SU(2)$ with $N_f\leq 4$ and $SU(2)$ with massive
adjoint \cite{Ferrari:1996sv, Bilal:1996sk, Bilal:1997st, Ferrari:1997gu}. 

Quite recently, Mathematician proposed a remarkable wall crossing formula
which constructed an invariant quantity from the BPS spectrum \cite{Kontsevich:2008fj, Joyce:arXiv0810.5645,Kontsevich:2010px}. Physical understandings of the wall crossing formula using the Hyperkahler metric of the Coulomb branch of the
corresponding three dimensional theory  are 
given in \cite{Gaiotto:2008cd}. However, the formula 
itself does not give us the answer of the BPS spectrum of a given $\mathcal{N}=2$ theory, and
finding the BPS spectrum is still a very difficult problem even with lots of exciting development in the
past few years \cite{Denef:2007vg, Gaiotto:2009hg,Cecotti:2009uf,Dimofte:2009bv,Cecotti:2010fi, Chuang:2010ii, Gaiotto:2010be,Gaiotto:2011tf,  Alim:2011ae,Alim:2011kw, Gaiotto:2012rg, Gaiotto:2012db}, 
in particular, very few information is known for the higher rank theory. The main purpose of this 
paper is to find the explicit BPS spectrum for a large class of higher rank theories. 

We are going to focus on a particular class of theory called theory of 
class ${\cal S}$ which can be engineered by compactifying six dimensional $(2,0)$ theory $A_{N-1}$ theory
on a Riemann surface with regular and irregular singularities \cite{Gaiotto:2009we,Gaiotto:2009hg,Xie:2012hs}. This class is very huge, for 
example, it includes  generalized superconformal quiver gauge theory\cite{Gaiotto:2009we,Nanopoulos:2010ga}, general Argyres-Douglas
type theories \cite{Argyres:1995jj,Argyres:1995xn,Xie:2012hs}, and lots of new asymptotical free theories \cite{Xie:2012hs}. Most of theories in this class are strongly coupled, but 
the remarkable geometrical construction can tell us many properties of them including superconformal index \cite{Gadde:2011uv},  3d mirror \cite{Benini:2010uu}, etc, and this class is a 
golden arena for studying the dynamics of the quantum field theory. 

In this paper, we would like to study the BPS spectrum of all kinds of  theories from class ${\cal S}$. Previous studies of BPS spectrum and wall crossing mainly focused on the $A_1$ theory, and the geometrical approach using the flow lines on the Riemann surface \cite{Gaiotto:2009hg} and the equivalent quiver 
approach \cite{Cecotti:2011rv, Alim:2011ae,Alim:2011kw} have the tremendous success in understanding the BPS spectrum of these theories, however 
very little is known for the higher rank cases \footnote{One chamber for $T_3$ theory and one single gauge group with various fundamentals are worked out in \cite{Alim:2011kw}, and some
weakly coupled chambers of pure $SU(N)$ theory are studied in \cite{ Chen:2011gk};
Two chambers for the sphere with one type of irregular singularity is worked out in \cite{Cecotti:2010fi}, and this class  is also considered in \cite{Gaiotto:2012db}.}.
 It seems that these higher rank theories are much more difficult 
than the $A_1$ theory since the underlying combinatorics is much harder.

\begin{figure}[htbp]
\small
\centering
\includegraphics[width=10cm]{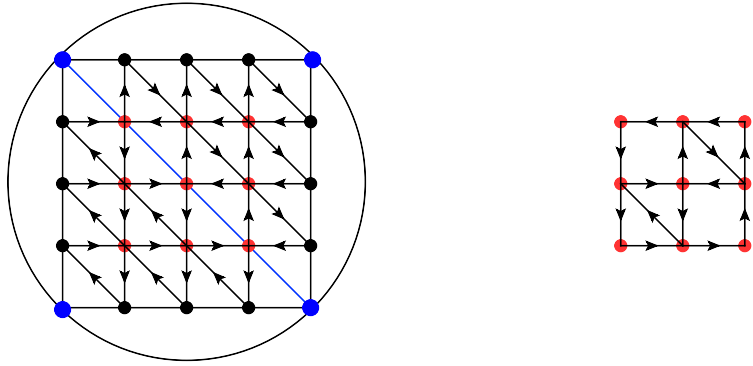}
\caption{The triangulation of fourth punctured disc of $A_3$ theory which represents the $(A_3, A_3)$ Argyres-Douglas theory. 
The quiver is shown on the right. }
\label{intro1}
\end{figure}

Our results presented in this paper  show that the finite chamber of the higher rank theory in this class can be easily worked out 
by combining the geometric construction and the quiver approach: we can find explicitly the charges and 
the order of phases of these particles. To achieve this goal, we have the following two 
main assumptions:

1. The triangulation and the network from the corresponding Riemann surface constructed in \cite{fock-2003,Xie:2012dw,Xie:2012jd} give
the BPS quiver of the theory, and the potential of the quiver is also given, see figure. \ref{intro1}.

2. The Donadson-Thomas invariant from the quiver with potential encodes the BPS spectrum, and the 
factorization depends on the $\theta$ stability condition of the quiver representation theory \cite{1232.53072}. 

\begin{figure}[htbp]
\small
\centering
\includegraphics[width=14cm]{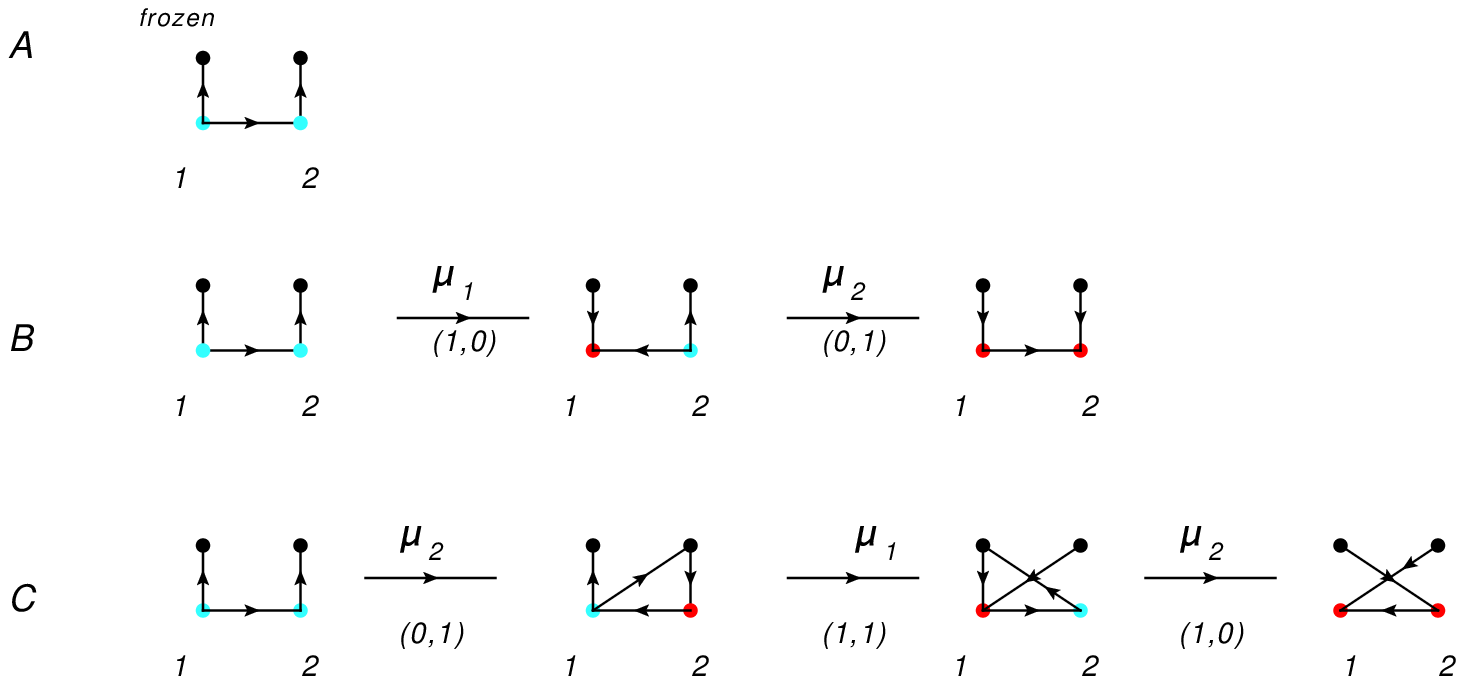}
\caption{A: The extended quiver derived by adding extra frozen node. B: A maximal green mutation sequence and the charge is indicated for each green mutation. C:
A different maximal green mutation sequence.  }
\label{intro2}
\end{figure}

The Donadson-Thomas invariant is in fact very complicated for the general quivers, and it is hard to extract 
the spectrum even the invariant is found. However, for the finite chamber, the BPS spectrum can be found using a remarkable combinatorial 
method called maximal green mutation introduced by Keller \cite{Keller:arXiv1102.4148} (which is equivalent to the mutation method proposed in \cite{Alim:2011kw} and is used implicitly in \cite{Gaiotto:2009hg}.). 
Basically, green mutation is defined by first extending the quivers by adding an extra frozen quiver node to 
each node, see figure. \ref{intro2}a. Now each node is defined as green (red) if it is the source (sink) of the frozen node. 
The \textbf{green} mutation sequence is defined as mutating only the green nodes, and the \textbf{maximal green mutation} 
is defined as the green mutation sequences such that no green node is left, see figure. \ref{intro2}b. The road map for finding the finite chamber is 
\begin{mydef3}
 Extend the BPS quiver by adding a frozen node to each original quiver node.
\end{mydef3}

\begin{mydef3}
Define a charge vector $\gamma_i$ for  the frozen nodes satisfying $<\gamma_i,\gamma_j>=\epsilon_{ij}$ with $\epsilon_{ij}$ the antisymmetric matrix for the quiver, 
 then each green mutation on a node $k$ probes a BPS hypermultiplet with charge 
$\alpha=\sum_i n_i\gamma_i$ where $n_i$ is the number of arrows from node $k$ to the $i$th frozen node.
\end{mydef3}

\begin{mydef3}
Find the maximal green mutation sequences such that no green node is left.
\end{mydef3}

The green mutation sequences give the phase order and the charge vector automatically. In practice, 
the Java program in \cite{Keller:2012} developed by Keller is extremely useful for us to find the maximal green mutation sequences for the complicated quiver. 

Now for each maximal green mutation sequences $\bold{k}=(k_1,\ldots, k_s)$, one can associate a quantum dilogarithm product \cite{Keller:arXiv1102.4148}
\begin{equation}
E(\bold{k})=E(X^{\alpha_1})\ldots E(X^{\alpha_s}),
\end{equation}
here $E(x)$ is the familiar quantum dilogarithm function and $X^{\alpha}$ is operator satisfying the noncommutative 
relation
\begin{equation}
 X^{\alpha}X^{\beta}=q^{1/2<\alpha,\beta>}X^{\alpha+\beta},
\end{equation}
with $<\alpha,\beta>$ the familiar Dirac product of two charges. If there is another sequences $\bold{k^{'}}=(k_1^{'},\ldots, k_r^{'})$ which represents the spectrum in another chamber, then the 
wall crossing formula is the following quantum dilogarithm identity $ E(\bold{k})=E(\bold{k^{'}})$. From two chambers shown in figure. \ref{intro2}, we have
\begin{equation}
E(X_1)E(X_2)=E(X_2)E(q^{-{1\over2}}X_1X_2)E(X_1),
\end{equation}
 which is the basic quantum dilogarithm identity found in \cite{Faddeev:1993rs}.
 
The above combinatorial method is very useful and in principle one could try all possible green mutations, but in practice
it has very limited uses if one does not know the mutation structure of the quiver. The  geometric 
picture of the triangulation of the bordered Riemann surface from which the quiver is derived turns out to be very useful 
in finding maximal green mutation sequence. 
We are going to give many concrete examples showing explicitly the  mutation 
sequences. In some cases, the result is very elegant, for example, we find that the minimal chamber of $T_N$ theory has the following number of states
\begin{equation}
N_{min}=2N(N-1)^2,
\end{equation}   
when $N=5$, the minimal chamber has 160 states, and we can find the charge vectors and the phase order explicitly which is very hard to derive
without knowing some structures of the quivers!

Another use of our result is that we can write down the spectral generator \cite{Gaiotto:2009hg} pretty easily for the higher rank theory since the final cluster coordinates (identified with the spectral generator)
associated with the quiver nodes can be derived given the mutation sequence. A remarkable feature of maximal 
green mutation sequence is that it tracks the permutation of the quiver nodes, so we can identify the final cluster 
coordinates of the original quiver node, which do not depend on the specific chamber. 

This paper is organized as follows: Section 2 reviews some backgrounds about the BPS particles, quivers, and cluster
algebra; Section 3 discusses how to use the quiver representation theory and the $\theta$ stability condition to find the 
BPS spectrum;  Section 4 discusses using the combinatorial tools called maximal green mutation to find the 
finite chamber; Section 5 and section 6  discuss the finite chamber of the $A_1$ theory and $A_{N-1}$ theory with many examples; Section 7 shows 
how to write the spectrum generator for the higher rank theory;
Section 8 discusses briefly the chamber with vector multiplets;  We give a short conclusion in section 9.

\section{Review}
\subsection{Generality about wall crossing}
The exact solution of Coulomb branch of  four dimensional $\mathcal{N}=2$  theory is solved  by Seiberg and Witten 
in \cite{Seiberg:1994rs, Seiberg:1994aj}.  Let's take pure $SU(2)$ theory as an example in which the gauge group is broken to $U(1)$ at a  generic point of Coulomb branch, 
and there are singularities on the Coulomb branch  where extra monopoles or dyons become massless. 

\begin{figure}[htbp]
\small
\centering
\includegraphics[width=10cm]{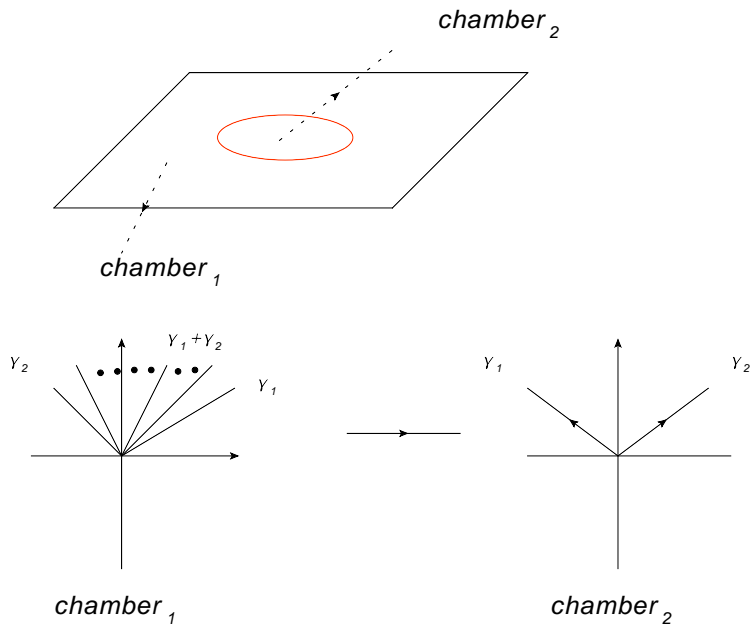}
\caption{Top: The marginal stability wall of pure $SU(2)$ theory. Bottom: The BPS spectrum in two chambers, and there are infinite number of states
in one chamber and finite states in another chamber.}
\label{wall}
\end{figure}

The $\mathcal{N}=2$ supersymmetry algebra allows 
a central charge extension, and the central charge of a BPS particle with charge vector $\gamma=(n_e,n_m)$ is described by
\begin{equation}
Z_{\gamma}(u)=n_ea(u)+n_ma_D(u),
\end{equation}
where $a$ is the scalar component in the vector multiplet and $a_D$ is the dual variable, and the central charge
depends on the coordinate  $u$ of the Coulomb branch. The BPS particle with charge $\gamma$ has mass $M(\gamma)=|Z(\gamma)|$.
As discussed in Seiberg and Witten's original paper, the wall crossing behavior of these BPS particles is important for the consistency of the solution. 
Basically, a BPS particle with charge $\gamma$ can decay to other BPS particles, say $\gamma_1$ and $\gamma_2$, in crossing the marginal stability wall.
Due to the BPS condition and charge conservation, this is only possible if  their central charges have the same phases: 
\begin{align}
&\gamma=\gamma_1+\gamma_2,  \nonumber\\
&\arg Z(\gamma)=\arg Z(\gamma_1)=\arg Z(\gamma_2).
\end{align} 
 For pure $SU(2)$ theory, there is only one marginal stability wall for all the particles, see figure. \ref{wall}. 
There are only two chambers: one  has finite number of hypermultiplets and the other chamber has a $W$ boson and an infinite number of dyons.
All of the particles but two from chamber $1$ become unstable in crossing the wall. 
The above wall crossing behavior for pure $SU(2)$ theory happens in the strongly coupled region. However, wall crossing behavior is not tied to the strongly coupled effect, 
and it can also happen in the weakly coupled region as discussed in \cite{Seiberg:1994aj}.  

The BPS particle for a general $\mathcal{N}=2$ theory 
has  charge vector $\gamma=(n_e^i,n_m^i, s^f)$, here $n_e^i$ and the $n_m^i$ are the electric and magnetic charge of the $i$th $U(1)$ gauge group, and 
$s^f$ is the flavor charge. Therefore the rank of the charge lattice is $R=2n_r+n_f$, where $n_r$ is the rank of the gauge group and 
$n_f$ is the number of mass parameters.  Two charge vectors have  a natural antisymmetric Dirac product 
\begin{equation}
<\gamma_1,\gamma_2>=n_e^ip_m^i-n_m^ip_e ^i;
\end{equation}
Note that the product does not depend on the flavor charge, so the rank of a matrix formed by an independent basis of BPS particles  is $2n_r$.
The central charge for a BPS particle with charge $\gamma$ is 
\begin{equation}
Z(\gamma)(u)=n_e^ia^i(u)+n_m^ia_D^i(u)+s_fm_f.
\end{equation}
This formula does not tell us which charge vectors are allowed as a possible BPS particles, and it also does not tell us which 
BPS particle is stable, so one need extra analysis to find the BPS spectrum.
By finding the BPS spectrum of a given $\mathcal{N}=2$ theory, we mean to find the  charge vectors and their order of phases of all the stable BPS particles. 
All BPS states in 3+1 dimensions have at least a half-hypermultiplet (i.e. a hypermultiplet without its CPT conjugate) of spin degrees of freedom.  The CPT conjugate has opposite
phase and same masses, so we will only consider half of the BPS particles in all later study.  

To count the number of BPS states, one can define a helicity supertrace  for a charge vector $\gamma$ ( see \cite{Kiritsis:1997gu} for details):
\begin{equation}
\Omega(\gamma, u)=-{1\over2}Tr_{{\cal H_{\gamma}}}(-1)^{2J_3}(2J_3)^2=(-1)^{2j}(2j +1).
\end{equation}
This index receives  contributions from BPS particles (short representation of the SUSY algebra), and
$\Omega(\gamma)=1$ for the hypermultiplet, and $\Omega(\gamma)=-2$ for the vector mulitplet.  
 To track the spin content, a refined index can be defined \cite{Gaiotto:2010be}:
\begin{equation}
\Omega(\gamma, y,  u)=Tr_{{\cal H_{\gamma}}} (-1)^{2J_3}y^{2J_3+2I_3}.
\end{equation}
The hypermultiplet contributes $\Omega(\gamma, y)=1$ and the value for the vector multiplet is $\Omega(\gamma, y)=y+y^{-1}$.  Since the BPS spectrum is 
only piece-wise constant across the Coulomb branch, the above index is not an invariant. 
Konsetvich and Soilbeman (KS) came up with a remarkable wall crossing formula which basically constructed a Donaldson-Thomas (DT) invariant  from the BPS spectrum.
They first associate a quantum torus algebra $e_{\gamma}$ on the
charge lattice 
\begin{equation}
<e_{\gamma_1}, e_{\gamma_2}>=(-1)^{<\gamma_1,\gamma_2>}<\gamma_1,\gamma_2>e_{\gamma_1+\gamma_2};
\end{equation}
and then define a group element for a BPS particle with charge $\gamma$:
\begin{equation}
U_\gamma=\exp({1\over n^2}e_{n\gamma}),
\end{equation}
and the invariant from the BPS spectrum are given by an ordered product:
\begin{equation}
A=\sum_{\gamma}U_\gamma^{\Omega(\gamma,u)}.
\end{equation}
The ordered product is taken over the particles with decreasing phases. The KS  wall crossing formula states that this product 
is independent of the BPS chamber!

The KS wall crossing formula is very beautiful. However, it is still not enough to find the BPS spectrum for 
a given $\mathcal{N}=2$ quantum field theory. To apply the wall crossing formula, one need to know at least the spectrum of one chamber and then 
apply the known wall crossing formula.  In practice, usually nothing is known for the BPS spectrum of a given theory, and even if 
we know the spectrum of one chamber, it seems hard to find other chambers using the wall crossing formula. 

The quiver approach, on the other hand, provides hope of solving the BPS spectrum in practice. The idea is to attach a unique quiver to a given 
$\mathcal{N}=2$ theory, and then use various tools attached to the quiver to study the BPS spectrum, since there are many 
wonderful properties about the quivers, the BPS counting problem is actually much easier.
The factorization of the DT invariant for a quiver with potential is an important class studied by Kontesvich and Soilbman \cite{Kontsevich:2008fj}. 
and the quiver approach to study the BPS spectrum has been used successfully in the early study of the wall crossing in the physics literature \cite{Douglas:2000ah,Denef:2002ru, Fiol:2006jz}.
In this paper, we will start with a quiver with potential for a large class of $\mathcal{N}=2$ theory, and since the BPS information is encoded in this quiver, it is  
called BPS quiver.

\subsection{BPS quiver for $\mathcal{N}=2$ theory}
 Let's review some background on BPS quiver which 
 could be understood directly from the spectrum at a given point on the Coulomb branch  \cite{Alim:2011kw}.
Given a  UV complete $\mathcal{N}=2$ field theory,  let's assume that we know explicitly the central charges 
and the stable BPS spectrum in one chamber, and each BPS particle can be represented by a ray in the complex plan. Since there is 
always an antiparticle whose phase is opposite to it, we only need to focus on a half plane. However, 
such choice is arbitrary which leads to many equivalent descriptions.

Let's take a half-plane $S_0$ and  all the BPS particles fall into this region. A \textbf{canonical basis} for these BPS states is defined as follows: the basis is chosen such that
the charge of any BPS state in this half plane is  expressed as a sum of this basis with positive integer coefficient
\begin{equation}
\gamma=\sum n_i\gamma_i,
\end{equation}
where $n_i$ is a non-negative integer. It is easy to show that such basis is unique (again using the positivity property). 
Now a BPS quiver can be formed by taking the Dirac product of this basis and forming an antisymmetric matrix
\begin{equation}
\epsilon_{ij}=<\gamma_i,\gamma_j>,
\end{equation}
which defines a quiver. From this definition, it is easy to see that the left and right-most of states should be included into this basis, since 
they can not be written as the sum of other charges with the positive coefficient. This basis is not always possible, and usually one need to 
turn on all the mass deformations of the theory and it turns out that a BPS quiver can be found for a large class of theories
considered in this paper, which will be discussed in more detail later.

\begin{figure}[htbp]
\small
\centering
\includegraphics[width=12cm]{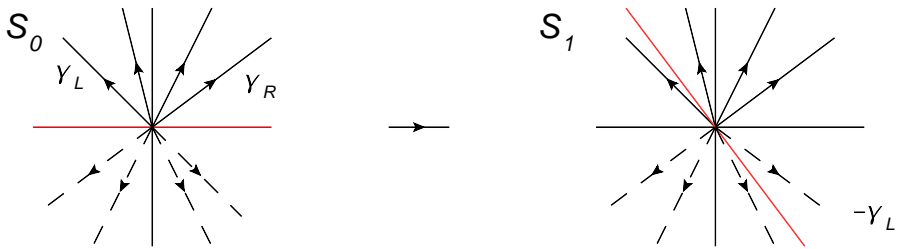}
\caption{By rotating the half plane, the canonical basis is changed. }
\label{rotation}
\end{figure}

The basis would be different if we choose a different half-plane $S_1$. If we slightly rotate the region $S_0$ clockwise and choose $S_1$ such that only the left-most
particle $\gamma_L$ drops out, then the charge vector $-\gamma_L$ which is in the right-most of $S_1$ has to be included in the new basis, see figure. \ref{rotation}, and 
the charge $\gamma_L$ is dropped out. Moreover, other charge vectors should also be changed, and the new basis is assumed to take the following form (see a proof in \cite{Alim:2011kw}): 
\begin{align}
&\gamma_{L}^{'}=-\gamma_L, \nonumber\\
&\gamma_{i}^{'}=\gamma_i+[\epsilon_{iL}]_+\gamma_L,
\end{align}
here $[\epsilon_{iL}]_+=max[0, \epsilon_{iL}]$. Now the new antisymmetric tensor built from the canonical basis is different and we have a new quiver, which 
is equally good to capture the BPS spectrum. 
So the BPS quiver is not just a single quiver but a family of quivers related by the above transformation on basis  called  quiver mutations (we will explain this term later), 
and by BPS quiver we really mean its mutation class.

The above definition of the BPS quiver is conceptually good but not useful for finding the BPS quiver since usually the BPS
spectrum is not known for any chamber. Other approaches are needed to attach a quiver to 
a given $\mathcal{N}=2$ theory.  For a large class of $\mathcal{N}=2$ theory engineered from compactifying six dimensional higher rank \footnote{The BPS quiver for $A_1$ theory is discussed in \cite{Gaiotto:2009hg, Cecotti:2011rv}.} $A_{N-1}$ theory
on a Riemann surface $\Sigma$ with defects, one could associate a quiver from the combinatorial data of the Riemann surface.
Our main conjecture in this paper is that 

\begin{mydef1}
 BPS quiver for these theories  is the one found
from the triangulation of a bordered Riemann surface as described  in \cite{fock-2003,Xie:2012jd,Xie:2012dw}. 
\end{mydef1}

There are many compelling evidence that this conjecture is true, i.e. the rank of the quiver is always equal to twice of 
the Coulomb branch dimensions, and the results are in agreement with the quiver found using other approaches like 2d-4d correspondence.

\subsection{Quiver mutation and cluster algebra}
\begin{figure}[htbp]
\small
\centering
\includegraphics[width=8cm]{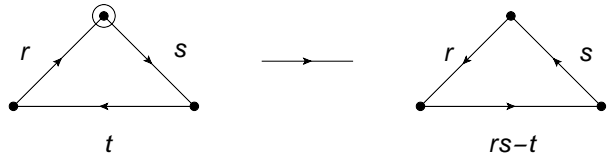}
\caption{The quiver mutation. }
\label{seiberg1}
\end{figure}

After finding a quiver, there are many tools one could use to study the BPS spectrum like the quiver representation theory, stability and quiver moduli space  \cite{Denef:2002ru}, etc, which
we will review in detail in next section. 
One could also attach new combinatorial structure called cluster algebra  \cite{Fomin2001} on the quiver which proves to be very powerful in BPS counting. There is 
a huge amount of literature on cluster algebra, here we only review some basic definitions which is sufficient for our purpose. The first element of the cluster
algebra is the quiver mutations, which is a combinatorial operation acting on quiver in following way:
\begin{equation}
\epsilon^{'}_{ij}=\left\{
\begin{array}{c l}
    -\epsilon_{ij}& if~i=k~or~j=k\\
    \epsilon_{ij}+sgn(\epsilon_{ik})[\epsilon_{ik}\epsilon_{kj}]_+ & otherwise
\end{array}\right.
\label{cluster1}
\end{equation}
Notice that this definition is the same as the change of the canonical basis shown in  last subsection.
The quiver mutations can be represented beautifully using the quiver diagram:  A quiver is a directed graph where multiple arrows between two vertices are allowed, which
is derived using $\epsilon_{ij}$ as follows: attach a quiver node for $i=1,\ldots n$, and there are $\epsilon_{ij}$ arrows between node $i$ and node $j$ \footnote{If $\epsilon_{ij}$ is positive, the quiver arrows are
pointing from node $i$ to node $j$; otherwise, they are pointing from node $j$ to node $i$.}.
The quiver mutation for a quiver without one and two cycles (such quiver is called 2-acyclic) is 
 defined as the following: Let Q be a quiver and k a vertex of Q. The mutation $\mu_k(Q)$ is the quiver obtained from Q as follows, see figure. \ref{seiberg1}:

1) for each sub quiver $i\rightarrow k\rightarrow j$, create a new arrow between $ij$ starting from $i$;

2) we reverse all arrows with source or target k;

3) we remove the arrows in a maximal set of pairwise disjoint 2-cycles. 
 
Formally, the quiver mutation is exactly like the Seiberg duality \cite{Seiberg:1994pq} for four 
dimensional $\mathcal{N}=1$ quiver gauge theory: the quarks are transformed to antiquarks and vice versa, and there is a new singlet for each meson field in the original quiver;
Finally the potential term is used to integrate out the massive fields. 
There are several obvious features about the quiver mutations: a. $\mu_k$ is invertible and $\mu_k^2=1$. b. If there are no quiver arrows between two quiver nodes $i$ and $j$, then $\mu_i$ and $\mu_j$ commute.

\subsubsection{Quiver with potential}
\begin{figure}[htbp]
\small
\centering
\includegraphics[width=10cm]{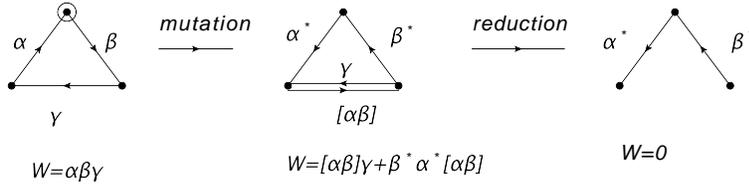}
\caption{The quiver mutation for the quiver with potential, and the final quiver is the reduced one. }
\label{seiberg}
\end{figure}

If there are oriented cycles in the quiver, one can define a potential term $W$ as familiar from the quiver gauge theory (a gauge invariant operator). 
The quiver mutation acting on the quiver itself is  the same form as the Seiberg duality, similarly, 
 the mutation action on the potential \cite{Derksen:arXiv0704.0649} is also the same as what is happening in the context of  Seiberg duality. If there is an oriented path $\ldots i\rightarrow^{\alpha} k \rightarrow^{\beta} j\ldots$ passing through
 the node $k$ under mutation, and 
 the potential involving this piece has the form
 \begin{equation}
 W=\ldots\alpha\beta\ldots+\ldots;
 \end{equation} 
 In doing the Seiberg duality, the bi-fundamental fields $\alpha$ and $\beta$ change the orientations, which are denoted
 as $\alpha^{*}$ and $\beta^{*}$ in the new quiver, and there is a new singlet bifundamental field $[\alpha\beta]$ between node $i$ and $j$. The potential changes in the following way:
  the  $\alpha\beta$ term in the original potential is replaced by the new field $[\alpha\beta]$, and there is an extra cubic potential
  term for the new quarks and the singlet:
   \begin{equation}
 W^{'}=\ldots[\alpha\beta]\ldots+\beta^{*}\alpha^{*}[\alpha\beta]+\ldots;
 \label{cluster2}
 \end{equation} 
 Now there might be a quadratic term in $W^{'}$ which means that there are two cycles in the new quiver, and the existence of 
 the potential can be used to integrate out these fields, and we get a reduced quiver $Q_{reduced}$ and  a  reduced potential $W_{reduced}$, see figure. \ref{seiberg}.
The quiver and potential after the mutation are always the reduced one!

\subsubsection{Cluster $X$ variable and cluster $A$ variable}
A complex variable $X_i$ can be defined on each quiver nod \footnote{In cluster algebra literature, this $X$ variable is called coefficient and denoted as $y$, and the 
cluster $A$ variable is denoted as $x$, here we follow the convention of Fock-Goncharov \cite{fock-2003}},
and its transformation behavior under the quiver mutation acting on node $k$ is
\begin{equation}
{X}_j^{'}=\left\{
\begin{array}{c l}
    {X}_k^{-1}& if~j=k\\
    {X}_j(1+{X}_k^{-sgn(\epsilon_{jk})})^{-\epsilon_{jk}} & if~j\neq k,
\end{array}\right.
\label{clusterX}
\end{equation}
namely, only the $X$ variable of the quiver nodes connected to node $k$ is changed. 

Similarly, another set of $A$ variable isdefined on  each quiver node, and the transformation behavior of the $A_k$ under the quiver mutations are
\begin{equation}
A^{'}_k= {\prod A_i^{[\epsilon_{ik}]_+}+\prod A_i^{[-\epsilon_{ik}]_+} \over A_k}
\label{clusterA}
\end{equation}
where $[\epsilon_{ik}]_+=max[\epsilon_{ik},0]$, and other $A$ variables are not changed.  There is an interesting duality between the $X$ and $A$ variables
\begin{equation}
X_i=\sum_j A_j^{\epsilon_{ij}}.
\end{equation}
Therefore cluster algebra is formed by a lot of seeds, and each seed comprises of quadruple $(\epsilon_{ij}, W, X_i A_i)$, and the seeds are related by 
the quiver mutation formula (\ref{cluster1},~\ref{cluster2},~ \ref{clusterX},~\ref{clusterA}). A degenerate two form can be defined on  $A$ space 
\begin{equation}
\omega=\epsilon_{ij}d\log A_i\wedge d \log A_j;
\end{equation}
and a poisson structure can be defined on the $X$ space:
\begin{equation}
\{X_i,X_j\}=\epsilon_{ij}X_iX_j.
\end{equation}
These structures are compatible with the cluster transformation, i.e. if you express $(X_i, \epsilon_{ij})$ in terms of $(X_i^{'}, \epsilon_{ij}^{'})$ using the 
cluster transformation rule, and you will get the same form expressed in terms of $(X_i^{'}, \epsilon_{ij}^{'})$. 

\section{Quiver representation theory}
\subsection{Acyclic quiver and BPS spectrum}

We review the idea of using quiver representation theory and $\theta$ stability condition to find the  BPS spectrum,  and this subsection
is mainly following \cite{1232.53072}.  The quiver considered in this section is assumed to be acyclic, i.e. there is no closed oriented path in the quiver.

Let's first discuss some backgrounds on the quiver representation theory and its moduli space. The content reviewed below is quite standard and 
more details could be found in many mathematical literature, i.e. the review by Reneke \cite{Reineke:arXiv0802.2147}.  Let's denote the quiver as $Q$, and 
$Q_0$ as the set of quiver nodes, $Q_1$ as the set of quiver arrows.
A representation $V$ of $Q$ consists of complex vector spaces $V_i$ for $i\in Q_0$ of dimension $d_i$, and 
of linear maps $V_{\alpha}:V_i\rightarrow V_j$ for every arrow $\alpha: i\rightarrow j$ in $Q$. 
Physically, a quiver representation can be thought of  as  assigning a $U(d_i)$ gauge group on each vertex and the linear map is the expectation value of the scalar in the bifundamental matter.
The homomorphism $\phi$ between two quiver representations are a set of linear maps 
\begin{equation}
\phi_i:V_i\rightarrow V_i^{'},
\end{equation}
which preserves the structure of the quiver representation, i.e $V_{\alpha}=\phi_iV^{'}_{\alpha}\phi_j^{-1}$. Again, when the dimension vectors of two representations are same, this is the familiar
gauge transformation acting on the bi-fundamental fields. The endomorphism of a quiver representation is the morphism between itself, and the set of all endomorphism of a representation is denoted as $End(V)$.
The automorphism is the  the endomorphism which is also invertible, and the set of all the automorphism of a representation is denoted as $Aut(V)$.

A representation $N$ is a \textbf{subrepresentation} of $M$  if $N_i\subset M_i$ for all the quiver nodes and the map $M_\alpha$ satisfies the condition $M_{\alpha}(N_i)\subset N_j$ for 
all the arrows.  A \textbf{simple} representation is the one whose sub representations are the zero and itself. The quiver representation with dimension vector $(0,\ldots, 1, 0,\ldots,0)$ is a simple representation, here
all the linear maps associated with the quiver arrows are trivial.
The direct sum of two representations $M\bigoplus N$ consists of the vector space $M_i\bigoplus N_i$ on each node, and the new
linear maps associated with the arrows are  $M_\alpha\bigoplus N_\alpha$ . 
The \textbf{indecomposable} representation is the one which can not be written as the direct sum of two representations. By definition, the simple representation is an indecomposable representation.

Every representations can be decomposed as a direct sum of the indecomposable representations, and
indecomposable representations are very important for the BPS counting problem. It is a very difficult problem to find all the \textbf{indecomposable} 
representations for a given quiver. However, it is 
possible to find the dimension vectors explicitly due to the Gabriel (Kac) theorem.

Quiver can be classified using the property of the  indecomposable representations of the quiver.
A quiver is of \textbf{finite} type if and only if the underlying undirect  graph
is of the ADE Dynkin type, and there are only finite number of indecomposable representations. For a Dynkin quiver Q, 
the dimension vectors of indecomposable representations do not depend on the orientation of the arrows in Q.
A quiver Q  is of \textbf{tame} type if and only if the underlying directed graph is 
an extended Dynkin graphs of type $\hat{A}$, $\hat{D}$, $\hat{E}$. The indecomposable representations of finite and tame quiver
are in one-to-one correspondence with the positive roots of the corresponding root system. 
All other quivers are called "wild" and the indecomposable representations are related to the roots of the quiver.

Let's review the Gabriel-Kac theorem in some detail. One can define a Euler form on the positive lattice  $\Lambda=Z_{+}^{Q_0}$:
\begin{equation}
\chi(\alpha,\beta)=\sum_{i\in Q_0}\alpha_i\beta_i-\sum_{\rho:i\rightarrow j}\alpha_i\beta_j.
\end{equation}
The Tits form is defined as $T(\alpha)=\chi(\alpha,\alpha)$. The antisymmetric form from the Euler form is defined as
\begin{equation}
<\alpha,\beta>=\chi(\beta,\alpha)-\chi(\alpha,\beta).
\end{equation}
Notice that $<e_i,e_j>$ count the number of arrows from $i$ to $j$ minus the number of arrows from $j$ to $i$ if 
$e_i=(0,\ldots, 1, 0,\ldots,0)$ with one on the $i$th quiver node.

The roots could be 
found explicitly just from the structure of the quivers. Let's denote $e_i=(0,\ldots, 1, 0,\ldots,0)$ as the simple roots and $I$ as the set
of all the simple roots, and 
define the reflection $s_i$ on lattice $Z^{Q_0}$:
\begin{equation}
s_i(d)=d-<e_i,d>e_i
\end{equation} 
The Weyl group $W(Q)$ is defined as the subgroup generated by $s_i$. The fundamental domain $F(Q)$ is defined as the set
of all non-zero dimension vector $d$ with connected support, i.e. the full sub quiver with nonzero $d_i$ is connected, such that 
$(e_i,d)\leq 0$ for all $i\in I$. The set of real roots are
\begin{equation}
\Delta^{re}(Q)=W(Q)I
\end{equation}
and the set of imaginary roots are
\begin{equation}
\Delta^{im}(Q)=W(Q)F(Q).
\end{equation}

There exists an indecomposable representation of $Q$ of dimension vector $d$ if and only if $d$ is a  positive root $d$, 
\begin{equation}
d=\sum_id_ie_i,~~~~~ d_i\geq0.
\end{equation}
In case $d\in \Delta^{re}(Q)$, there exists a unique indecomposable (up to isomorphism) of dimension vector $d$. In case $d\in\Delta^{Im}(Q)$, the number of parameters of 
the set of indecomposable representations is $1-\chi(d,d)$.  For the real roots $d$, we have $\chi(d,d)=1$.  Alternatively, the positive real  roots 
are the positive integer solution of the following equation $\chi(d,d)=1$, and the positive imaginary roots are the positive integer solutions of the 
quadratic equation $T(d,d)\leq n$ with $n\leq 0$.

From physics' perspective, the number of parameters of an indecomposable representation is equal to the dimension of the Higgs branch of the quiver gauge theory with gauge group $U(d_i)$: each arrow contributes $d_id_j$ and 
the dimension of the gauge group is $d_i^2$, and an overall $U(1)$ is decoupled, so the dimension of the Higgs branch (assume that all the gauge symmetry is broken) is
\begin{equation}
Dim(d)=\sum d_i d_j-\sum_i d_i^2+1=1-\chi(d,d).
\end{equation} 

Let's now give the identification between the possible BPS particles and the indecomposable  representations: 

\begin{mydef2}
 Each indecomposable representation \footnote{More precisely it is the Shur representation whose endmorphism is $End(V)=C$.} (up to isomorphism) with dimension vector $d$ represents a possible BPS particle, and the charge vector $\gamma$ of this state is the following
\begin{equation}
\gamma=\sum d_ie_i.
\end{equation}
Here $e_i$ is the dimension vector for the simple representation associated with the quiver node, which represent the elementary BPS particle.
\end{mydef2}

\begin{mydef2}
 The spin of the BPS particle is equal to $1-\chi(d,d)$. The real roots give the hypermultiplet and the imaginary roots give the higher spin states.
  \end{mydef2}

\subsubsection{Stability condition and stable BPS particle}
After identifying the possible BPS states with the indecomposable representation, it is time to study the stability condition from which  one can judge whether 
an indecomposable representation is stable or not.  The $\theta$ stability condition on quiver representation is the one which we
are going to use. Let's briefly review those concepts below.

Let's fix the dimension vector $d_i$ and denote the complex vector space at each quiver node $as$ $M_i$. Consider the following space 
\begin{equation}
R_d=\bigoplus _\alpha Hom(M_i,M_j),
\end{equation}
obviously each point of $R_d$ parametrizes a representation. The linear group $G_d=\prod_{i}GL(M_i)$ acts on $R_d$ via the following gauge 
transformation on an element $M_a \in Hom(M_i, M_j)$:
\begin{equation}
M_a \rightarrow g_jM_a g_i^{-1}.  
\end{equation}
Therefore each $G_d$ orbit parametrizes an isomorphism class of the quiver representation with dimension vector $d$. Physically each orbit of $G_d$ parametrizes 
all the gauge equivalent scalar field configuration ( the gauge group action is the complex one).  The moduli space $R_d/G_d$ parameterizes all the representations with dimension vector
$d$, however, this space is very complicated. The space is much more simpler if we concentrate on a subspace which covers almost all the quiver representation, this is 
the place where the stability condition plays an important role.

 The stability condition (central charge) $Z(d)$ for the quiver representation is a linear functional acting on  the lattice of dimension vectors which is generated by $Z(e_i)$ defined on 
 each vertex. Then for a representation with nonnegative dimension vectors $d$, the central charge is 
 \begin{equation}
 Z(d)=\sum_id_iZ(e_i).
 \end{equation}
 The slop of a representation is defined as $\mu(d)=\arg Z(d)$ \footnote{Usually the slop for a representation is defined as $\mu_{\theta}(d)={\theta_id_i\over \sum d_i}$, where $\theta_i$ is 
 defined on each quiver node, and this gives the name for the $\theta$ stability. Our definition is slightly different but actually equivalent.}, and the definition of the slop depends only on the dimension vector but not on the quiver arrows. The quiver information enters into
the characterization of the stability condition though.
We say a representation semistable (reps. stable) if for any proper subrepresentation \footnote{The zero representation and the representation itself is not included into the proper subrepresentation.} $N$, we have $\mu(N)\geq \mu(M)$
(resp. $(\mu(N)> \mu(M)$).   All the simple representations are stable since it has no proper subrepresentation, and all the stable representations 
are indecomposable, which can be seen as follows: if a representation $P$ is decomposable as $P=\sum_iM_i$ in which $M_i$ is the subrepresentation
of $M$, and the dimension vector is decomposed as $d=d_1+d_2+....+d_s$, then there is at least one of representation say $M$
whose slop is bigger than $P$, which implies the representation $P$ is unstable.

We denote $M_{\theta}^{ss}(Q,d)$ (resp $M_{\theta}^{s}(Q,d)$) as the moduli space of  semistable
(reps. stable) representations. It was shown in \cite{King:1994qz} that each point in $M_{\theta}^{s}(Q,d)$ parametrizes a solution to the $D$ term equations
modulo complex gauge group transformation of the quiver gauge theory:
\begin{equation}
\sum_{a:i\rightarrow *}\phi^{a+}\phi^{a}-\sum_{a: *\rightarrow i}\phi^{a}\phi^{a+}=\theta_i I_{d_i},
\end{equation} 
here $\theta_i$ is related to the stability condition, so $\theta_i$  is just the Fayet-Iliopolous (FI) term for the quiver gauge theory defined using the dimension vector $d$.  
$M_{\theta}^{s}(Q,d)$ simply parameterizes the moduli space of the gauge theory with FI terms turned on. 

Now we are coming to the third identification of the BPS spectrum and quiver representation from the $\theta$ stability condition:
\begin{mydef2}
The $\theta$ stability condition is the stability condition \footnote{More precisely, we only consider the discrete 
stability condition in this paper, which means that only one BPS particle is allowed for each slop.} for the BPS particles: the stable indecomposable 
representation represents the stable BPS particle.
\end{mydef2}
One can immediately derive some general features about the BPS spectrum:  first, there  are at least $n$ stable BPS particle since every 
simple representation associated with the quiver node is stable regardless of the chosen stability condition;  second,
 if there is a BPS particle with charge $\gamma$, then $k\gamma$ with $k\geq 2$ can not be the charge vectors for the BPS particle.

If we change the stability condition, and some of the stable representations would become unstable, which leads to the wall crossing behavior. 
The following two examples are  very suggestive to  us.  Consider the $A_2$ quiver which is the BPS quiver for $(A_1, A_2)$ Argyres-Douglas theory:
\begin{equation}
\bullet\rightarrow \bullet
\end{equation}
The simple representations $S_1$ and $S_2$ corresponding to two quiver nodes are indecomposable. The only other indecomposable representation $P$
is 
\begin{equation}
1\overset{1}{\rightarrow}1
\end{equation}
whose only subrepresentation is  $S_2$. If the stability condition is chosen such that $arg(Z(S_1))>arg (Z(S_2))$, then $P$ is not stable, and there
are only two stable BPS particles correspond to $S_1$ and $S_2$. If the stability condition is taken such that $arg(Z(S_1))<arg(Z(S_2))$, then
$P$ is stable and we have three stable BPS particles. This exhausts the two possible chambers for this theory. 
 \begin{figure}[htbp]
\small
\centering
\includegraphics[width=10cm]{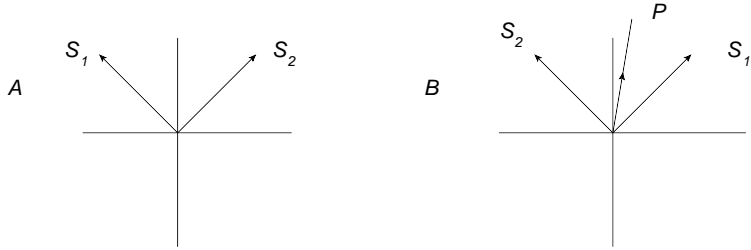}
\caption{The two chambers for $A_2$ quiver depending on different choices of the stability condition. }
\label{green}
\end{figure}

Let's consider the following affine $A_1$ quiver which is the BPS quiver for pure $SU(2)$ SYM:
\begin{equation}
\bullet \Rightarrow \bullet,
\end{equation}
The real roots for this quiver are $e_1$ and $e_2$, $f_n=(n,n+1)$ and $d_n=(n,n-1)$.  Unlike 
the previous examples, this quiver has an imaginary root $Q=(1,1)$ which represents 
the W boson.  These are all possible indecomposable 
representations for this quiver, and $e_2$ is the subrepresentation of $d_n, f_n$ and $Q$.

If the stability conditions are taken such that  $arg(Z(S_1))>arg (Z(S_2))$, then $S_1$ 
and $S_2$ corresponding to roots $e_1$ and $e_2$ are the only stable representations. If $arg(Z(S_1))<arg (Z(S_2))$, then all
the indecomposable representations are stable!
 \begin{figure}[htbp]
\small
\centering
\includegraphics[width=10cm]{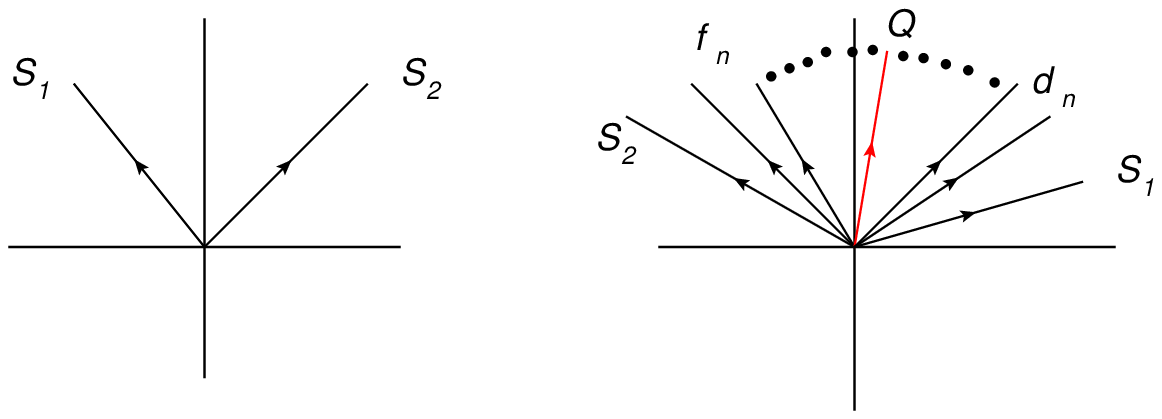}
\caption{ }
\label{green}
\end{figure}
 These two chambers also successfully recover the BPS spectrum of the two chambers of 
 the pure $SU(2)$ theory.
 
 This method can be  generalized to the other simple ADE quivers, however, it would 
 be a formidable problem for very complicated BPS quiver of general $\mathcal{N}=2$
 theory. In next section, we are going to use other combinatorial methods to deal with the BPS counting problem, but the quiver 
 representation theory is always conceptually useful.

\subsubsection{Quantum dilogarithm identity}
If there are two stability conditions  and two different BPS chamber, then what is the invariant constructed from them? Kontestvich-Soilbeman wall crossing formula provides 
an invariant, and Reneke derived the similar  formula using the Hall algebra and Harder-Narasimhan filtration \cite{Reineke:arXiv0802.2147}, see also the exploration of the quiver invariant 
and the BPS spectrum in physics literature \cite{Dimofte:2009tm, Kim:2011sc,Alexandrov:2011ac,Lee:2012sc, Lee:2012naa}.

Reneke's construction starts with a quantum algebra on the lattice $Z_{+}^{Q_0}$ for the quiver: one associate a quantum operator to each dimension vector, and they satisfy the condition:
\begin{equation}
y^\alpha y^\beta=q^{{1\over 2}<\alpha, \beta>}y^{\alpha+\beta},
\end{equation}
here $<\alpha, \beta>$ is the antisymmetric form we defined earlier.

The construction of the invariant uses the Hall algebra and Harder-Narasimhan filtration in an essential way.  Let's first review the Harder-Narasimhan filtration:
a filtration $0=X_0\subset X_1\subset \ldots \subset X_s=X$ of a representation is called Harder-Narasimhan (HN) if 
all the sub quotients $X_i/X_{i-1}$ is semistable and $\mu(X_1/X_0)>\mu(X_2/X_1)>\ldots > \mu(X_s/X_{s-1})$. Every 
reprŽsentation $X$ poses a unique HN filtration.

A Hall algebra can be defined on the isomorphic classes of quiver representations
\begin{equation}
[M].[N]=\sum_{[X]} F_{M,N}^{X} .[X]
\end{equation}
where $F_{M,N}^X$ denotes the number of sub representations $U$ of X which are isomorphic to $N$, with the 
quotient $X/U$ isomorphic to $M$. This coefficient is finite and the sum is also finite, and he dimension
of $[X]$ is equal to the sum of dimension $[M]$ and $[N]$.
Let's consider the following special elements in the Hall algebra
\begin{equation}
e_d=\sum_{dimM=d}[M],~~~~~~~~~e_d^{sst}=\sum_{dimM=d}^{M~semistable}[M],
\end{equation}
here $e_d$ contains all the quiver representation isomorphism class with dimension vector $d$, and $e_d^{sst}$
contains all the semi-stable isomorphism class with dimension vector $d$.
Now because of the uniqueness of HN filtration, $[M]$ appears with coefficient $1$ in the product
\begin{equation}
\prod e_{d_s}^{sst}.\ldots e_{d_2}^{sst} e_{d_1}^{sst}.
\end{equation}
and we have 
\begin{equation}
e_d=\sum_{*}e_{d_s}^{sst}\ldots e_{d_1}^{sst},
\end{equation}
where the sum is running over all decompositions $d_1+\ldots+d_s=d$ such that $\mu(d_1)>\ldots>\mu(d_s)$.  In particular, we have the following identity
\begin{equation}
\sum_d e_d=\prod_{\rightarrow} 1_\mu^{sst} 
\label{identity}
\end{equation}
where the right rand side is the ordered product on the semistable quiver representations, and $1_\mu^{sst} =1+e_\mu^{sst}$.
The left-hand side does not depend on 
the stability condition, so this identify shows that  there is an invariant for each stability condition.

Now use the evaluation map which maps an element from the Hall algebra to the power series in torus algebra
\begin{equation}
[M]\rightarrow {(q)^{1/2 \chi(\alpha,\alpha)} \over \prod (q^{-1})_{\alpha_i}}y^{\alpha}
\end{equation}
 where $(q^{-1})_n=\prod_{i=1}^n (1-q^{-i})$.  
 
Let's consider the case where all the stable representation correspond to the real roots, which implies $\chi(\alpha,\alpha)=-1$.  If ${M}$ is semistable, then all the power $M^n$ would be also semistable and  the charge vectors of $M^n$ is $n\alpha$ whose Tits form is  $\chi(n\alpha,n\alpha)=-n^2$.  Therefore, for a single stable representation, we have the following series in the torus algebra
  \begin{equation}
[1+\sum e_\mu^{sst}]=1+{q^{-1/2} \over \prod (q^{-1})_{\alpha_i}}y^{\alpha}+\ldots+{q^{-n^2/2} \over \prod (q^{-1})_{n\alpha_i}}y^{n\alpha}+\ldots
 \end{equation}
Let's consider $\alpha$ as the simple representation associated with a quiver node, then $\alpha=(0,\ldots,1,\ldots,0)$ and $y^\alpha=y_i$;  the above series 
becomes
\begin{equation}
E(y_i)=1+{q^{1\over 2}\over q-1}y_i+\ldots+{q^{n^2/2}y_i^n\over (q^n-1)(q^n-q)\ldots(q^n-q^{n-1})}+\ldots,
\end{equation}
which is the famous quantum dilogarithm function. In general for an stable indecomposable representation from the real root $\alpha$, one can associate 
a quantum dilogarithm function  $E(y^\alpha)$, and the identity (\ref{identity}) implies that 
\begin{equation}
E(y^{\alpha_1}) \ldots E(y^{\alpha_s}) =E(y^{\beta_1}) \ldots E(y^{\beta_r}),
\end{equation}
and the product is taken over the stable representation with the decreasing order of the slop, here $(\alpha_1, \ldots, \alpha_s)$ are the 
stable representations of on stability condition, and $(\beta_1, \ldots, \beta_r)$ are the stable representations of the other stability condition.
One could write a similar formula for the stability condition involving higher spin states.

\newpage
\subsection{Quiver with potential}
\begin{figure}[htbp]
\small
\centering
\includegraphics[width=10cm]{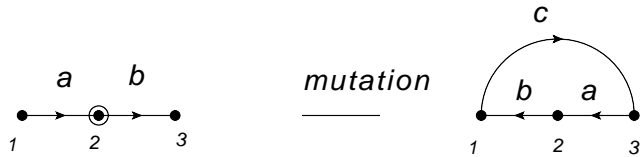}
\caption{$A_3$ quiver and affine $A_2$ quiver which are related by quiver mutations. }
\label{pots}
\end{figure}

Remember that the BPS quiver for a theory is really a class of quivers related by the quiver mutations. The quiver 
representation theory is definitely very different for the quivers related by the quiver mutations. For instance, consider
the $A_3$ quiver and the affine $\tilde{A}(3,0)$ quiver which are related by quiver mutations, see figure. \ref{pots}.  There are only finite number of indecomposable representations for the $A_3$
quiver, but there are infinite many for affine $\tilde{A}(3,0)$ quiver. 

To solve this problem ,  one need to add constraints to the affine quiver to kill many representations. 
In the original context of Seiberg duality, to match the moduli space, Seiberg introduced a new superpotential 
term to the quiver. In the same spirit, the addition of the potential will kill these extra representations.
In fact, it is possible to add a unique superpotential $W$ term to the affine $\tilde{A}(3,0)$ quiver. Now a representation 
of the quiver with potential should satisfy the extra condition
\begin{equation}
{\partial W \over \partial \phi_i}=0,
\end{equation}
for all the fields $\phi_i$ appearing in the potential. The general analysis of the indecomposable representations of the quiver with potential is quite complicated,
and we would like to introduce an algebraic approach.

To deal with this case, let's first introduce the path algebra associated with a quiver.
The path algebra $CQ$ is generated by the quiver arrows, moreover we need to add the length 
zero generator $e_i$  attached on the quiver node, so the elements for the path algebra are

$\bullet$ Path  $\phi_{ij}$ in the quiver going from $i$ to $j$.

$\bullet$  The length zero element $e_i$.

The product between two elements in the path algebra is very simple: the only nonzero products 
are $\phi_{ij}\phi_{jk}, e_i\phi_{ij}=\phi_{ij}, e_i^2=e_i$.   
For example, for the quiver in left of figure. \ref{pots}, the path algebra has
elements
\begin{equation}
[e_1, e_2, e_3, a, b, ab]
\end{equation}
The path algebra is finite if and only if there is no oriented cycle in the quiver.
The nice thing about the path algebra is that 
the category of quiver representation is the same as the category of $CQ$ left modules \footnote{
A left CQ-modulue consists an 
an abelian group $(M, +)$ and operation $CQ\times M\rightarrow M$ such that for all $r,s$ in $CQ$ and $x, y$ in $M$, we have
following condition 
\begin{eqnarray}
1. ~r(x+y)=rx+ry \nonumber\\
2. ~(r+s)x=rx+sx \nonumber\\
3. ~(rs)x=r(sx) \nonumber\\
4.~1.s=s
\end{eqnarray}
}.  One could also define the direct sum of two modules, submodule, and the indecomposable modules in a similar way 
as we did for the quiver representation theory. Similarly, one can define the stability conditions, etc.
The special module $P_i$ attached to a vertex plays an important role, here 
$P_i=(CQ)e_i$ consists all the paths ending at node $i$.  The nice thing about the $P_i$ module is 
that: the $P_i$ are projective modules, and every projective module is a direct sum of $P_i$. The dimension of 
the homomorphism $Hom(P_i, P_j)$  are the number of independent paths from node $i$ to node $j$.
$P_i$ is the left-module representing the indecomposable representation corresponding to the simple root.

Now if we add a potential to the quiver, the path algebra is modified and becomes the so-called Jacobi algebra. The potential will give 
zero relations which will kill some of the generators in the original path algebra. For example, consider the quiver 
on the right of figure. \ref{pots} in which there is a potential term  $W=abc$, and the $F$ term equation from the potential would be 
\begin{equation}
ab=0,~~bc=0,~~ca=0.
\label{zero}
\end{equation}
Now the Jacobi algebra is generated by 
\begin{equation}
[e_1, e_2, e_3, a, b, c],
\end{equation}
and it is finite dimensional. The rule for the product in the algebra is the same as the one defined for the path algebra, and they should satisfy the 
 relation in doing the product (\ref{zero}), say $ab=0$, etc. Notice that although the two quivers in figure. \ref{pots} are related by quiver mutation, the two 
 Jacobi algebra is not equivalent, which is natural since the quiver representation theory is not the same even with the inclusion of the 
 potential. They however define the same quiver invariant which is formed from the BPS spectrum. 
 
After defining the Jacobi algebra, one can similarly define the modules, direct sum of modules, and indecomposable modules, etc; 
Now the possible BPS states are represented by the indecomposable module of the Jacobi algebra, and 
the stability conditions can be similarly defined, everything is kind of similar.
The mathematical results about the representation theory of quiver with potential is fruitful and they play an important role in the studying 
of the BPS spectrum and wall crossing, see \cite{Kontsevich:2008fj, Kontsevich:2010px, Nagao:arXiv1103.2922,Mozgovoy:arXiv1103.2902}. However, the mathematical descriptions are quite complicated and we do not 
really use these descriptions in our later description, so we will not discuss them in details and leave 
it to other occasions in the future.

\section{Hypermultiplets: Maximal green mutation}
\subsection{Maximal green mutation: Definition}

The detailed analysis of the quiver representation theory is rather complicated, especially when 
the nontrivial potential exists. There is a remarkable combinatorial method called green mutation \cite{Keller:arXiv1102.4148}
which  will make the task of finding the finite chamber much easier.

Let's  first  modify the quiver as follows: introduce an extra frozen node \footnote{A frozen node can never be mutated, which could be though of as the flavor group.} 
 for each quiver node, and there is a quiver arrow pointing into the frozen node, see figure. \ref{green}A.  
The original quiver and the extended quiver are denoted as $Q$ and $\tilde{Q}$ respectively.
A non frozen node is called \textbf{green} if it is the source to the frozen nodes, and  called \textbf{red} if it is the sink to the frozen nodes. A green mutation sequence is 
the one where only green nodes can be mutated, and the mutation rule for the extended quiver is the same as the ordinary one.  
Several features of the green mutation is immediately clear from the definition:

 \begin{figure}[htbp]
\small
\centering
\includegraphics[width=14cm]{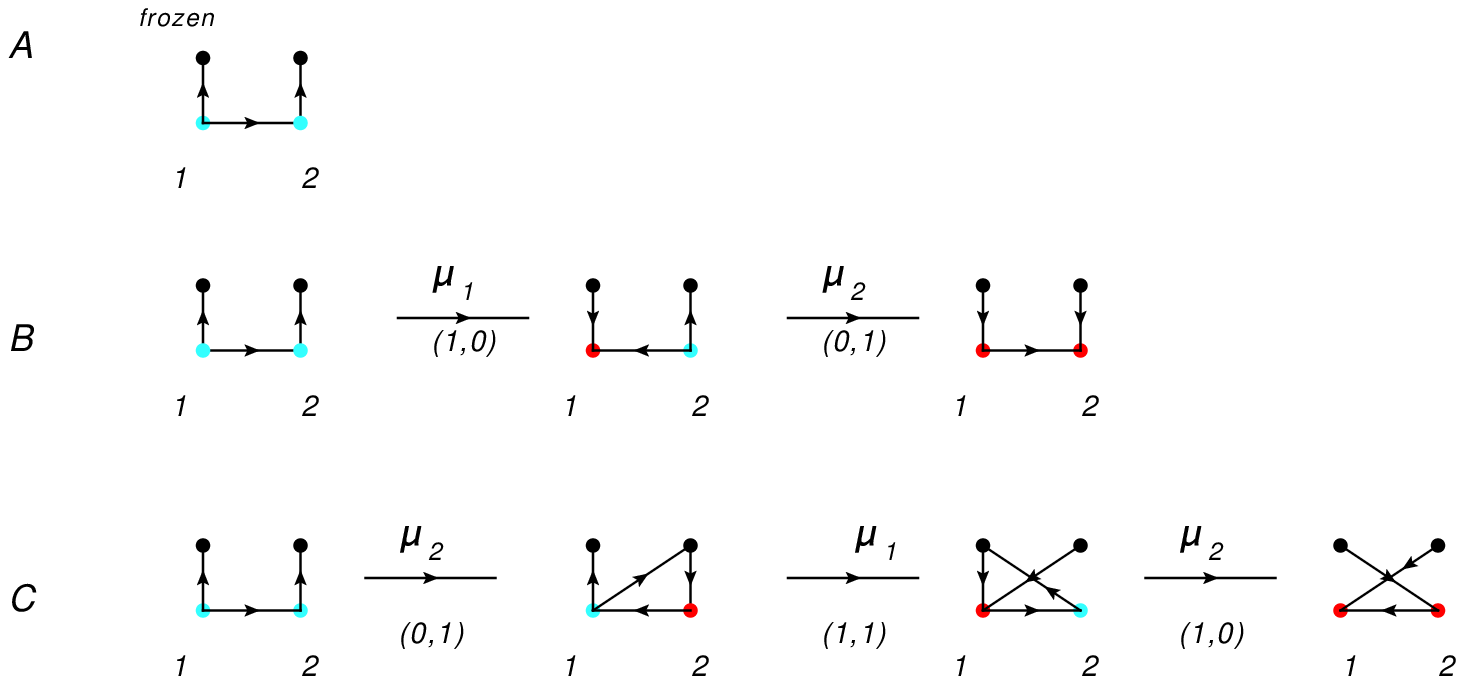}
\caption{A: The extended quiver derived by adding extra frozen nodes. B: A maximal green mutation sequences. C: Another maximal green mutation sequences of the same quiver. }
\label{green}
\end{figure}

$\bullet$ If we assign the charge vector $\gamma_i$ on each quiver node, and the Dirac product of these charges satisfies the condition
\begin{equation}
<\gamma_i,\gamma_j>=\epsilon_{ij},
\end{equation}
(this can be done explicitly by taking $\gamma_i=(0,\ldots,1,\ldots,0)$).
The charge vector of the quiver node during the quiver mutation is determined by the quiver arrows connected with the frozen nodes, i.e. 
the charge vector is 
\begin{equation}
\lambda=\sum m_i\lambda_i,
\end{equation}
where $m_i$ is positive (resp. negative) if the quiver node is the source (resp. sink) for the frozen nodes. For the green node, the subquiver formed by 
positive $m_i$ is connected.
Let's denote the charge vector as $\alpha$, then after the mutation on node $k$, the new charge vector is changed as follows:
\begin{align}
&\alpha_{k}^{'}=-\alpha_{k}, \nonumber\\
&\alpha_{i}^{'}=\alpha_i+<\alpha_i,\alpha_k>\alpha_k~~if~~~<\alpha_i,\alpha_k>0,
\end{align}
this formula can be checked by looking at the green mutation. The new quiver is formed by doing the Dirac product  using $\alpha^{'}$.

$\bullet$ Each node is either green or red at any step of the green mutation.  This can be seen as follows: let's mutate a green node $k$, by the rule 
of mutation, all the quiver arrows including the arrows to the frozen node are reversed, then it becomes red after the mutation. The color of the other
quiver nodes $i$ would not change if there is no quiver arrow to node $k$ or $i$ is the sink of the arrows between $i$ and $k$. If $i$ is the 
source of the arrows between $i$ and $k$, then we have two choices to consider: A. $i$ is green, then $i$ is also green after the mutation; B. $i$ is red, this case 
is a little bit complicated, but it can be proven that it is either green or red after we identify the charge vector as the $c$ vector appearing in the study of the cluster algebra.
So we have the following conclusion: The green node is still green after a green mutation on other nodes, and the red node can become either red or
green.

$\bullet$ If one green node with charge vector $m_i$ is mutated,  the quiver moduli space with the assignment of dimensions $m_i$ has 
dimension zero, and therefore this corresponds to a hypermultiplet!

Now let's introduce the definition of maximal green mutation sequence:
\begin{mydef4}
A maximal green mutation sequence is the finite green mutation sequence such that all the nodes are red at the end. It is not hard 
to see that the final quiver has basis $-\gamma_1,-\gamma_2,\ldots,$ and the quiver (non frozen part) is isomorphic to the original quiver.  
\end{mydef4}

Now each maximal green mutation sequence represents a chamber with finite number of  (hypermultiplets) states! This method is essentially equivalent to that proposed 
in \cite{Alim:2011kw} \footnote{In that paper, one only mutate the quiver nodes with positive coefficient of the original charges, which means only mutating the green node
in our language, moreover, each mutation corresponds to rotating the half-plane of the central charges, and when the plane is rotated by 180 degree, the canonical 
basis is $-\gamma_i$, which essentially means that no green node is left.}
, but the combinatorial way presented here make the calculation much easier.
 An example of maximal green mutations is shown in figure. \ref{green}B, and this chamber has two BPS particles with charge $\gamma_1$ and $\gamma_2$.

\begin{figure}[htbp]
\small
\centering
\includegraphics[width=6cm]{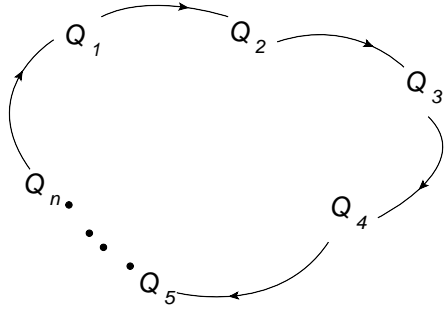}
\caption{The quivers connected by the green mutation. }
\label{quiverequivalence}
\end{figure}

In the above procedure, we have fixed a quiver from the very beginning, now according to our previous discussion on the BPS quiver, all the quivers appearing in this sequence are equally 
good for describing this particular BPS chamber, see the illustration in figure. \ref{quiverequivalence}. Let's denote this subset as $[Q]_G$, 
apparently only a subset of the quivers in the mutation equivalence class would appear, and this subset depends on a particular maximal green mutation sequence. 
Different green mutation sequence defines different subsets. If a quiver appears in $[Q]_G$, then there will be a maximal green mutation sequence with length $|G|$. 

 There are several observations about the feature of maximal green mutations:

1. The length of the maximal green mutations is at least $|Q|$, where $|Q|$ is the number of quiver nodes, and each 
node is mutated at least once. If there is a BPS particle with charge $\alpha_i$ in the spectrum, then the charge vector $k\alpha_i,~k\geq2$ will not appear. 
 Notice that this is consistent with the quiver representation theory by taking discrete stability condition.

2. The final quiver is isomorphic to the original quiver, and each red node is only connected to one frozen node.

\begin{figure}[htbp]
\small
\centering
\includegraphics[width=4cm]{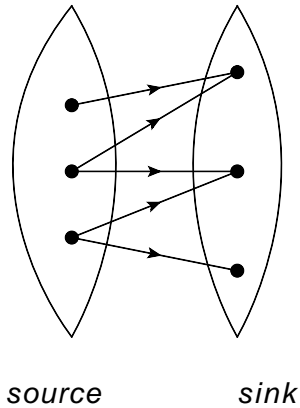}
\caption{The bipartite quiver has two types of quiver nodes: sink and source nodes. }
\label{source}
\end{figure}

\paragraph{Special quiver}
Let's consider a  bipartite quiver which has only two types of quiver nodes: source and sink (see figure. \ref{source}), and the following two special mutation sequences
 \begin{eqnarray}
 \mu_{+}=\mu_{i_1}\ldots\mu_{i_m},~~\mbox{$i_k$ is the source node} \nonumber\\
 \mu_{-}=\mu_{j_1}\ldots\mu_{j_n},~~\mbox{$j_k$ is the sink node} 
 \end{eqnarray}
The order of mutations in $\mu_{+}$ and $\mu_{-}$ does not matter since there are no arrows between the source nodes (the same is true for the sink nodes).
Due to the special structure of the quiver, there is no cycle in the quiver and therefore no potential is allowed. The mutation sequences $\tau=\mu_{+}^2$ is a
 maximal green mutation sequence as can be easily checked from the definition. 

The above observation can be generalized to an acyclic quiver which  always has at least a source node.  Let's do the green mutation 
on this source node, and  the new quiver is still acyclic and there is again a source node. Continuing mutating the source node, we 
are going to find the maximal mutation sequence with $|Q|$ steps.

The above mutation method only gives us the hypermultiplet. The vector multiplet does not correspond to the quiver mutation, but it can be taken as limit of 
a infinite number quiver mutation sequence. If there is only one vector multiplet,  one might be able to 
probe the existence of the vector multiplet by doing the \textbf{maximal red mutation}: one add a frozen node and quiver arrow for each quiver node, but the arrow 
is pointing into the quiver node. The red mutation sequence is the one which only red node is mutated. The interested reader can check that the  red 
mutation  has quite  similar property as green mutations.  
This red mutation sequence corresponds to rotating the half plane in counterclockwise direction. The infinite chamber with one vector multiplet can found as follows: first do the maximal green mutation and  
then do the red mutation carefully to make everything consistent. If these two sequences has a common limit, then we conclude that there is a vector multiplet.

\subsection{Finding maximal green mutation sequence: a clue}
There is one serious question about the use of the maximal green mutation: there is no information 
on the order of green mutations and random green mutations usually would not 
stop at finite steps.  The quiver representation theory and $\theta$ stability
do tell us something about the green mutation sequences, we have the following conjecture:

\begin{mydef1}
There are infinite number of BPS states if there are stable higher spin states.
\end{mydef1}

According to this conjecture, the stability condition should be chosen such that no higher spin stable state exists. 
For simplicity  let's assume that the vector multiplet is the possible higher spin states, and  denote the corresponding representation as $V_d$, and its 
subrepresentation as $V_1, V_2, \ldots V_s$. According to the $\theta$ stability, $V_d$ is stable if and only if its slop is smaller than all of its sub representations:
\begin{equation}
\mu(V_i)>\mu(V_d),~for~all~i,
\end{equation}
So to have a finite chamber, we must ensure that such situation will not happen in our mutation process.  

\begin{figure}[htbp]
\small
\centering
\includegraphics[width=12cm]{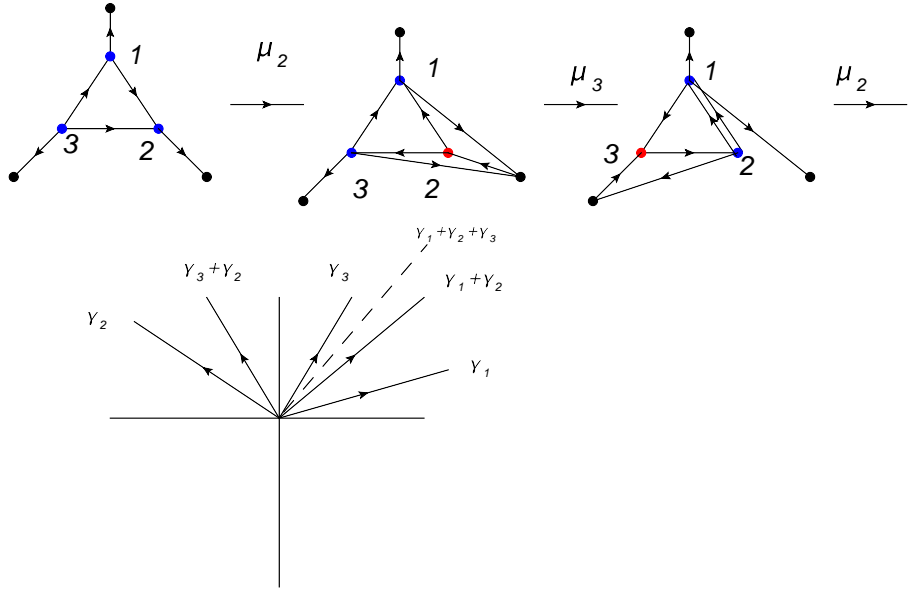}
\caption{ Up: the maximal green mutation sequences. Bottom: the charge vector for the various BPS particle, the dash line represent
the vector boson which is not stable, since the slop of its subrepresentation $\gamma_1+\gamma_2$ is smaller than it. }
\label{cluegreen}
\end{figure}

Let's look at an example to see how to use the stability condition to determine maximal green mutation sequences. The BPS quiver is the one  representing  $SU(2)$ theory with one flavor, see figure. \ref{cluegreen}. The gauge boson
corresponds to the representation $d=(1,1,1)$, and $S_2=(0,1,0)$, $P=(1,1,0)$ are the sub representations of it. Now, let's  
start doing mutation on node $2$ in step $1$, and probe a BPS particle with charge $S_2$ on the far left. The node one 
and node three are green nodes now. In this second step, if mutate node $1$ which has charge $\gamma_1+\gamma_2$ and probe 
the BPS particle $P$, then the vector multiplet is definitely stable since it would  have smaller slop than $S_2$ and $P$. So to find a 
finite chamber, we can only mutate node $1$. In step $3$, similar analysis forces us to mutate node $2$, etc. At the end, 
we find the following maximal mutation sequences
\begin{equation}
\mu_2,\mu_3,\mu_2,\mu_1,\mu_3.
\end{equation}
Similar analysis can be done on the mutation sequences starting from other nodes.
A simple but useful fact is that if in the mutated quiver $Q_1$, and there are two green nodes connected by double arrows, then 
we can not mutate the sink nodes of this subquiver which follows directly from the representation theory of affine $A_1$ quiver.  This  follows from 
the fact the mutated quiver is equally good for describing this particular chamber. Use this observation, it is easy to 
see that in step $3$, we can only mutate quiver node $2$.

\newpage
\subsection{Wall crossing: Quantum dilogarithm identity}
Usually there are more than one maximal green mutation sequences for a given quiver. It would 
be nice to have an invariant built from the BPS spectrum, which can be written down using the quiver representation theory. For the finite
chamber, it is actually very easy to write the invariant using the maximal green mutation data as shown in  \cite{Keller:arXiv1102.4148}. Let's fix a quiver and consider the quantum algebra on charge lattice, and the commutation relation is
\begin{equation}
X^{\alpha}X^{\beta}=q^{{1\over 2}<\alpha,\beta>}X^{\alpha+\beta},
\end{equation}
where $<\alpha, \beta>$ is the familiar Dirac product.

Assume that we find a maximal green mutation sequence $\bold{k}=(k_1,k_2,\ldots,k_s)$ and the charge vector in step $i$ is $\alpha_i$, 
we can form a quantum dilogarithm function for $i$th mutation:
\begin{equation}
E(X^{\alpha_i}).
\end{equation}
Given a maximal green mutation sequences, we can form a quantum dilogarithm product
\begin{equation}
E(k)=E(X^{\alpha_1})E(X^{\alpha_2})\ldots E(X^{\alpha_s}),
\end{equation}
If there is another maximal green mutation sequences $k^{'}$ whose length is $r$, then this represents another chamber and we can form another quantum dilogarithm product $E(k^{'})$,
then these two quantum dilogarithm products are the same 
\begin{equation}
E(k)=E(k^{'}).
\end{equation}
which could be interpreted as the wall crossing formula, which has been proved using quiver representation theory.

\textbf{Example}: Let's consider our familiar $A_2$ AD theory. There are two maximal green mutation as shown in figure. \ref{green}. The first chamber has two BPS hypermultiplets 
with charges $\gamma_1, \gamma_2$, and the other chamber has three BPS hypermultiplets with charge $\gamma_2, \gamma_1+\gamma_2,\gamma_1$ (the charges are listed in the order of decreasing phase angle).  The quantum dilogarithm
identity from this theory is
\begin{equation}
E(X^{\gamma_1})E(X^{\gamma_2})=E(X^{\gamma_2})E(X^{\gamma_1+\gamma_2})E(X^{\gamma_1}).
\end{equation}
The above identity has the familiar form if we use the generator $X_i=X^{\gamma_i}$:
\begin{equation}
E(X_1)E(X_2)=E(X_2)E(q^{-{1\over2}}X_1X_2)E(X_1).
\end{equation}

\subsection{Quantum dilogarithm identity from quantum cluster algebra}
In this part, a proof of the quantum dilogarithm identity is given using the quantum cluster algebra.

\subsubsection{Charge vector as $c$ vector}
The charge vector appeared in previous section has a nice interpretation from the cluster algebra. 
Let's first introduce some background on tropical semi-field. 
Assume the semi field is generated by generators $y_i, i=1,\ldots n$, and each element has the form $\prod y_i^{a_i}$,  where $a_i$ is an integer.
The tropical sum is defined as 
\begin{equation}
\prod y_i^{a_i}+\prod y_i^{b_i}=\prod y_i^{min(a_i,b_i)}.
\end{equation}

\begin{figure}[htbp]
\small
\centering
\includegraphics[width=16cm]{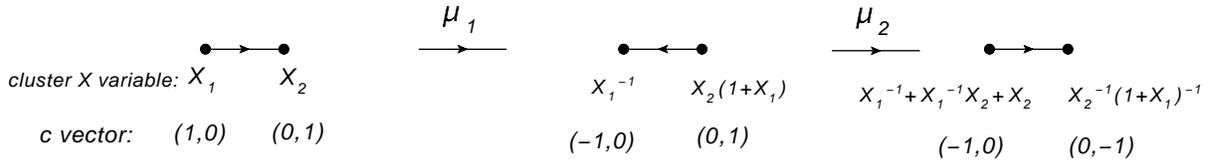}
\caption{The $X$ variable and $c$ vectors of the mutations.  }
\label{sign}
\end{figure}

Given a quiver and consider the mutation rule of the cluster $X$ coordinates, 
\begin{eqnarray}
X_k^{'}= X_k^{-1} \nonumber\\
X_i^{'}=X_i(1+X_k^{-sgn(\epsilon_{ik})})^{-\epsilon_{ik}}. 
\end{eqnarray}
If we replace the ordinary sum in the new $X$ variables with the tropical sum, then the cluster coordinates has the following form
\begin{equation}
[X]=\prod X_i^{c_i},
\end{equation} 
where $c_i$ is an integer vector whose entries are all nonpositive or nonnegative \cite{ Fomin:math0602259}, and $X_i$ is the original cluster variable.  These $c$ vectors are not the new stuff, and they are just  the charge vector 
appearing in the study of green mutation.  The initial $c$ vector has only one entry $1$ with all the other entries zero, and the final $c$ vector has only one entry $-1$, see an example in figure. \ref{sign}.

The $c$ vector has the following simple transformation rule from the definition if a mutation is done on the vertex $k$:
\begin{align}
&c_k^{'}=-c_k \nonumber\\
&c_i^{'}=c_i~~~~if~\epsilon_{ik}c_k\leq 0 \nonumber\\
&c_i^{'}=c_i+\epsilon_{ik}c_k~~~~if~\epsilon_{ik}c_k>0 \nonumber\\
\end{align}
Since we always mutate on green nodes which means the $sign(c_k)$ is always positive, the above formula simplifies as 
\begin{align}
&c_k^{'}=-c_k \nonumber\\
&c_i^{'}=c_i~~~~if~\epsilon_{ik}< 0 \nonumber\\
&c_i^{'}=c_i+\epsilon_{ik}c_k~~~~if~\epsilon_{ik}>0 \nonumber\\
\end{align}

Let's now use the $c$ vector analysis to prove the maximal green sequences for the bipartite quiver.  Let's consider one of the source node whose initial c vector is 
$(1,0,\ldots,0)$, after the mutation $\mu_{+}$, $c^{'}=(-1,0,\ldots,0)$. The $c$ vector of the sink nodes is not changed. After mutation $\mu_{-}$ on the sink node, all the $c$ vector of 
the sink nodes becomes $c^{'}=(0,\ldots,-1,0)$, and the final quiver is a quiver whose nodes are all red.

\subsubsection{Quantum cluster algebra}
We have actually seen the deformations in defining a quantum torus algebra on the charge lattice, and the cluster algebra has a quantum deformation too \cite{Fock:math0311245,Fock:math0702397}, here we only use the quantum version of 
the cluster $X$ variable, and keep the $A$ variable as classical.
There is a quantum $X$ variable associated with each vertex and the noncommutative relations are
\begin{equation}
X_{i}X_{j}=q^{\epsilon_{ij}}X_{i}X_{j},
\end{equation} 
and the transformation rules for the quantum cluster variables under the mutation on node $k$ are
\begin{align}
& X_k\rightarrow X_k^{-1} \nonumber\\
& X_i\rightarrow X_i(\prod_{a=1}^{|\epsilon_{ik}|}(1+q^{a-1/2}X_k^{-sgn(\epsilon_{ik})}))^{-sgn(\epsilon_{ik})},
\end{align}
this transformation preserves the form of the commutation relation $X_{\alpha}^{'}X_{\beta}^{'}=q^{\epsilon^{'}_{\alpha\beta}}X_{\beta}^{'}X_{\alpha}^{'}$.  The mutation can be 
decomposed into two steps $\mu_k=\mu_k^{'}*\tau_{k,+}$, where $\tau_{k,+}$ is defined as the transformation
\begin{align}
&X_k^{'}=X_k^{-1} \nonumber \\
&X_i^{'}=q^{1/2\epsilon_{ik}[\epsilon_{ik}]_{+}}X_iX_k^{[\epsilon_{ik}]_{+}}~if~i\neq k.
\end{align}
Here $[\epsilon_{ik}]_{+}=max(0,\epsilon_{ik})$, and the transformation $\mu_k^{'}$ is given  by the adjoint action $Ad(\Psi_q(X_k))$:
\begin{align}
&X_i^{'}=Ad(\Psi_q(X_k))X_i=\Psi_q(X_k)X_i^{'}\Psi_q(X_k)^{-1}=X_i\Psi_q(q^{\epsilon_{ik}}X_k)\Psi_q(X_k)^{-1}= \nonumber\\
&~~X_i(\prod_{a=1}^{|\epsilon_{ik}|}(1+q^{-sgn(\epsilon_{ik})(a-1/2)}X_k))^{\epsilon_{ik}}.
\end{align}
 here $\Psi_q$ is the familiar quantum dilogarithm function. What is important is that there are another kind of decomposition using $\mu_{k,-}^{'}$ and $\tau_{k,-}$, where $\tau_{k,-}$
is defined by replacing $\epsilon_{ik}$ by $-\epsilon_{ik}$, and $\mu_{k,-}^{'}$ is defined by using the adjoint action
\begin{align}
&Ad(\Psi_q(X_k^{-1})^{-1})X_i=\Psi_q(X_k^{-1})^{-1}X_i^{'}\Psi_q(X_k^{-1})= \nonumber\\
&~~X_i(\prod_{a=1}^{|\epsilon_{ik}|}(1+q^{sgn(\epsilon_{ik})(a-1/2)}X_k^{-1})^{-\epsilon_{ik}}.
\end{align}
It can be checked that  $\mu_{k,-}^{'}*\tau_{k,-}=\mu_{k,+}^{'}*\tau_{k,+}$.

Here comes the crucial point: if there is a sequence of mutations $(\mu_1,\mu_2,\ldots, \mu_s)$ 
such that the final cluster coordinates are the same to the original one up to the permutation, there is a quantum dilogarithm identity associated with 
this sequence. If the $c$ vector is $(\alpha_1, \alpha_2, \ldots, \alpha_s)$ and 
denote the sign of the $c$ vector as $(\epsilon_1,\epsilon_2, \ldots, \epsilon_s)$ \footnote{Each $c$ vector is either all nonnegative or nonpositive, $\epsilon$ is 
the sign of the $c$}, then the quantum dilogarithm identity is 
\begin{equation}
\Psi_q(X^{\epsilon_1\alpha_1})^{\epsilon_1}\ldots\Psi_q(X^{\epsilon_s\alpha_s})^{\epsilon_s}=1.
\end{equation}

The proof is the following \cite{Kashaev:arXiv1104.4630}: since the cluster coordinates come back to itself up to permutation, we have the following identity
\begin{equation}
Ad(\Psi_q(X_{1}^{\epsilon_1})^{\epsilon_1})\tau_{1,\epsilon_1}\ldots Ad(\Psi_q(X_{s}^{\epsilon_s})^{\epsilon_s})\tau_{s,\epsilon_s} \nu=1
\end{equation}
Now move all the $\tau_{i, \epsilon_i}$ to the far left, and we get 
\begin{equation}
Ad(\Psi_q(X^{\epsilon_1\alpha_1})^{\epsilon_1})\ldots Ad(\Psi_q(X^{\epsilon_s\alpha_s})^{\epsilon_s})\tau_{1,\epsilon_1}\ldots \tau_{s,\epsilon_s} \nu=1
\end{equation}
Using the relation $\tau_{1,\epsilon_1}\ldots \tau_{s,\epsilon_s}\nu=1$, we get the quantum dilogarithm identity in the desired form.

The relation to the wall crossing formula is the following: the sign of the $c$ vector are assembled into two groups, say $\alpha_1, \alpha_2,\ldots, \alpha_r$ has positive sign
and the remaining ones has negative sign, we have
\begin{equation}
\Psi_q(X^{\alpha_1})\ldots \Psi_q(X^{\alpha_r})=\Psi_q(X^{-\alpha_{r+1}})\ldots\Psi_q(X^{-\alpha_s}),
\end{equation}
which gives the wall crossing formula for the $A_2$ quiver.

\section{BPS spectrum for $A_1$ theory}
Although the above mutation method is remarkably powerful, this approach has severe limitations without further knowledge of the quiver mutation structure. 
Without such knowledge, the above method is also kind of largely constrained to the ADE  quiver class or its special generalization. The reason for the difficulty is due to  
the following two necessary conditions for doing green mutations:

1. A quiver with $2n_r+n_f$ quiver nodes for a given $\mathcal{N}=2$ theory, here $n_r$ is the rank of the gauge group at the generic point of Coulomb branch and $n_f$ is 
the number of mass deformations.

2. A mutation sequence whose final quiver is isomorphic to the original one.

The first condition is by itself already highly non-trivial since usually there is no information about the BPS spectrum for most of theories. 
Our first claim is that the quivers and potential constructed in \cite{Xie:2012jd,Xie:2012dw} are the BPS quiver for the corresponding field theory.
The second condition is even more difficult since the quiver for higher rank theory is of mutation infinite class, and it is very difficult to find a mutation sequence whose final quiver is isomorphic to the original one. 
Luckily, the construction given in \cite{Xie:2012jd,Xie:2012dw} also gives us the above very needed mutation sequence, which we will review in more detail in later sections. 

In this section, we will mainly study the theory engineered using six dimensional $A_1$ theory which has been discussed in \cite{Gaiotto:2010be,Cecotti:2011rv, Alim:2011ae,Alim:2011kw,Saidi:2012pf}. The study will be very 
useful for our later applications  to the higher rank theory. Moreover, Using our method, it is pretty easy 
to recover the results in the literature, and in fact we can get a lot more finite BPS spectrums starting with arbitrary quivers, which is new.

\subsection{Definition of the theory and BPS geometry}
In this section, we will consider  four dimensional $\mathcal{N}=2$ theories derived by compactifying six dimensional  $(2,0)$ 
 $A_1$ theory on a Riemann surface with regular singularity and irregular singularity.  To study the BPS spectrum, 
 each irregular singularity is replaced by a disc with marked points, and the number 
of marked points depend on the specific form of the irregular singularity, and the regular singularity is interpreted as 
puncture in the bulk. There is only one type of marked point since there
is only one type of Young Tableaux of $A_1$ group. The BPS geometry is therefore a bordered Riemann surface with marked points on 
the boundary and punctures in the bulk.

\subsubsection{Ideal triangulation}
The BPS quiver is derived from the ideal triangulation of the corresponding bordered Riemann surface.
Let us begin with a Riemann surface with boundaries and specify a finite set of points $M_{\rm boundary}$,
called boundary marked points, on the boundary circles of $\Sigma$.
Each connected component of $\partial \Sigma$ has at least one
boundary marked point;  The bulk puncture is not blown up and remained as a point in the interior of the Riemann surface.
The defining data of our theory is a triple $(\Sigma, M_{\rm boundary},p)$. For notational convenience we
sometimes denote this triple simply by $\Sigma$.  In other words, $\Sigma$ is defined by following data:

a. the genus $g$ of the Riemann surface;

b: the number of bulk punctures $p$.

d. the number $b$ of boundary components;

d. the number of marked points $h_i$ on each boundary.

Each puncture represents the regular singularity while the boundary with marked points 
means an irregular singularity, all the marked points have a Young Tableaux label. The punctures and the marked
points are all called marked points for simplicity in the following, and one should be careful about whether it is in the bulk or one the boundary.
One can define a combinatorial object called ideal triangulation on above Riemann surface.
An ideal triangulation is defined using arcs \cite{fomin-2008-201}.  A simple arc $\gamma$ in $\Sigma$ is a curve such that 

1. the endpoints of $\gamma$ are marked points; 

2. $\gamma$ does not intersect itself, except at the endpoints;

3. $\gamma$ is disjoint from the marked points and the boundary.

We also require the arc $\gamma$ is not contractible into the marked points or onto the boundary. Each arc is considered up to
isotopy. Two arcs are called compatible if they do not intersect in the interior of $\Sigma$.  A maximal collection of distinct pairwise arcs 
is called an {\it ideal triangulation}. An edge is called external if it is isotopic to a segment of the boundary, otherwise
it is called internal. It is not hard to get the following formula for the number of internal edges:
\begin{equation}
6 g+3b+3p+\#\left| M_{\rm boundary }\right|-6 \ ,
\end{equation}
where as defined previously $g$ ($b$) is the genus (the number of boundary components) of
$\Sigma$, respectively. There are 
a total of $\#\left|M_{\rm boundary}\right|$ external edges. Several examples for the ideal triangulations for various
bordered Riemann surface are shown in figure. \ref{ideal}

 \begin{figure}[htbp]   
\small
\centering
\includegraphics[width=14cm]{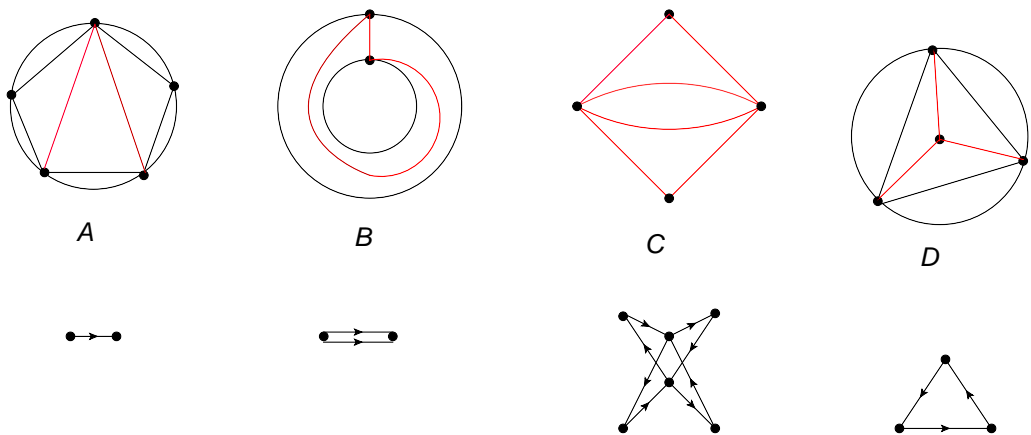}
\caption{The triangulation of various BPS geometry. A: Disc with five punctures which represents $A_2$ Argyres-Douglas theory. B: Annulus with
one marked point on each boundary, this is pure $SU(2)$ theory. C: Sphere with four punctures which is $SU(2)$ theory with four fundamental flavors.
D:  Disc with three marked points and a bulk puncture, which is the $D_3$ Argyres-Douglas theory.}
\label{ideal}
\end{figure}

We always start with an ideal triangulation without self-folded triangles, and
the BPS quiver can be read from the triangulation in the following way:

1. Assign a quiver node to each internal edge of the triangulation.

2. There is a quiver arrow for two nodes if the two corresponding edges are in the same triangle. The total 
quiver arrows are the signed sum of the quiver arrows if the two nodes are in more than one triangles. 

There are two special features for the quivers from the triangulated surface: first the maximal number of 
arrows are two for any two quiver nodes, second the quiver is in finite mutation class, namely the quiver 
will come back to itself after a finite number of mutations.

The quantum field theory is formed by gauging two kinds of matter together \footnote{The one disc with marked points represents the $A_n$ type Argyres-Douglas theory.}: trifundamental of $SU(2)$ which 
is represented by the three punctured sphere and the $D$ type AD theory represented by a sphere with 
one irregular singularity and one regular singularity. Each bulk puncture contributes $3$ to the charge lattice (one electric, one magnetic and one flavor charge), and
each boundary with $n_i$ marked points contributes  $(n_i+3)$ (when $n$ is even, there is a mass parameter), so the total dimension of the charge lattice is
\begin{equation}
3p+3b+\#\left| M_{\rm boundary }\right|-6+6g,
\end{equation}
which is equal to the number of the internal edges of the triangulation, so the BPS quiver from the triangulation has the right dimensions, and it can be checked that 
the rank of the quiver matrix is equal to $2n_r$.

\subsubsection{Potential}
The potential for the quiver arising from the triangulation of the bordered Riemann surface is given in \cite{labardinifragoso-2008}.
There is one potential term for the quiver arrows in each triangle and one term for each puncture. If
there is a quadratic term in the potential, then the two quiver arrows are massive and can be integrated out and one
get a reduced  quiver and potential. 
 \begin{figure}[htbp]   
\small
\centering
\includegraphics[width=10cm]{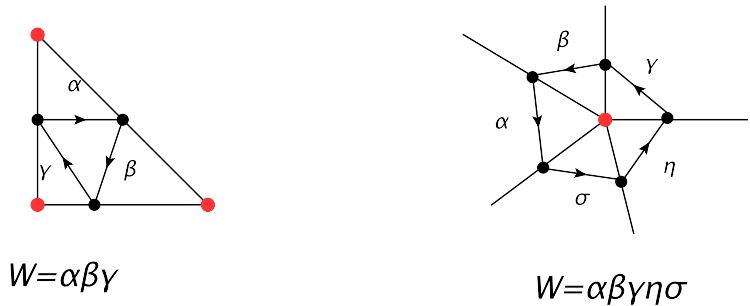}
\caption{There is a potential term for each triangle, and a potential term for each internal puncture.}
\label{ideal}
\end{figure}

Mathematically, the above integrating out process corresponds to removing  two cycles in 
the quiver. The quiver defined in previous paragraph is  actually the reduced one. In this paper, the quiver with potential $(Q,W)$ for
a triangulation is always the reduced one. 

\begin{figure}[htbp]   
\small
\centering
\includegraphics[width=10cm]{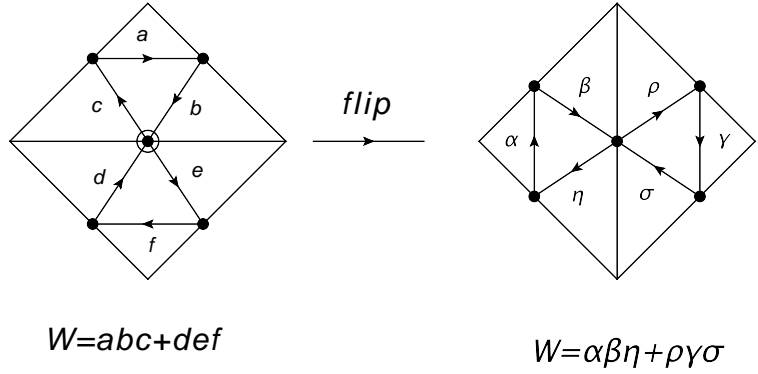}
\caption{The flip which relates two triangulations of the quadrilateral. The quivers are related by quiver mutation, and 
the potentials are also related by the mutation rule.}
\label{flip}
\end{figure}

The triangulation of the BPS quiver is not unique and two different of triangulations are related by a sequence local 
moves called flips. It can be checked that  the two quivers are related by mutations, and
the reduced potential of two triangulations are also related exactly by the 
mutation rules, see figure. \ref{flip}. 

\newpage
\subsection{Indecomposable objects: a geometric representation}
As we discussed earlier, the indecomposable representations of the quiver with potential are the possible BPS states. 
One could find them either from the indecomposable  modules of the Jacobi algebra, or from  the  representation
theory of $(Q,W)$. In this subsection, we are going to provide a geometric representation for these objects.

Let's consider the Riemann surface without the punctures, the Jacobi algebra defined from the triangulation is the
so-called string algebra. As discovered in \cite{Assem:arXiv0903.3347}, the indecomposable modules of a string algebra are represented by the strings and bands. 
Let's fist give a definition of the strings and bands. Given an arrow $\beta$, and let $S(\beta)$ be its starting point and 
$e(\beta)$ its ending point. We denote $\beta^{-}$ as the formal inverse of $\beta$ with $s(\beta^{-1})=e(\beta)$ and
$e(\beta^{-1})=s(\beta)$, notice that $(\beta^{-})^{-}=\beta$. A sequence of the quiver arrows (and their formal inverses)
 $\omega=\alpha_1\alpha_{2}\ldots \alpha_n$ is called a string if they satisfy the following two conditions
 
 1. The ending point of $\alpha_i$ is the starting point of $\alpha_{i+1}$.
 
 2. The quiver arrows and its formal inverse are not appearing in sequel, i.e. $\alpha_i\neq \alpha_{i+1}^{-}$.

Thus a \textbf{string} $w$ for the Jacobi algebra is defined as a walk in the quiver avoid the zero relations from the potential:
\begin{equation}
\omega:x_1\overset{\alpha_{1}}{-}x_2\overset{\alpha_{2}}{-}\ldots x_{n-1}\overset{\alpha_{n-1}}{-}x_n\overset{\alpha_{n}}{-}x_{n+1}.
\end{equation}
Namely, there is no subsequence in $\omega$ which appears in the ideal $I$ defined by the potential.  A string is called 
cyclic if $x_1=x_{n+1}$. A $\textbf{band}$ is a simple cyclic string (the starting point and the ending point are the same for the string), i,e, it is not a power of any string. The dimensional 
vector of the string or band module is defined as
\begin{equation}
d_i=\sum_{x\in \omega}\delta_{x,x_i},
\end{equation}
namely, the $i$th component of the dimension vector is equal to the number of times the quiver nodes $x_i$ appears in 
the $\omega$.

The string and band module have a very nice geometric interpretation as the curves on the Riemann surface. The curve is required to be not 
homotopy to the curve of the triangulation and boundary component.
The end points of open  curve $\gamma$ without self-intersection are either on the puncture or on the boundary, such curves are representing the string module. 
A simple closed curve $l$ represents the band module.  For two curves $\gamma$ and $\gamma^{'}$ in $\Sigma$, we denote by 
$I(\gamma, \gamma^{'})$ as the minimal intersection number of two representatives of the homotopic classes of $\gamma^{'}$ 
and $\gamma$. Now the dimension vector of the module associated with a curve has the dimension vector associated with a triangulation
\begin{equation}
 d=\sum_{\gamma^{'}\in\Gamma}I(\gamma, \gamma^{'}).
\end{equation}

It is not hard  to see that for the quiver from triangulated surface, the module from the open curve has no parameters and therefore
represent the hypermultiplet, and the module from the closed curve has $1$ parameter which then represents the $W$ boson. 
These curves then represent the possible BPS states, and our result is in perfect agreement with that found in \cite{Gaiotto:2009hg}. The association 
of the vector multiplet from the closed curve is actually natural from the $M$ theory point of view, i.e the self-dual string wrapping on the 
closed curve gives the $W$ boson.

\begin{figure}[htbp]
\small
\centering
\includegraphics[width=4cm]{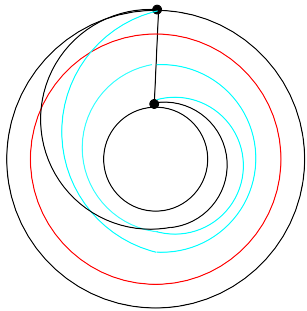}
\caption{The red closed curve represents the vector boson, and the green curve represents the hypermultiplet. }
\label{curve}
\end{figure}

Finally, let's give a simple application of the above example.  The BPS geometry is a annulus with one marked points on each boundary, 
the triangulation and one closed curve is shown in figure. \ref{curve}. The dimension vector for this curve is $(1,1)$ from the intersection number.
The other string module has dimension $(n,n\pm1)$ as seen from the open curves. These results 
match the result from the quiver representation theory.

The situation is more complicated for the Riemann surface with punctures, and the Jacobi algebra in that cases usually is not 
a string algebra. But we will make the following conjecture: the indecomposable modules associated with the $W$ boson is 
still represented by the closed curve, and one can read the subquiver from the intersection pattern. That is actually all we need for 
the later analysis.

\newpage

\subsection{Finite cases: Disc and Disc with one puncture}
\begin{figure}[htbp]
\small
\centering
\includegraphics[width=8cm]{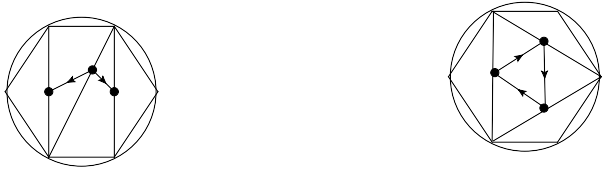}
\caption{Two  ideal triangulations  of a disc with 6 marked points, which represents $A_3$ Argyres-Douglas theory. The BPS quiver is given too. }
\label{triangulation}
\end{figure}

Let's consider a disc or a disc with one puncture in the bulk. When the BPS geometry is just a disc with $n+3$ marked points, it represents 
the $A_{n}$ Argyres-Douglas theory. The quiver from one of the ideal triangulation 
is of the $A_n$ shape which gives the name, see the left of  figure. \ref{triangulation}. The orientation of the quiver arrows is not important since they are all in the same quiver mutation class.  
These quivers have special property that there are finite number of indecomposable representations, and all of them represents the hypermultiplet.
This fact means that all the chambers have finite number of hypermultiplets.

We are going to use the maximal green mutation to find the BPS spectrum of these theories by starting with the corresponding Dynkin diagram, since there is 
only finite number of indecomposable object, one can do random green mutations and would definitely find a maximal green mutation at the end.
The minimal chamber has $r$ BPS states, where $r$ is the number of simple roots
of the corresponding lie algebra, and the number of BPS states in the maximal chamber is equal to the number of positive roots. The BPS states and 
charges are in one to one correspondence with the positive roots of lie algebra. 
The minimal chamber and maximal chamber is easy to find. Since the $A$ type quiver is acyclic, there is always a source node and sink node, the minimal 
chamber is found by always mutating the source node in each step, and the maximal chamber is found by mutating sink node only in each step, see table. \ref{A3spectrum}.
Moreover, for every integer $l$ satisfying $l_{min}\leq l \leq l_{max}$,
there is a finite chamber with $l$ states. The interested reader can work out the charge vectors and the corresponding ordering of phase using the green mutations. 
The results of the minimal and maximal chamber of the ADE quiver are summarized in table. \ref{ADEdata}.

\begin{table}[h]
\centering
    \begin{tabular}{|c|c|c|c|c|c|}
        \hline
        ~ & $A_n$ & $D_n$ & $E_6$ & ~$E_7$& ~$E_8$\\ \hline
        Minimal & n & n & 6 & 7&8 \\ \hline
        Maximal& ${n(n+1)\over 2}$ & n(n-1) & 36 & 63 & 120 \\
        \hline
    \end{tabular}
    \caption{The number of BPS states in the minimal and maximal chamber for the ADE  quiver.}
    \label{ADEdata}
\end{table}

\begin{mydef}
 Let's consider the $A_3$ quiver with orientation $1\rightarrow 2 \leftarrow 3$. We could easily list the chamber with three, four, five, and six BPS states, see table. \ref{A3spectrum}.
\begin{table}[h]
\centering
    \begin{tabular}{|c|c|c|}
        \hline
        ~& Maximal green mutation &charge vectors \\ \hline
        3 & $\mu_1, \mu_3, \mu_2$  & $\gamma_1, \gamma_3, \gamma_2$ \\ \hline
         4 &  $\mu_1, \mu_2, \mu_3, \mu_2$ & $\gamma_1, \gamma_2,  \gamma_2+\gamma_3, \gamma_3$ \\ \hline
           5 & $\mu_2, \mu_3, \mu_2, \mu_1, \mu_3$& $\gamma_2, \gamma_2+\gamma_3, \gamma_3, \gamma_1+\gamma_2, \gamma_1$ \\ \hline
            6 & $\mu_2, \mu_3, \mu_1, \mu_2, \mu_3, \mu_1$& $\gamma_2, \gamma_2+\gamma_3, \gamma_1+\gamma_2, \gamma_1+\gamma_2+\gamma_3, \gamma_1,\gamma_3$\\ \hline 
                         \end{tabular}
    \caption{The maximal green mutation sequences and charge vectors for $A_3$ quiver.}
    \label{A3spectrum}
\end{table}
 \end{mydef}
\newpage

If we start with a quiver which is mutation equivalent to the $A_n$ quiver, one can do random green mutations too. This fact is rather interesting since 
the maximal green mutations actually knows about the potential, although we did not specify the potential in the definition of green mutations. In fact, the green mutation 
is desired for the generic potential of the quiver, namely, one can mutate the quiver and always get a 2-acyclic quiver due to the potential. 

\begin{figure}[htbp]
\small
\centering
\includegraphics[width=4cm]{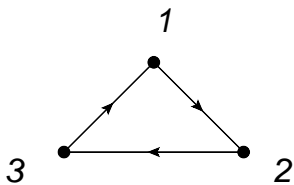}
\caption{Quiver which is mutation equivalent to the $A_3$ quiver, and there is a potential term for the cyclic path in the quiver. }
\label{A3cyclic}
\end{figure}

 The new issue is that the number of  BPS states in the minimal chamber and the maximal chamber do not necessarily equal to the number given by the Dynkin diagram.
For example, consider the quiver in figure. \ref{A3cyclic}, the minimal chamber has 4 BPS states instead of 3. The mutation sequence corresponding
to the minimal chamber is
\begin{equation}
\mu_1,\mu_2,\mu_3,\mu_1.
\end{equation}
The maximal chamber of this quiver has $5$ states. So the number of states in minimal chamber and maximal chamber is not an quiver invariant. This is natural from 
the green mutation subset point of view: the quiver subset of the minimal and maximal chamber of $A_n$ quiver is always acyclic and so it would 
not include the above cyclic quiver, so the minimal chamber or maximal chamber is not realized in the cyclic quiver.

\begin{figure}[htbp]
\small
\centering
\includegraphics[width=8cm]{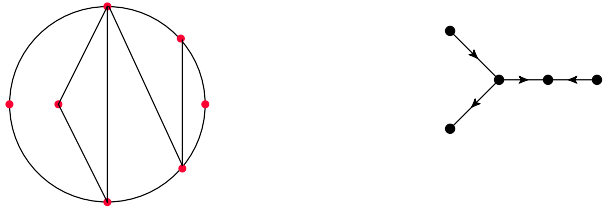}
\caption{The triangulation of the disc with one puncture in the bulk. The quiver is of the D shape. }
\label{DN}
\end{figure}

The BPS geometry of $D$ type AD theory is realized as a disc with one puncture on the bulk. The BPS quiver can be found easily from the triangulation
and it is indeed of the $D$ shape in one triangulation, see figure. \ref{DN}. This is of the finite type and one could easily find the finite chamber and the charge vectors
by doing random green mutations. 

The $E$ type AD theory can be found  using $A_2$ theory compactified on a sphere with one specific irregular singularity as shown in \cite{Xie:2012hs}.
The BPS spectrum can be similarly  found using random green mutations, and the minimal and maximal chamber is listed in table. \ref{ADEdata}.

\subsection{Riemann surface without punctures}
In this subsection, we are going to study theory whose BPS geometry is a bordered Riemann surface without bulk punctures, and
the Jacobi algebra of this class of theories are string algebra, and in particular the vector boson and its corresponding quiver representation can be easily found from the closed curves on the BPS geometry. 
By inspection, the subquiver  for the vector boson is always acyclic and there are definitely a source node and a sink node. 
Let's denote the charge vectors of the source node as $\gamma_{source}$ and the sink node as
$\gamma_{sink}$. We now argue that the sufficient condition for the $W$ boson to be unstable is that the source node has larger slop than the the sink node.
\begin{equation}
\mu(\gamma_{source})>\mu(\gamma_{sink}).
\end{equation}
Let's prove the above statement using the quiver representation theory.  The $W$ boson representation $P$ has the following two special subrepresentations
\begin{equation}
P_{1}=\gamma_{sink}, ~~P_2=\sum_{i-source}\gamma_i.
\end{equation}
The condition for $P$ to be stable is that  the slop of $P$ should be smaller than all of its subrepresentation, namely $P_{1}$ and $P_{2}$ should be on the left of $P$ on the half plane:
\begin{equation}
\mu(P)<\mu(P_{1})~\&~\mu(P)>\mu(P_2).
\end{equation}
If $P$ is stable, since $dim(P)=dim(P_2)+dim(P_{source})$ and $P_2$ is on the left of $P$,  then $P_{source}$ is definitely on the right of $P$, this implies that
\begin{equation}
\mu(P_{source})<\mu(P)<\mu(P_{sink}),
\end{equation}
 In another word,  as long as the slop of the source node is larger than the sink node,  the $W$ boson is definitely unstable. 
 Notice that this is not the necessary condition to find the finite spectrum. In practice, this means that if the source charge
 appears before the sink charge in the green mutation sequence, then this $W$ boson would be killed.

For simple BPS geometry, one can find all possible finite chambers using the random mutations without worrying about the $W$ boson.
Some simple BPS geometry of this sort and the minimal plus maximal chamber are listed 
in table. \ref{Su2finite}. There is a finite chamber for all the integers  $l_{min}\leq l\leq l_{max}$.
The minimal chamber is easy to find: there is a source node in the quiver, and one mutate
the source node in each step. 

\begin{table}[h]
\centering
    \begin{tabular}{|c|c|c|c|}
        \hline
        Quiver & theory &minimal & maximal \\ \hline
           $\tilde{A}(1,1)$ & $SU(2)$  & $2$& $2$ \\ \hline
           $\tilde{A}(2,1)$ & $SU(2)$ with $N_f=1$ & $3$& $5$ \\ \hline
             $\tilde{A}(2,2)$ & $SU(2)$ with $N_f=2$ & $4$& $10$ \\ \hline
             $\tilde{D}_4$ & $SU(2)$ with $N_f=3$ & 5&22 \\ \hline 
                          figure.  \ref{four}A & $SU(2)$ with $N_f=4$ & $12$& $46$ \\ \hline
           \end{tabular}
    \caption{The minimal and maximal chamber for $SU(2)$ with $N_f\leq4$.}
    \label{Su2finite}
\end{table}

For more complicated geometry, one can either find the finite chamber using the random green mutations, or use the following steps:

a. Identify the subquiver corresponding to the $W$ boson which is represented by the closed curve in the Riemann surface, 
and identify the source node and sink node of this subquiver.

b. Do the green mutation such that the charge of the source node appear before the charge of the sink node. The easiest thing you can do is 
to mutate source node first.

\begin{figure}[htbp]
\small
\centering
\includegraphics[width=10cm]{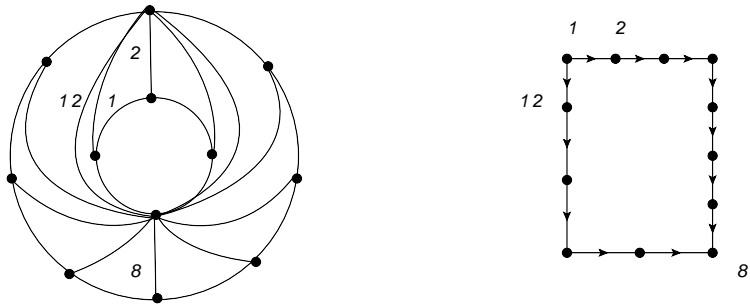}
\caption{The triangulation of an annulus with $4$ points on one boundary and $8$ points on another boundary;
The quiver is shown on the left which is of the affine Dynkin diagram $\tilde{A}(4,8)$.  }
\label{twobound}
\end{figure}

Let's consider an annulus with $n_1$ marked points and $n_2$ marked points on each boundary.  The BPS quiver from one specific triangulation has the form of 
affine $\tilde{A}(n_1,n_2)$ shape where there are $n_1$ arrows in one direction, and $n_2$ arrows in another direction, see figure. \ref{twobound}.  There is only one closed 
curve and so only one $W$ boson whose corresponding quiver representation $P$ has dimension vector $(1,1,\ldots, 1)$.  Since node $1$ is the 
source node of this subquiver,  $W$ boson is definitely unstable if node 1 is mutated first, and we will always find a finite chamber by doing 
random green mutation afterwards.

\begin{figure}[htbp]
\small
\centering
\includegraphics[width=10cm]{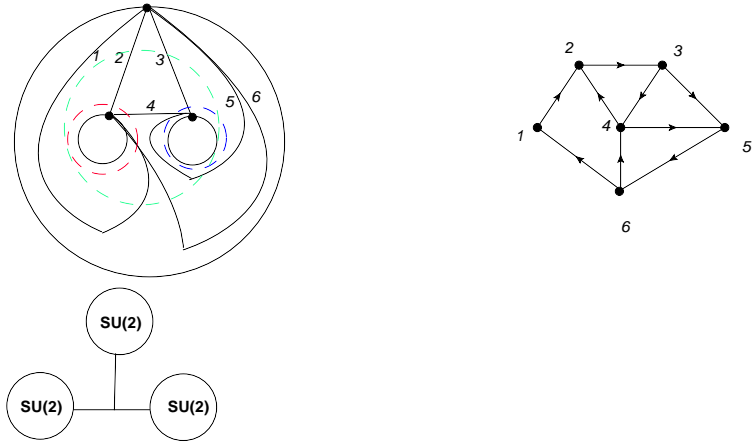}
\caption{The triangulation of a sphere with three boundaries each with one marked point, and the quiver gauge theory underlying this BPS geometry is shown too.  The three closed 
curves representing the $W$ bosons are drawn from which one can read the subquiver for them.
The BPS quiver is shown on the right.  }
\label{threebound}
\end{figure}

\begin{mydef}
Here is another theory whose BPS geometry is a sphere with three boundary each with a marked point, and the four dimensional theory is  three 
$SU(2)$ gauge groups coupled through a trifundamental, so the BPS quiver should have six quiver nodes. The triangulation and 
quiver are shown in figure. \ref{threebound}, and there are three closed curves which represent three $W$ bosons. The sink-source analysis
of these bands are listed in table. \ref{threeb}. So if we mutate node $1$ and node $3$ first, then the three $W$ bosons would be 
unstable, and random green mutations can be done later. A maximal green mutation sequence is 
\begin{equation}
\mu_1,\mu_3, \mu_2,\mu_4,\mu_6,\mu_5,\mu_3,\mu_4.
\end{equation}

\end{mydef}

\begin{table}[h]
\centering
    \begin{tabular}{|c|c|c|c|}
        \hline
        Band & quiver nodes & source & sink \\ \hline
           $B_1$ & $(1,2,4,6)$ & $1$& $2$ \\ \hline
           $B_2$ & $(1,2,3,5,6)$ & $1$& $6$ \\ \hline
                      $B_3$ & $(3,4,5)$ & $3$& $5$ \\ \hline
              \end{tabular}
    \caption{The source and sink nodes for the three bands from the quiver in figure. (\protect  \ref{threebound}).}
    \label{threeb}
\end{table}
 
\newpage
\subsection{Riemann surface with punctures}
\subsubsection{Asymptotical free theory}
\begin{figure}[htbp]
\small
\centering
\includegraphics[width=10cm]{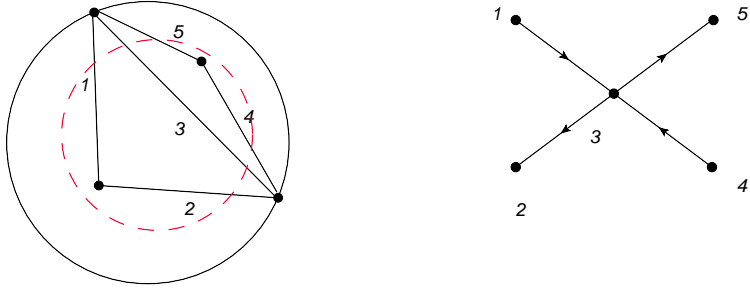}
\caption{An ideal triangulation of twice punctured disc and the corresponding quiver.  }
\label{affineD4}
\end{figure}

The BPS geometry for $SU(2)$ with $N_f=3$ requires the bulk puncture: it is a disc with two marked points and two bulk punctures. One triangulation and 
the BPS quiver is shown in figure. \ref{affineD4}.  This quiver has the affine $\tilde{D}_4$ shape, and more generally the quiver
is $\tilde{D}_{n+2}$ in one of the triangulation if there is two bulk punctures and $n$ marked point on boundary of the disc. 
There is one closed curve representing the $W$ boson for the gauge group, and its dimension vector is 
$(1,1,2,1,1)$.  Since the affine $\tilde{D}_4$ quiver is acyclic, the number of BPS states in minimal chamber has $5$ states and can be found using the source sequences. 
The  maximal chamber has $22$ states as found by doing all possible green mutations.

Let's now give a conjecture on the number of states in the maximal chamber from the the following assumption: the number of states is a continuous function
of the parameters of the BPS geometry. Consider affine $\tilde{D}_n$ quiver, we further conjecture that it is of the order of $n^2$:
\begin{equation}
f(n;2)=an^2+bn+c,
\end{equation}
when $n=0$, the geometry becomes a three puncture sphere, but 
only the mass deformations corresponding to bulk punctures are allowed, therefore there are only two BPS states. When $n=1$, the underlying theory is 
a $SU(2)$ with two flavors, so $f(1)=10$, finally $f(2)=22$ as from the experimental study. Using these initial data, we conjecture the maximal chamber has the following number of states:
\begin{equation}
f(n;2)=2n^2+6n+2.
\end{equation}
which is in agreement with the result found from computer scanning  in \cite{BrŸstle:arXiv1205.2050}.

The above analysis can actually be generalized to other type of BPS geometry.  Let's an annulus with $1$ and $n$ marked points on each boundary, and the BPS quiver
in one particular triangulation is the affine $\tilde{A}(n,1)$ quiver.  The minimal 
chamber has $n+1$ states,  let's  use $f(n,1)$ to represent the number of states in maximal chamber and again assume the number of states is a quadratic function of $n$. 
 Using the initial data $f(0,1)=0$ (since this is a trivial theory), $f(1,1)=2$ for pure $SU(2)$ theory, and $f(2,1)=5$ for $SU(2)$ with one flavor, we find
\begin{equation}
f(n,1)={n^2+3n\over 2}.
\end{equation}

\begin{table}[h]
\centering
    \begin{tabular}{|c|c|c|c|c|}
        \hline
        BPS geometry&Quiver & theory &minimal & maximal \\ \hline
         Annulus: $B_1$, $B_n$&  $\tilde{A}(n,1)$ & $SU(2)-AD$ &n+1 &${n(n+3)\over2}$ \\ \hline
           $B_n$, two punctures&$\tilde{D}_{n+2}$ & $2-SU(2)-AD$& $n+3$& $2n^2+6n+2$ \\ \hline
           \end{tabular}
    \caption {The minimal chamber and maximal chamber for two class of BPS geometry.}
    \label{general}
\end{table}

More generally, consider an annulus with $n_1$ and $n_2$ marked points on each boundary,  and the BPS quiver is the affine quiver $\tilde{A}(n_1,n_2)$ from 
one particular triangulation. The number of states in maximal chamber should be invariant under exchange of $n_1$ and $n_2$, so
\begin{equation}
f(n_1, n_2)=f(n_1+n_2, n_1n_2), 
\end{equation}
and it should be a quadratic polynomial from our assumption.  Denote $x=n_1+n_2, y=n_1n_2$, then the general expression reads
\begin{equation}
f(n_1, n_2)=ax^2+by^2+cxy+dx+ey+f.
\end{equation}
Using the result  for $f(n,1)$, $f(2,2)=10$, $f(3,0)=5$, and $f(4,0)=9$ \footnote{Notice that disc with one bulk puncture is representing  D type Argyres-Douglas theory, but the quiver 
is of the cyclic affine type, so the maximal number of states is different from the one derived from the D type Dynkin diagram as listed in table. \ref{ADEdata}.}, the number of states in maximal chamber is 
\begin{equation}
f(n_1, n_2)={1\over 2}x^2+{1\over 4}y^2-{1\over 4}xy+{1\over2}x+{1\over 4}y-1.
\end{equation} 
Immediately, we make the prediction that the maximal chamber of the quiver $\tilde{A}(n,0)$ has states
\begin{equation}
f(n,0)=1/2 (-2 + n + n^2).
\end{equation}
It would be interesting to prove our this  conjecture  using computer scanning, etc.

\begin{figure}[htbp]
\small
\centering
\includegraphics[width=8cm]{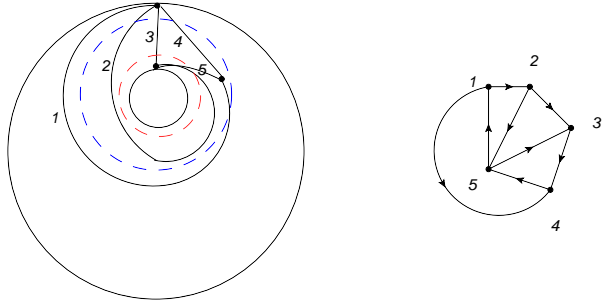}
\caption{The quiver of two boundaries with a single marked point plus a bulk puncture, the $W$ boson is represented by closed curve.  }
\label{2gauge}
\end{figure}

Now let's give an example showing how to use the generalized source-sink sequence to find the finite chamber for the theory defined using the bulk puncture.
\begin{mydef}
Consider a Riemann surface with two boundaries with a single marked point  and a bulk puncture. The $\mathcal{N}=2$
theory is $SU(2)\times SU(2)$ asymptotical quiver gauge theory.  One triangulation and the quiver are shown in figure. \ref{2gauge}.
\end{mydef}

There are two closed curves representing $W$ bosons for two $SU(2)$ gauge groups. The subquvier corresponding to these two bands
and the source-sink nodes are shown in table. \ref{SU22}.
\begin{table}[h]
\centering
    \begin{tabular}{|c|c|c|c|}
        \hline
        Band & quiver nodes & source & sink \\ \hline
           $B_1$ & $(2,3,5)$ & $2$& $3$ \\ \hline
           $B_2$ & $(1,2,3,4)$ & $1$& $4$ \\ \hline
              \end{tabular}
    \caption{The source and sink nodes for the two bands of the quiver shown in figure. (\protect \ref{2gauge}).}
    \label{SU22}
\end{table}
Since $2$ and $1$ are the source nodes for two bands, and if we mutate node $1$ and node $2$ first, then the $W$ bosons will be 
disabled and a finite chamber can be found by doing random green mutations.  We just list one chamber below
\begin{equation}
\mu_1,\mu_2,\mu_5,\mu_1,\mu_3,\mu_4,\mu_1.
\end{equation}
There are many other possibilities which could be easily found from the green mutations.

\subsubsection{Fourth punctured sphere}
\begin{figure}[htbp]
\small
\centering
\includegraphics[width=12cm]{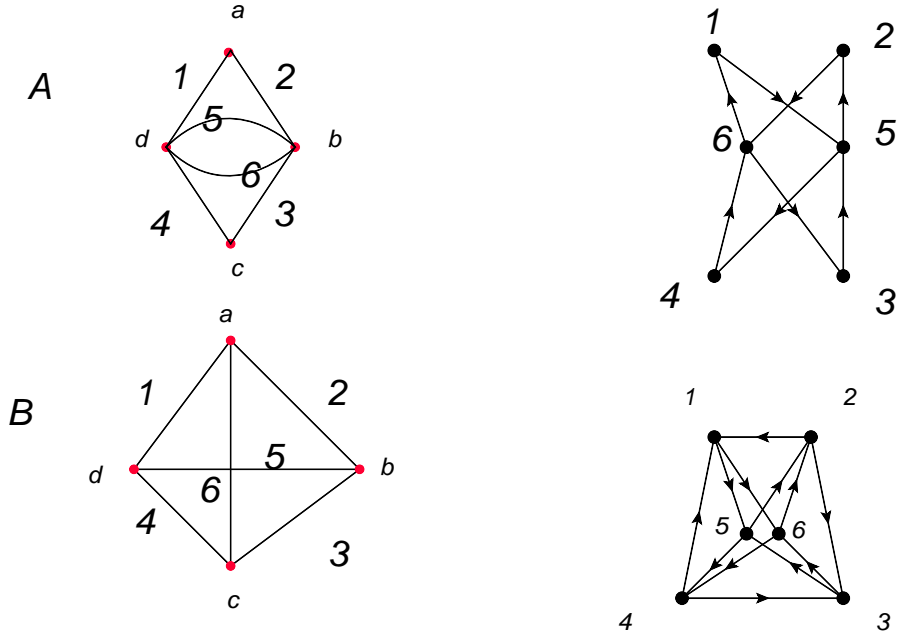}
\caption{A: one triangulation of fourth punctured sphere and the quiver. B: another triangulation of fourth punctured sphere and the quiver. }
\label{four}
\end{figure}

\begin{mydef}
Consider the fourth punctured sphere which represents  $SU(2)$ theory with four flavors.  The BPS quiver from one triangulation is shown in figure. \ref{four}A. 
Now there are three closed curves and we list the source  and sink nodes in table. \ref{SU4flavor}.
\end{mydef}
\begin{table}[h]
\centering
    \begin{tabular}{|c|c|c|c|}
        \hline
        Band & quiver nodes & source & sink \\ \hline
           $B_{a,b}$ & $(2,5,4,6)$ & $5$& $6$ \\ \hline
           $B_{a,c}$ & $(1,5,3,6)$ & $6$& $5$ \\ \hline
           $B_{a,d}$ &$(5,1,2,3,4)$& $(1,3)$&$(2,4)$\\ \hline
              \end{tabular}
    \caption{Source and sink nodes for three bands of the quiver of $SU(2)$ with four flavors, see figure. (\protect \ref{four})A.}
    \label{SU4flavor}
\end{table}

We want charge vectors $\gamma_5,\gamma_6, \gamma_1$ to appear first in doing green mutations \footnote{Equivalently, we can do the green mutation such that charge $\gamma_5, \gamma_6$ and $\gamma_3$ appear first.} ,
 which can be done by doing the mutation sequences $\mu_6,\mu_1,\mu_5$, which will ensure all the $W$ bosons are unstable. 
 Then we can do green mutation in a random way.  One sequence is the following
 \begin{align}
 &\mu_6,\mu_1,\mu_5, \mu_2,\mu_3,\mu_4, \mu_3,\mu_1, \nonumber \\
  &\mu_4,\mu_6,\mu_5, \mu_3,\mu_5,\mu_4, \mu_2,\mu_3, \mu_5.
 \end{align}
and this chamber has $17$ states.
There are many other possibilities for the maximal green mutations, and this example has been studied extensively in \cite{BrŸstle:arXiv1205.2050}. The result is that the maximal chamber has $46$ states and the minimal 
chamber has $12$ states.  The minimal chamber can be found using the following mutation sequences
\begin{equation}
(\mu_5, \mu_6),~(\mu_1,\mu_2,\mu_3,\mu_4),~(\mu_5, \mu_6),~(\mu_1,\mu_2,\mu_3,\mu_4).
\end{equation}
If we start with another triangulation as shown in figure. \ref{four}B, then there are also three subquivers for $W$ bosons and the source, sink nodes are shown in table. \ref{four2}.
\begin{table}[h]
\centering
    \begin{tabular}{|c|c|c|c|}
        \hline
        Band & quiver nodes & source & sink \\ \hline
           $B_{a,d}$ & $(4,6,2,5)$ & $(5,6)$& $(2,4)$ \\ \hline
           $B_{a,b}$ & $(3,6,1,5)$ & $(1,3)$& $(6,5)$ \\ \hline
            $B_{a,c}$ & $(1,2,4,3)$ & $(2,4)$& $(1,3)$ \\ \hline
             \end{tabular}
    \caption{The source and sink analysis for the bands of another quiver from $SU(2)$ with four flavors.}
    \label{four2}
\end{table}
We found the following special beginning sequences which will make $W$ boson  unstable:
\begin{equation}
\mu_5, (\mu_2, \mu_1), \mu_5,
\end{equation}
because this sequence will produce the following charge vectors,
\begin{equation}
\gamma_5,~\gamma_2,~\gamma_1+\gamma_5,~\gamma_1.
\end{equation}
which will ensure the source charges of each band appear before that of the sink node.  After these steps, one can do random green mutations, i.e. the following mutation sequences 
are the maximal one:
\begin{equation}
\mu_5,\mu_2,\mu_1,\mu_5,\mu_6,\mu_3,\mu_6,\mu_4,\mu_6,\mu_2,\mu_1,\mu_2,\mu_5,\mu_2.
\end{equation}

\paragraph{The evidence for $S$ duality}
Notice there are three  closed curves and therefore three possible $W$ boson, which indicate that there are three duality frames and
each $W$ boson represents a duality frame.  However, a natural question for the consistency is whether  they can appear in a single chamber.
We will prove that this can not happen by studying the quiver representation of quiver in figure. \ref{four}A. Let's first list all the sub representations 
of the three bands, see table. \ref{SU4sub}.

First of all, band $B_{a,b}$ and $B_{a,c}$ can not be coexisting since the necessary condition for the stability of the bands is that 
the slop of the source node is smaller than the sink node, this is  can not be satisfied simultaneously for two bands since node $5$ and
node $6$ exchange the role of source and sink in two bands.  Next 
consider the pair of bands $B_{a,d}$ and $B_{a,b}$, they share a comment subquiver $2\leftarrow 5 \rightarrow 4$, now 
to make $B_{a,d}$ stable, we need to mutate this subquiver first such that all the charge vectors $\gamma_2,\gamma_4, \gamma_5$ appear, but his 
automatically will make $B_{ab}$ unstable since $\mu_(\gamma_5)>\mu(\gamma_6)$ (the slop of the source charge is bigger than the sink charge)!  Similar analysis applies to the bands $B_{a,d}$ and $B_{a,c}$.

\begin{table}[h]
\centering
    \begin{tabular}{|c|c|c|}
        \hline
        Band & quiver nodes & Subrepresentation \\ \hline
           $B_{a,b}$ & $(2,5,4,6)$ &$\gamma_2+\gamma_4+\gamma_6, \gamma_2+\gamma_6, \gamma_4+\gamma_6, \gamma_6$ \\ \hline
           $B_{a,c}$ & $(1,5,3,6)$ & $\gamma_1+\gamma_3+\gamma_5, \gamma_1+\gamma_5, \gamma_3+\gamma_5, \gamma_5$  \\ \hline
           $B_{a,d}$ &$(5,1,2,3,4)$&$\gamma_2+\gamma_3+\gamma_4+\gamma_5, \gamma_2+\gamma_5+\gamma_4, \gamma_2, \gamma_4, \gamma_1+\gamma_2+\gamma_5+\gamma_4$ \\ \hline
              \end{tabular}
    \caption{Subrepresentations of various bands of $SU(2)$ with four fundamentals.}
    \label{SU4sub}
\end{table}

It is possible to generalize the above consideration to other theories defined from the bordered Riemann surface. Our conjecture is that there are only 
a maximal set of non-intersected closed curves whose $W$ boson can become simultaneously stable. 

\newpage
\subsubsection{Sphere with five punctures}

\begin{figure}[htbp]
\small
\centering
\includegraphics[width=8cm]{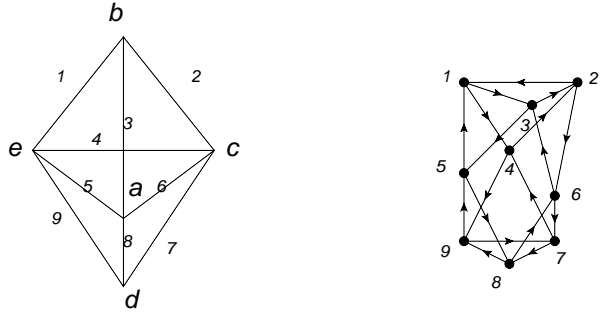}
\caption{A triangulation and quiver of the sphere with five punctures.}
\label{SU5triangualtion}
\end{figure}

\begin{mydef}
Let's consider a sphere with five punctures, which represents the conformal quiver gauge theory $2-SU(2)-SU(2)-2$. A
 triangulation and the corresponding quiver are shown in figure. \ref{SU5triangualtion}. The subquiver and source-sink analysis 
 of  various bands are listed in the following table. \ref{SU5puncture}.
\end{mydef}

\begin{table}[h]
\centering
    \begin{tabular}{|c|c|c|c|}
        \hline
        Band & quiver nodes & source & sink \\ \hline
           $B_{a,b}$ & $(1,5,8,6,2)$ & $(2,5)$& $(1,6)$ \\ \hline
           $B_{a,c}$ & $(7,8,5,3,2,4)$ & $(3,7)$& $(8,2)$ \\ \hline
             $B_{a,d}$ & $(5,3,6,7,9)$ & $(6,9)$& $(7,5)$ \\ \hline
             $B_{a,e}$ & $(9,8,6,3,1,4)$ & $(8,1)$& $(3,9)$ \\ \hline
          $B_{b,c}$ & $(7,6,3,1,4)$ & $(6,1)$& $(3,4)$ \\ \hline
        $B_{b,e}$ & $(9,5,3,2,4)$ & $(3,4)$& $(2,5)$ \\ \hline
             $B_{c,d}$ & $(2,4,9,8,6)$ & $(4,8)$& $(6,9)$ \\ \hline
             $B_{c,e}$ & $(7,6,2,1,5,9)$ & $(2,9)$& $(1,7)$ \\ \hline
             $B_{d,e}$ & $(1,4,7,8,5)$ & $(5,7)$& $(4,8)$ \\ \hline
        \end{tabular}
    \caption{The source and sink analysis of the bands of the sphere with five punctures.}
    \label{SU5puncture}
\end{table}

Now let's take two disjoint triangles, say $\Delta_{123}$ and $\Delta_{789}$,  and do the mutation sequences 
\begin{eqnarray}
Step 1:~\mu_1, (\mu_2,\mu_3),\mu_1, \nonumber\\
Step 2:~\mu_7, (\mu_8,\mu_9),\mu_7.
\end{eqnarray}
This mutation will create the charge vectors 
\begin{equation}
\gamma_1, \gamma_1+\gamma_2, \gamma_2, \gamma_3, \gamma_7, \gamma_7+\gamma_8, \gamma_8, \gamma_9,
\end{equation}
and they will kill all the $W$ boson!  For example, for the band $B_{ab}$, the source charge vector $\gamma_2$ appears before the sink  $\gamma_6$, so 
it is unstable. After the above step, we will always find the finite chamber by doing random mutation sequences.  A maximal green mutation sequences are
\begin{align}
&\mu_1, (\mu_2,\mu_3),\mu_1,\mu_7, (\mu_8,\mu_9),\mu_7, \mu_6,\mu_3,\mu_8,\mu_5,\mu_8,  \nonumber\\
&\mu_2, \mu_6, \mu_7,\mu_4, \mu_3, \mu_6,\mu_7, \mu_4, \mu_3, \mu_1, \mu_8, \mu_9, \mu_4, \mu_3.
\end{align}

One can do similar analysis for sphere with more punctures, although the analysis would become very 
tedious. In next section, we are going to describe another simpler method to find the finite chamber 
using the higher rank realization of the same theory.

\section{BPS spectrum for $A_{N-1}$ theory}
A large class of four dimensional  $\mathcal{N}=2$ field theory can be engineered by compactifying six dimensional $A_{N-1}$  $(2,0)$ theory on 
a Riemann surface with regular singularity and irregular singularity \footnote{Roughly speaking, regular singularity means first order pole while
the irregular singularity means the higher order pole.}. The geometric data  defining the theory is: 

1. A Riemann surface $M_{g,p_i,b_j}$, where $g$ is the genus, $p_i$ is the regular singularity, and $b_j$ is the irregular singularity.

2. $p_i$ is classified by the Young Tableaux \footnote{We call a puncture full if the Young Tableaux has the form $[1,1,\ldots,1]$, and simple if the Young 
Tableaux is $[N-1, 1]$.}\cite{Gaiotto:2009we,Nanopoulos:2009uw}, and $b_j$ is classified by a Newton Polygon \cite{Xie:2012hs} \footnote{The degenerating case needs further data, i.e. a sequence of Young Tableaux.}.

Let's describe a little bit about the the four dimensional theory defined by various geometries.
The Riemann surface $M_{g,p_i,0}$ defines a four dimensional superconformal field theory whose gauge coupling constants  are
identified by the complex structure moduli of $M$. Different duality frames are realized as different degeneration limits of the same 
Riemann surface. Weakly coupled gauge theory description in each duality frame is completely determined by the genus and the 
Young Tableaux type, and generically the theory is formed by gauging the flavor symmetries of the strongly coupled isolated 
SCFT defined by the three punctured sphere. Many properties of these theories including S duality \cite{Argyres:2007cn, Gaiotto:2009we}, Seiberg-Witten curve, 
3d mirrors \cite{Benini:2010uu}, central charges, and superconformal index \cite{Gadde:2011uv} can be understood from  this beautiful geometric construction. 

The Riemann surface $M_{0,p,b}$ (one regular and one irregular singularity on the sphere) and $M_{0,0,b}$ (only one irregular singularity on the sphere) define another type of SCFT called 
Argyres-Douglas theory, which is typically an isolated theory (without marginal deformations) and has fractional scaling dimension for the operator spectrum.
 Lots of properties regarding these type of theories are studied in detail in \cite{Xie:2012jd}. 
 
 In general, $M_{g, p_i, b_j}$ defines a four dimensional $\mathcal{N}=2$ 
theory which in each duality frame is formed by gauging the flavor symmetries of the following two types of matters:
 AD type theory represented by a sphere with one irregular singularity and one regular singularity, and the isolated SCFT represented 
 by a sphere with three regular singularities.

\begin{figure}[htbp]
\small
\centering
\includegraphics[width=10cm]{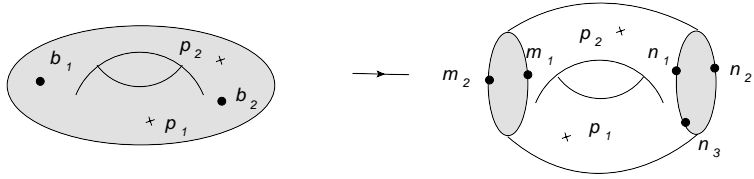}
\caption{Left: The Riemann surface with regular and irregular singularity defines a four dimensional UV complete $\mathcal{N}=2$ theory; Right: The BPS geometry for 
the corresponding 4d theory is derived by replacing each irregular singularity with a boundary with marked points labeled by Young Tableaux.  }
\label{blowup}
\end{figure}

The BPS geometry of the underlying field theory can be derived by blowing up the irregular singularity, and 
each irregular singularity is replaced by  a boundary with marked points labeled also by Young Tableaux. The detailed 
map between the irregular singularity and the corresponding marked boundary is worked out in \cite{Xie:2012jd}. The bordered Riemann surface 
relevant for the construction of ideal triangulation is depicted in the figure. \ref{blowup}. Once an ideal triangulation is given, 
the BPS quiver can  be found from this geometric data by introducing more structures into each triangle of the ideal triangulations. 
In next subsection, we will describe the construction of the BPS quiver  in more detail.

\newpage

\subsection{Dot diagram, network and quiver}
Let's review the results presented in \cite{Xie:2012dw} which discussed how to find the BPS quiver for the higher rank $\mathcal{N}=2$ theory defined by $M_{g,p_i, b_j}$. The construction  starts 
with an ideal triangulation of the bordered Riemann surface and a choice of cyclic path connecting all the punctures in the triangulation. The difference from the $A_1$ case is that
 more structures are needed to put on the edges  and inside each triangle: there are more than one quiver node on each 
edge and there are quiver nodes inside each triangle.  

The basic ingredient is attach a quiver to a single triangle with different Young Tableaux at the vertex, and the full quiver is derived by gluing the triangle quivers together. 
Given a triangle labeled by three Young Tableaux $(Y_1, Y_2, Y_3)$ in a cyclic order, one could find a dot diagram and
 a tessellation of the triangle using the brane construction proposed in \cite{Benini:2009gi}.  Let's put the triangle inside a two dimensional lattice with unit
 spacing, and put the three vertices at positions $(N,0), (0,0)$,  and $(0,N)$.  The dot diagram for the lattice points bounded by the triangle (including the points on the boundary) is found as follows:
 
a. Decorating the boundary edge of the triangle with black dots and white dots using $Y$ vertex right ahead of it in the clockwise direction: If the Young Tableaux $Y$  has partition $[n_1,n_2, \ldots, n_s]$, 
then first  put $n_1-1$ white dots and  a black dots to represent the $n_1$ column, and the second step is done by putting $n_2-1$ white dots and $1$ black dot, and 
continue this way until the whole Young Tableaux is represented by the black-white pattern on this boundary edge.

b. Constructing the dot diagram inside the triangle using only following two types of polygons whose edge is formed by lines \footnote{The lines should be parallel  with the boundary edges.} connecting two black dots. 1: Triangles whose edges have the same lengths. 2: Trapeziums whose 
parallel sides have lengths $n_1, n_2$ and the other two sides have length $n_1-n_2$ \footnote{This constraint is from the supersymmetric condition on the brane configuration.}.

\begin{figure}[htbp]
\small
\centering
\includegraphics[width=10cm]{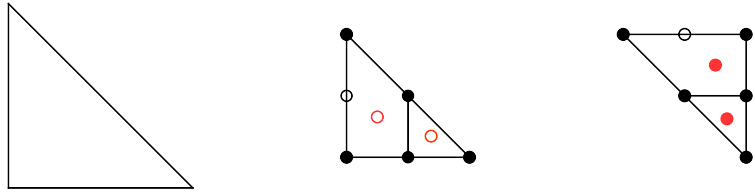}
\caption{Left: The orientation of the big triangle.  Middle: Put a white vertex to each polygon whose triangle completion has  the same orientation as 
the big triangle. Right: Put a black vertex to each polygon whose triangle completion has opposite orientation.}
\label{vertex}
\end{figure}

There are two types of polygons in the dot diagram:
The type A polygon is the one whose triangle completion has the same orientation as the big triangle, and
the type B polygon has opposite orientation.  A bipartite  \footnote{ A bipartite network has vertices colored with the black or white, and there are 
no edges connecting the vertices with the same color.} network and the quiver can be constructed directly from the dot diagram \footnote{Roughly speaking, the network is the $(p,q)$ five brane web with two types of decoration on brane junctions.}: we put a colored vertex inside each polygon using the following rule (see figure. \ref{vertex}):

a: Assign a white vertex to each type A polygon. 

b: Assign a black vertex to each type B polygon.

\begin{figure}[htbp]
\small
\centering
\includegraphics[width=10cm]{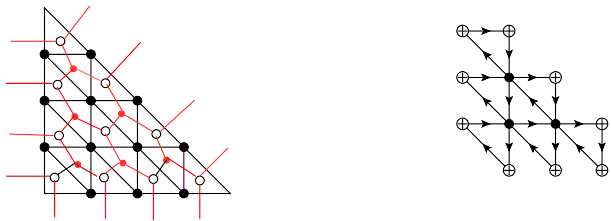}
\caption{Left: The dot diagram and the bipartite network. Right: The quiver from the network: the black dot is the gauged node while the white one is the flavor node.}
\label{quiverassignment}
\end{figure}

 A bipartite network is formed by connecting the white vertex and black vertex if there is a common edge between two corresponding polygons (vertices with the same color are never connected). 
  Moreover, an extra  line is coming out of the boundaries for the boundary polygon. The network formed in this way is always bipartite but there may be vertices with only two edges. 
 We can use the following moves to get rid of degree two vertices and get another bipartite network: Remove degree two vertices and then use the contraction to merge the line connecting 
 the vertices with the same color. After this reduction, one can find a quiver from the network using the following rule:
 Assign a quiver node to each surface and the quiver arrows are determined by the black vertices, namely there is 
 a clockwise closed circles around it, see figure. \ref{quiverassignment} \footnote{The quiver diagram is very much like the dot diagram itself, however, usually more 
 than one black dot represent a single quiver nodes. It is easy to read the quiver directly from the dot diagram without drawing the network after some practices.}. 

 When there are more than three punctures, we start with a regular ideal triangulation and take a closed loop connecting all
 the punctures, and the boundary edges on this closed loop  are decorated using the information of the Young Tableaux of 
 the punctures in the same way as the triangle.  The decoration of the other internal edges are automatically determined by the $S$ duality property
which is studied in detail in  \cite{Nanopoulos:2010ga}.
 After the decorations on the edges of all the triangles in the triangulation,  one can do the tessellations on each triangle using the minimal polygon
  and find the network, quiver, etc.  Many examples would be given in following sections..

 Remember that we have taken a clockwise convention in doing the decoration of the boundary edges and the 
 quiver arrows. One can take the anti-clockwise orientation for the decoration and the choice of the quiver arrows, and they
 will give the equivalent result.

\subsubsection{Quiver with potential and mutation}
\begin{figure}[htbp]
\small
\centering
\includegraphics[width=8cm]{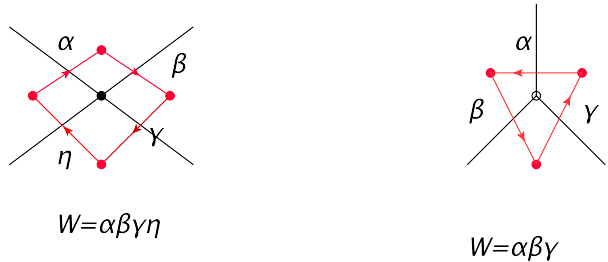}
\caption{ The rules for defining the potential for the quiver from bipartite network. }
\label{potential}
\end{figure}

The BPS quiver of the $\mathcal{N}=2$ theory is derived from the bipartite network as described in some detail above. The potential $W$ can also 
be read pretty easily from the network, see figure. \ref{potential}: 

1.  There is a potential term for each vertex of the network.

2. Each edge attached to this vertex represents a quiver arrow, and the potential term is a cyclic product of all the 
edges (the quiver arrows) attached on the vertex. 

This assignment of the potential for the quiver is exactly like the rule for the corresponding $\mathcal{N}=1$ 
quiver gauge theory from the network.  The BPS quiver is actually a pair $(Q, W)$ and the above rule makes sure
that under the square move, the two quivers with potentials $(Q,W)$ and $(Q^{'},W^{'})$ are related by 
the mutation rules.  This can be checked easily if  the quiver mutations are represented by the square moves.

Now a major difference with the $A_1$ case is that not all of our quivers are acyclic, namely, there are 
two cycles in the quiver. To eliminate these 2-cycles, one need the quadratic superpotential term associated
with them, but these super potentials are sometimes missing. This fact is actually important for the consistency
of our construction. Such quiver is not suitable for our later study of the spectrum using the quiver mutations which
require the absence of 2-cycles, and one need to use the quiver representation theory directly to study them.

The BPS quiver $(Q,W)$ constructed above have the following features:

a. The total number of quiver nodes are equal to $2n_r+n_f$, where $n_r$ is the rank of the gauge group and 
$n_f$ is the number of the mass deformations.

b. The rank of the quiver is equal to $2n_r$, this has been checked in many cases, it would be 
nice to have a general proof though.

c. The maximal number of arrows between two quiver nodes are two.

\subsubsection{Two flips}
\begin{figure}[htbp]   
\small
\centering
\includegraphics[width=4cm]{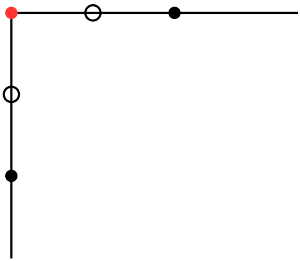}
\caption{The network would be non-minimal if the boundary of the dot diagram has this form at any vertex.}
\label{nonmini}
\end{figure}

 Different triangulations of the same bordered Riemann surface are related by  a sequence of local move called flip which relate 
  two triangulations of the quadrilateral.  In \cite{Xie:2012dw} we proved that if the dot diagram for  the quadrilateral
   does not have the  "bad" configuration shown in figure. \ref{nonmini}, the corresponding quivers of two triangulations (include the flavor nodes)
   are related by quiver mutation (or equivalently the two networks are related by square moves).  
  Even for the "bad" corner,  if we consider only the quiver nodes represented by the closed surfaces, 
  the quivers from different triangulations are still related by quiver mutations. This is good enough for us, since only the quiver 
  nodes associated with the closed surfaces are included into the BPS quiver.
  
Moreover, a sequence of quiver mutations acting only on the quiver nodes inside the triangle is also 
very useful for our later study of the BPS states counting. We call such quiver mutation sequences as 
"triangle" flip. In the following, we will provide some details on these two types of flips.

\begin{figure}[htbp]
\small
\centering
\includegraphics[width=10cm]{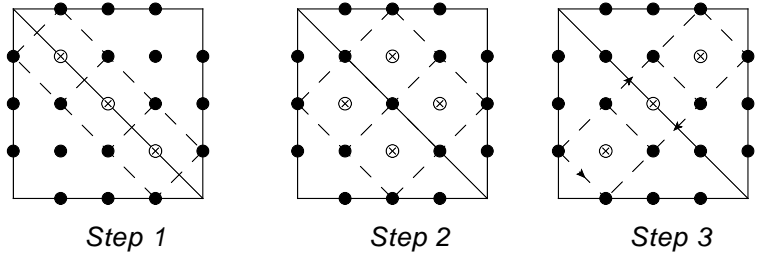}
\caption{Three steps for quiver mutations representing the flip for the quadrilateral with full punctures.}
\label{Gmutation1}
\end{figure}

\paragraph{Quadrilateral flip}
This sequence of quiver mutations representing the quadrilateral flip are first discovered  by Fock-Goncharov (FG) for the full puncture case in \cite{fock-2003}, and it is later generalized to  general cases in \cite{Xie:2012dw}.
The FG rules is best described using the dot diagram on the quadrilateral in which black dots are the quiver nodes.
The quiver mutations representing flip can be done in $N-1$ steps: in step $i$, we inscribe a rectangle with lengths $i\times (N-i)$ \footnote{We ignore an irrelevant normalization factor here.}  along with the diagonal edge, i.e. 
inside the quadrilateral (the sides with length $(N-i)$ is in parallel with the diagonal edge), 
then further decompose the rectangle into unit squares and we mutate the quiver nodes at the center of each 
little square at this step, see figure.~\ref{Gmutation1} for the description of  $A_3$ theory.
The quiver after these sequence is the same as the quiver from the quadrilateral derived by flipping the diagonal edge of the original one. 
Notice that in each step the mutated quiver node has four  quiver arrows. The total number of quiver mutations for one flip is 
\begin{equation}
N_m=\sum_{i=1}^ki(k-i)={1\over 6}(N^3-N).
\end{equation}

\begin{figure}[htbp]
\small
\centering
\includegraphics[width=10cm]{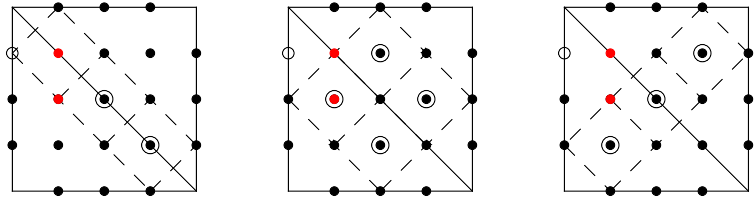}
\caption{Three steps for quiver mutations representing the flip with non-full puncture, and the two red nodes are identified as a single quiver node. In first step, the quiver
nodes represented by the red dot has six arrows on it, so we do not mutate it; In second step, the quiver node represented by the red node  has 
four arrows, and we mutate it.}
\label{Gmutation2}
\end{figure}

The quiver mutation sequences for the flip are found  for the non-full puncture case if the glued network is minimal. In
this case, usually a quiver node is represented by more than one black dots in the dot diagram.
One still have the same rectangle and unit square decomposition for each step, but  we only do the quiver mutations for 
the quiver nodes with four arrows.   See figure. \ref{Gmutation2} for the quiver mutation sequences representing the flip of the quadrilateral with a non-full puncture puncture.

The number of quiver mutations realizing the flip in the case of the non-full puncture can be counted explicitly if there 
is only one non-full puncture with partitions $[n_1, n_2,\ldots n_r]$, and the total number of flips would be
\begin{equation}
N_m-\sum_{i=1}^r{1\over 6}(n_i^3-n_i).
\end{equation}
One can count the number of quiver mutations representing the flips case by case for more general quadrilaterals.

There is a very nice heuristic way of explaining why mutating the quiver nodes with four quiver arrows in each step. Let's regard the 
quiver from the bipartite network as a four dimensional $\mathcal{N}=1$ quiver gauge theory by assigning $SU(N)$ to each quiver nodes (including 
the nodes associated with the open surfaces, which are the flavor groups.). Then the theory is anomaly free, and the flavors for each
gauge group are $N_f=2N, 3N, etc$ for the minimal network we considered thus far.  Only for $N_f=2N$ or the quiver nodes with four arrows, the gauge group is in 
conformal window, and one could do Seiberg duality (or quiver mutations) on it!

If the glued network for the quadrilateral is not minimal (there is a "bad" corner for the dot diagram and there are quiver nodes with only two quiver arrows on it.), the two networks
associated with two triangulations of the quadrilateral are not related by square moves.  We now state that the quivers associated with the closed surfaces (consider only the gauged quiver nodes)
 are still related by the quiver mutations. 
 
\begin{figure}[htbp]
\small
\centering
\includegraphics[width=10cm]{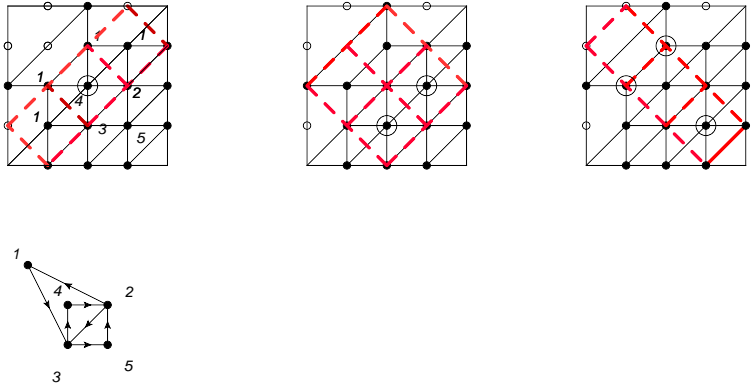}
\caption{The quiver mutation sequences representing the flip for the quadrilateral with a ``bad" corner.}
\label{Gmutation3}
\end{figure}

One need special treatment for the quiver nodes with two arrows, namely those quiver nodes with $N_f=N$. By analogy with the Seiberg
duality, one can still  do the Seiberg duality on the quiver nodes with $N_f=N$ if we assign the rank $N$ to all the quiver nodes,
 but the rank of the gauge group becomes $0$ after the Seiberg duality, so this quiver .
node is frozen after the mutation. Then some other nodes may have five arrows where one arrow is connected with this zero rank gauge group,  and we count the effective
arrows of the quiver nodes by ignoring such type of arrows. The mutations sequences for the quadrilateral flip are found in the following steps: 
still use the inscribed rectangular in each step, and mutate the non-frozen quiver nodes with effective number of flavors $N_f=N$ or $N_f=2N$.
For example, the quiver mutations representing the flip of quadrilateral in figure. \ref{Gmutation3} are:
\begin{equation}
(\mu_4),(\mu_2,\mu_3),(\mu_1, \mu_5).
\end{equation}
 
\begin{figure}[htbp]
\small
\centering
\includegraphics[width=4.5cm]{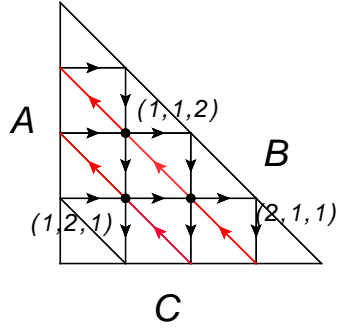}
\caption{ Each quiver node inside the triangle can be labeled by three integer numbers $(a,b,c)$ satisfying the relation $a+b+c=N$, here $a$ is the 
distance to the sides labeled by $A$, etc.  }
\label{label}
\end{figure}

\paragraph{Triangle flip}
There are another flip called "triangle" flip acting only on the  quiver nodes inside each triangle, which turns out  to be very important for 
our later study of BPS states counting.  Let's first
study a triangle with three full punctures and we will describe the mutation sequences representing the triangle flip. To describe the sequences, 
it is useful to label the quiver nodes inside each triangle with three non-negative integers $(a,b,c)$ such 
that the distances to the sides $A, B, C$ satisfying the following relation (see figure. \ref{label}):
\begin{equation}
a+b+c=N. 
\end{equation}
Let's take side $C$ as a reference side, and the mutation sequences for the triangle flip could be described in $(N-2)$ steps: in each step $1 \leq i\leq N-2$,  there 
are $N-1-i$ ordered sub steps: in each substep $(1\leq j\leq N-1-i)$ starting with $j=1$, we mutate quiver nodes with label $(a, b, N-j-1)$.
The total number of quiver mutations is 
\begin{equation}
N_s=\sum_{i=1}^{N-2}\sum_{j=1}^{N-i-1}j={1\over 6} (N^3-3 N^2+2N).
\end{equation} 

\begin{figure}[htbp]
\small
\centering
\includegraphics[width=12cm]{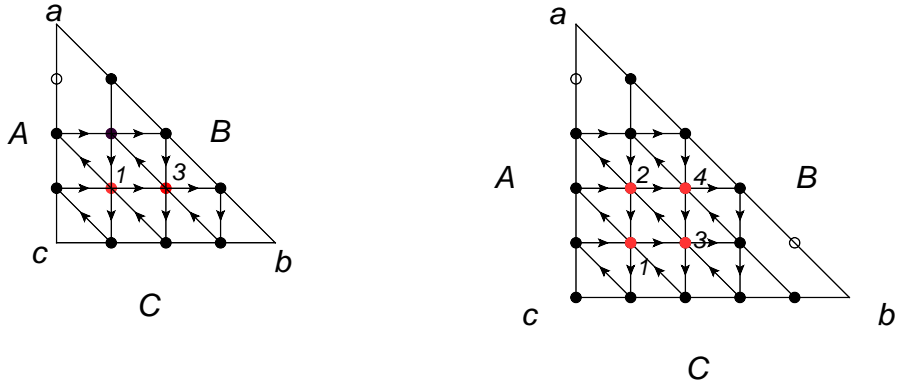}
\caption{Left: A  triangle with a non-full puncture for $A_3$ theory. Right: A  triangle with two non-full punctures for $A_4$ theory.  }
\label{tflip}
\end{figure}

The triangle flip for the non-full puncture can be defined in a similar way.  Let's start with the simplest case where  the only non-full puncture has
 partition $[n_1,1,1,\ldots, 1]$, see figure. \ref{tflip}.  The only difference from the full puncture case is that
some of the inside nodes are missing (the boundary node are filling inside though which we will not count  as the inside node.). 

There are two edges ($ab$ and $bc$) representing full punctures  which could be used to glue other triangles, and we would like to describe 
the triangle flip relative to these two edges. The labeling of the  remaining quiver nodes inside the triangle are the same as the full puncture case: 
they are labeled by $(a,b,c)$  with $1\leq c\leq N-n_1-1$. The triangle flip with respect to edge $B=ab$ is represented by the following mutation sequences:

Step 1: Mutate quiver nodes with label $(a,i, c)$ starting from $i=N-2$ and ending with $i=1$.

Step 2: Mutate quiver nodes with label $(a,i,c)$ staring from $i=N-2$ and ending with $i=2$, here only the quiver nodes with coordinate $c\leq N-n_1-2$ are mutated.

Step j: Mutate quiver nodes with label $(a,i,c)$ staring from $i=N-2$ and ending with $i=j$, moreover, we only mutate quiver nodes with $c\leq N-n_1-j$.

The above steps stop for $c=1$. According to above procedure, the "triangle" flip for the quiver on the left of figure. \ref{tflip} has the following mutation sequences:
\begin{equation}
\mu_1,\mu_3.
\end{equation}

If there is another non-full puncture $b$ with partition $[m_1, 1,1,\ldots,1]$, then some of the inner quiver nodes around puncture $b$ will be missing as 
shown in figure. \ref{tflip}. The "triangle" flip is implemented using the similar quiver mutation sequences:

Step j: Mutate quiver nodes with label $(a,i,c)$ staring from $i=N-2$ and ending with $i=j$, moreover, we only mutate quiver nodes with $c\leq N-n_1-j$.

So the mutation sequences implementing the "triangle" flip of the right triangle is 
\begin{equation}
[\mu_1, (\mu_2, \mu_3), \mu_4], [\mu_1, \mu_3].
\end{equation}

For more general configurations, one can also find similar mutation sequences for the 
triangle flip, since it seems that there is  not a uniform formula, we choose not to present the details here.

\subsection{One boundary}
In this subsection, we are going to use  maximal green mutation to find the finite spectrum of  a theory whose BPS geometry
has only one boundary, i.e. a disc with several marked points. 
The underlying $\mathcal{N}=2$ theory is a general Argyres-Douglas theory  as discussed in detail in \cite{Xie:2012hs}.

\subsubsection{Disc with full punctures}
Let's consider the higher rank generalization of a disc with $n$ marked points whose  Young Tableaux are all full. The corresponding 
irregular singularities are identified in \cite{Xie:2012jd} and many other properties of these theories are studied in \cite{Xie:2012hs}.  In particular, if there 
are $2(k+1)$ marked points, the theory is the so-called $(A_{N-1}, A_{kN-1})$ theory \footnote{The meaning of this label is that the BPS quiver
is a direct product of the two corresponding Dynkin diagrams.}.

We are going to use the quadrilateral flip and the triangle flip to find the finite BPS chamber of this class of theories. 
Let's define an internal edge as green if the quiver nodes on this edge are all green in the green mutation. 
The idea for finding the BPS spectrum  is the following:  Do the quadrilateral flips on the green edge in a random way, and stops if  no internal green edge is left;
finally the triangle flip are done for  the quiver nodes inside each triangle.
One could also do the triangle flip first and then do the quadrilateral flip, which is actually equivalent to the previous prescriptions.
In the following, we are going to do the quadrilateral first, and let's discuss several simple examples in the following part of this subsection.

\begin{figure}[htbp]
\small
\centering
\includegraphics[width=10cm]{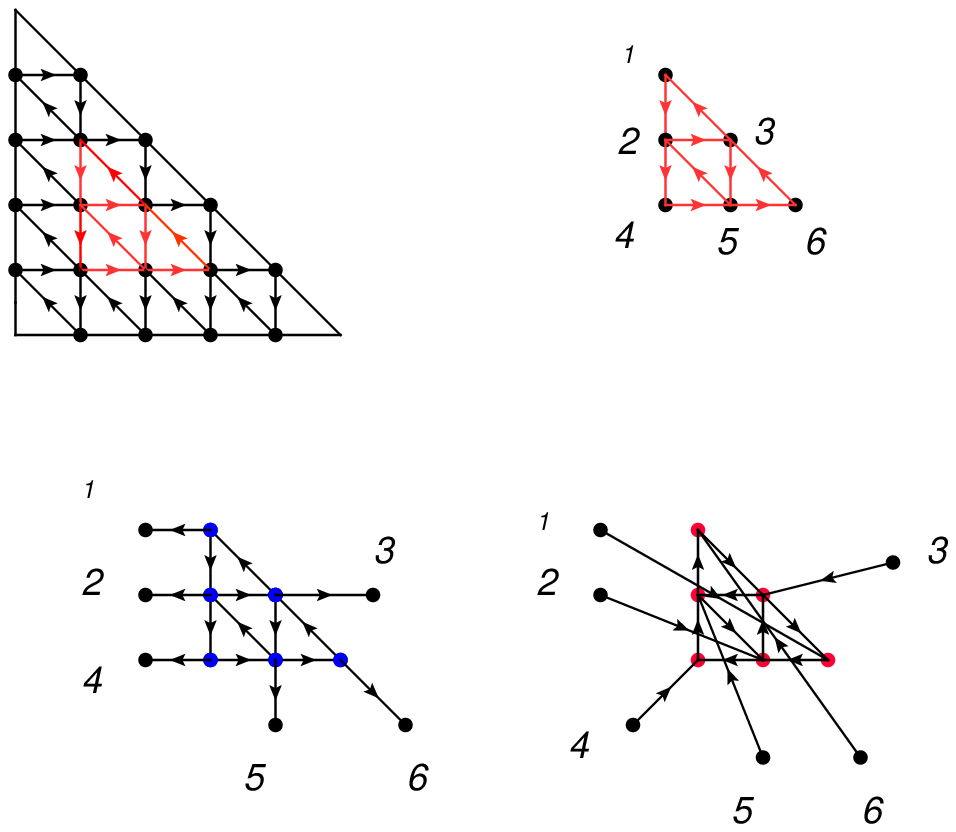}
\caption{Top: The BPS quiver from the triangulation of a disc with three full punctures. Bottom: The initial and final configuration of the maximal green mutation.  }
\label{3puncture}
\end{figure}

\begin{mydef}
 Consider a disc with three full punctures, and the triangulation is just a triangle and the quiver is given in figure. \ref{3puncture}. Since there is no 
 quadrilateral here, we only need to do the triangle flip, and the maximal green mutation sequence is the same as the triangle flip. The 
 number of BPS states in this chamber is
\begin{equation}
N_{bps}=N_s={1\over 6} (N^3-3 N^2+2N).
\end{equation}
\end{mydef}
We conjecture that this is the minimal chamber, and there might be other chambers which could be found using random green mutations.
The quiver shown in figure. \ref{3puncture} has $N=5$, and the maximal mutation sequences are
\begin{align}
&\mu_1, (\mu_2, \mu_3), (\mu_4,\mu_5,\mu_6) \nonumber,\\
&\mu_1, (\mu_2, \mu_3) \nonumber,\\
&\mu_1.
\end{align}
The mutation orders in each bracket is irrelevant. 

The number of BPS states in the minimal chamber might be derived in the following simple way. Assume that the number of BPS states is a smooth function of $N$, and it scales 
as $N^3$. Since  for $N\leq2$, the theory is trivial and the number of BPS states is zero, the number of states in the minimal chamber takes the following form
\begin{equation}
f(N)=aN(N-1)(N-2).
\end{equation}
There is only one quiver node for $N=3$ and the number of BPS states for this theory is one.
Using the simple data $f(3)=1$, we find $a={1\over 6}$ which is exactly the answer derived from the mutations. Amazingly, we could 
get the right number of the BPS states by using this simple assumptions.

\begin{mydef}
Consider a disc with four full punctures which is the BPS geometry for the  $(A_{N-1}, A_{N-1})$ theory. The maximal green mutation sequences involves 
one quadrilateral flip and two triangle flips, and the number of BPS states in this chamber is
\begin{equation}
N_{bps}=N_m+2N_s={1\over 6}(N^3-N)+{1\over 3} (N^3-3 N^2+2N)={1\over2} N(N-1)^2.
\end{equation}
The charge vector of the BPS states can be easily found from the green mutation sequences.
\end{mydef}

Let's write explicitly the mutation sequences for $N=4$. It is important to track the position of the  quiver nodes, i.e.  
whether it is inside the triangle or on the diagonal edge, since we need to do the triangle flip at the end. In the example, 
the quiver nodes $1,4,6$ are grouped inside one triangle after the flip, and $3,5,7$ is put inside another triangle.  
So the mutation sequences are 
\begin{eqnarray}
quadrilateral~flip:~(\mu_1,\mu_2,\mu_3)~(\mu_4,\mu_5,\mu_6,\mu_7)~(\mu_8,\mu_9,\mu_2), \nonumber\\
triangle~flip~ 1:~\mu_1, (\mu_4,\mu_6),\mu_1, \nonumber\\
triangle~flip~ 2:~\mu_3,(\mu_5,\mu_7),\mu_3.
\end{eqnarray}
\begin{figure}[htbp]
\small
\centering
\includegraphics[width=10cm]{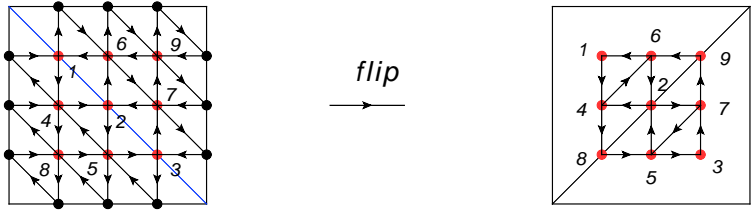}
\caption{ Left: The BPS quiver from a disc with four full punctures. Right: The  position of the original quiver nodes after the quadrilateral flip.}
\label{4puncture}
\end{figure}

\begin{mydef}
Consider a disc with five full punctures which is either a type I AD theory if N is even, or type II AD theory if 
N is odd \cite{Xie:2012jd}. One chamber involves two quadrilateral flips and  three triangle flips, so the number of states are
\begin{equation}
N_{1}=2N_m+3N_s=1/6 N (N-1)(5N-4).
\end{equation}
In another chamber, there are three big flips and also three triangles, and the number of the states are
\begin{equation}
N_{2}=3N_m+3N_s=1/2 N (N-1)(2N-1).
\end{equation}
\end{mydef}

\begin{figure}[htbp]
\small
\centering
\includegraphics[width=10cm]{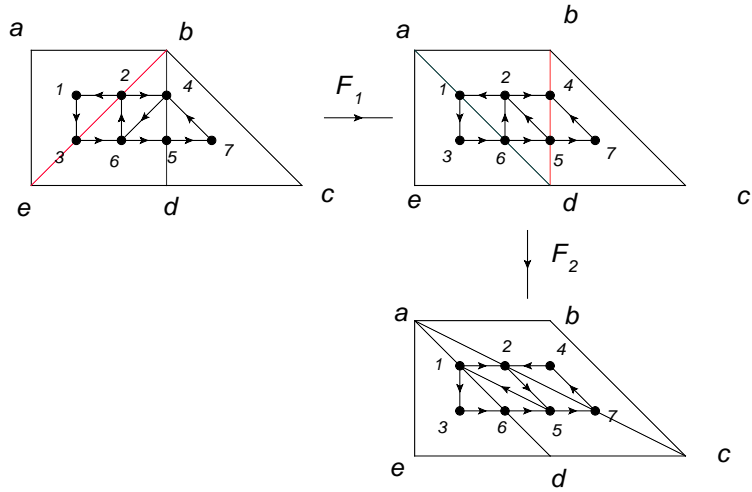}
\caption{ The flip sequences and the quiver positions for chamber found using two quadrilateral flips.}
\label{pentagon1}
\end{figure}

\begin{figure}[htbp]
\small
\centering
\includegraphics[width=10cm]{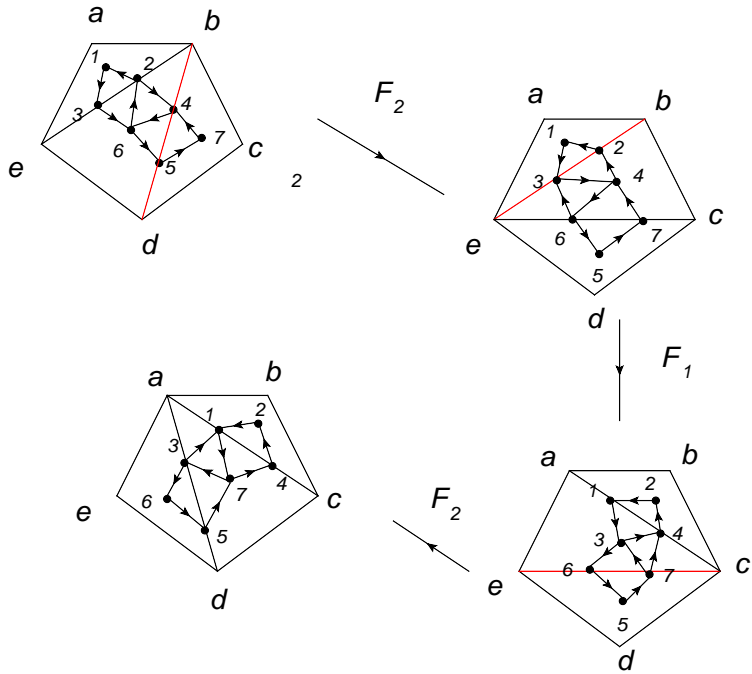}
\caption{The flip sequences and the quiver positions for chamber involving three quadrilateral flips.}
\label{pentagon2}
\end{figure}

Let's give the mutation sequences for these two chambers for $N=3$ (higher rank cases are exactly the same). The mutation sequences involving two quadrilateral flips  are ( see figure. \ref{pentagon1}):
\begin{eqnarray}
F_1: ~ (\mu_2, \mu_3),~(\mu_1, \mu_6), \nonumber\\
F_2: ~ (\mu_4, \mu_5),~(\mu_2, \mu_7), \nonumber\\
triangle~flip:~  (\mu_3, \mu_4, \mu_5).
\end{eqnarray}
The mutation sequences for the chamber involving three quadrilateral flips are (see figure. \ref{pentagon2})
\begin{eqnarray}
F_2: ~ (\mu_4, \mu_5),~(\mu_6, \mu_7), \nonumber\\
F_1: ~ (\mu_2, \mu_3),~(\mu_1, \mu_4), \nonumber\\
F_2:~(\mu_6, \mu_7),~(\mu_3, \mu_5), \nonumber \\
triangle~flip:~  (\mu_2, \mu_6, \mu_7).
\end{eqnarray}
Although the final quivers for two flip sequences are the same, the positioning of the  original quiver nodes are quite different. For example, 
quiver nodes  $(3,4,5)$ are  inside the triangle in first chamber, but   quiver nodes $(2, 6, 7)$ are playing this role in another chamber. 

For a disc with $n+3$ full punctures, the minimal flip number are $n$ and the maximal number of flips are ${n(n+1)\over 2}$, so 
we find the following chambers:
\begin{align}
&f_{min}=nN_m+(n+1)N_s, \nonumber \\
&f_{max}={n(n+1)\over 2}N_m+(n+1)N_s. 
\end{align}
Notice that $f_{max}$ is not necessarily the number of BPS states in the maximal chamber.  

\newpage
\paragraph{Generalized pentagon identity I}
The pentagon identify for the $A_1$ theory is nicely represented by the five flip sequences of the pentagon which actually
implies the basic quantum dilogarithm identify. As we show earlier using the maximal 
green mutation, this quantum dilogarithm identity is the wall crossing formula for the $A_2$ Argyres-Douglas theory:
The product on one side of the identify corresponds to the chamber involving two flips $F_1, F_2$ while the product on the right is derived from
the flip sequences $F_2, F_1, F_2$. 

Now similar quantum  dilogarithm identity  can be found using wall crossing formula for  the higher rank theory which is also represented 
by a pentagon. Now  each quadrilateral flip is represented by a sequence 
of quiver mutations.  However, unlike the $A_1$ case,  the triangle flips are also important for finding the BPS spectrum. 
 We need to list the charge vectors of two chambers for $N=3$, which are  crucial for writing down the 
 quantum dilogarithm identity. Let's call the chamber involving 2 flips (resp. 3 flips) as 
chamber I (chamber II), and the charge vectors for chamber I are
\begin{align}
&F_1: ~ (\gamma_2, \gamma_3),~(\gamma_1+\gamma_3, \gamma_2+\gamma_6), \nonumber\\
&F_2: ~ (\gamma_4, \gamma_5),~(\gamma_5+\gamma_6, \gamma_4+\gamma_7), \nonumber\\
&triangle~flip:~  (\gamma_1, \gamma_6, \gamma_7);
\end{align}
Similarly, the charge vectors for chamber II are
\begin{align}
&F_2: ~ (\gamma_4, \gamma_5),~(\gamma_6+\gamma_5, \gamma_4+\gamma_7), \nonumber\\
&F_1: ~ (\gamma_2+\gamma_4, \gamma_3+\gamma_5+\gamma_6),~(\gamma_1+\gamma_3+\gamma_5+\gamma_6, \gamma_2+\gamma_4+\gamma_7), \nonumber\\
&F_2:~(\gamma_3, \gamma_2)~(\gamma_1+\gamma_3, \gamma_2+\gamma_6), \nonumber \\
&triangle~flip:~  ( \gamma_1, \gamma_6, \gamma_7).
\end{align}

A quantum dilogarithm function $E(y^\gamma)$ is associated for each BPS particle with charge $\gamma$ and all the BPS particle from one chamber form
a ordered product. The wall crossing formula means that the two products from two chambers are the same.
Using the mutation sequences and the charge vectors, we have 
\begin{align}
&E(y^{\gamma_2})E(y^{\gamma_3})E(y^{\gamma_1+\gamma_3})E(y^{\gamma_2+\gamma_6})E(y^{\gamma_4})E(y^{\gamma_5})E(y^{\gamma_5+\gamma_6})E(y^{\gamma_4+\gamma_7})E(y^{\gamma_1})E(y^{\gamma_6})E(y^{\gamma_7}) \nonumber \\
&=E(y^{\gamma_4})E(y^{\gamma_5})E(y^{\gamma_5+\gamma_6})E(y^{\gamma_4+\gamma_7})E(y^{\gamma_2+\gamma_4})E(y^{\gamma_3+\gamma_5+\gamma_6})E(y^{\gamma_1+\gamma_3+\gamma_5+\gamma_6})E(y^{\gamma_2+\gamma_4+\gamma_7})  \nonumber \\
&E(y^{\gamma_3})E(y^{\gamma_2})E(y^{\gamma_1+\gamma_3})E(y^{\gamma_2+\gamma_6})E(y^{\gamma_1})E(y^{\gamma_6})E(y^{\gamma_7}).
\end{align}
It is interesting to note that the charge vectors from the triangle flip are the same for two chambers and they 
are living at the far left of the quantum dilogarithm product, so they  do not participate in the wall crossing process and they can be cancelled out in the quantum dilogarithm identity \footnote{We thank A.Neitzke for the helpful discussion on this point.}:
\begin{align}
&E(y^{\gamma_2})E(y^{\gamma_3})E(y^{\gamma_1+\gamma_3})E(y^{\gamma_2+\gamma_6})E(y^{\gamma_4})E(y^{\gamma_5})E(y^{\gamma_5+\gamma_6})E(y^{\gamma_4+\gamma_7}) \nonumber \\
&=E(y^{\gamma_4})E(y^{\gamma_5})E(y^{\gamma_5+\gamma_6})E(y^{\gamma_4+\gamma_7})E(y^{\gamma_2+\gamma_4})E(y^{\gamma_3+\gamma_5+\gamma_6})E(y^{\gamma_1+\gamma_3+\gamma_5+\gamma_6})E(y^{\gamma_2+\gamma_4+\gamma_7})  \nonumber \\
&E(y^{\gamma_3})E(y^{\gamma_2})E(y^{\gamma_1+\gamma_3})E(y^{\gamma_2+\gamma_6}).
\end{align}

\begin{figure}[htbp]
\small
\centering
\includegraphics[width=12cm]{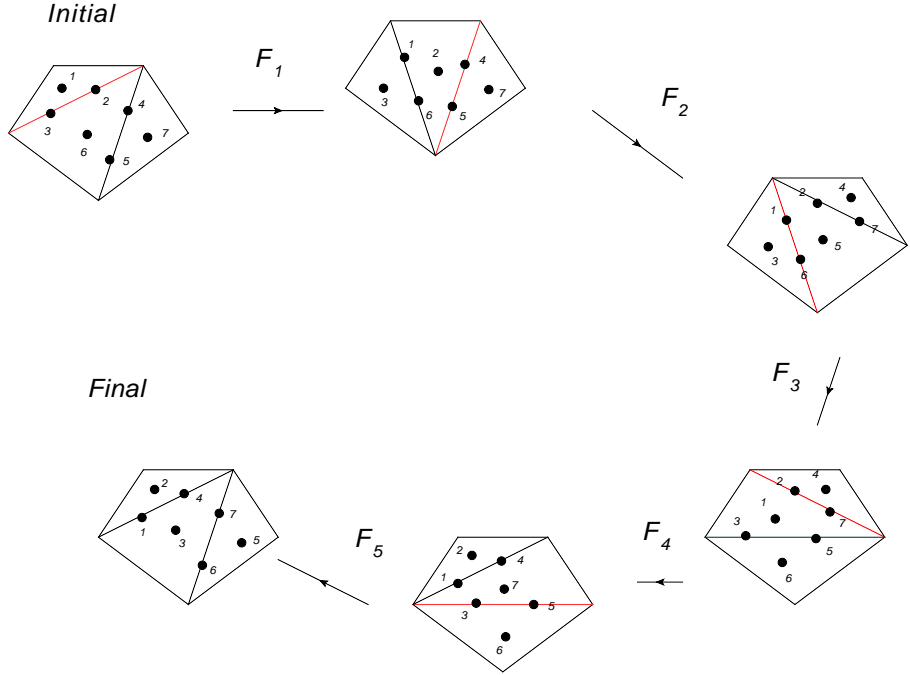}
\caption{The pentagon identity for higher rank pentagon.}
\label{5term}
\end{figure}

In fact, the cancellation of the BPS states from triangle flip in the quantum dilogarithm identity is true for any $N$ by explicitly  checking  the charge vectors. 
This is not so surprising from the quantum cluster algebra point of view, since the five quadrilateral flips would bring the coordinates back to their original values up to a permutation, so 
the quantum dilogarithm identity should only involve the mutations from the quadrilateral flips, see figure. \ref{5term}.
The mutation sequences and $c$ vector is shown in table. \ref{5termtable} for the five flips, and we can write the quantum dilogarithm identity 
\begin{align}
&E(y^{\gamma_2})E(y^{\gamma_3})E(y^{\gamma_1+\gamma_3})E(y^{\gamma_2+\gamma_6})E(y^{\gamma_4})E(y^{\gamma_5})E(y^{\gamma_5+\gamma_6})E(y^{\gamma_4+\gamma_7}) \nonumber \\
&E(y^{\gamma_2+\gamma_6})^{-1}E(y^{\gamma_1+\gamma_3})^{-1}E(y^{\gamma_2})^{-1}E(y^{\gamma_3})^{-1}E(y^{\gamma_2+\gamma_4+\gamma_7})^{-1}E(y^{\gamma_1+\gamma_3+\gamma_5+\gamma_6})^{-1} \nonumber \\
&E(y^{\gamma_3+\gamma_5+\gamma_6})^{-1}E(y^{\gamma_2+\gamma_4})^{-1}E(y^{\gamma_4+\gamma_7})^{-1}E(y^{\gamma_5+\gamma_6})^{-1}E(y^{\gamma_5})^{-1}E(y^{\gamma_4})^{-1}=1,
\end{align}
which is the same quantum dilogarithm identity derived from the wall crossing formula.

\begin{table}[h]
\centering
    \begin{tabular}{|c|c|>{\small}c|c|c|}
        \hline
                  Flips & Mutation & $c$ vector& sign\\ \hline
        $F_1$ &$ (\mu_2,\mu_3), (\mu_1,\mu_6)$&$(\gamma_2, \gamma_3),(\gamma_1+\gamma_3, \gamma_2+\gamma_6)$  &$(+,+),(+,+)$\\ \hline
          $F_2$ & $(\mu_4,\mu_5), (\mu_2,\mu_7)$ &$(\gamma_4, \gamma_5),(\gamma_5+\gamma_6, \gamma_4+\gamma_7)$ & $(+,+),(+,+)$ \\ \hline
           $F_3$ &$ (\mu_6,\mu_1), (\mu_5,\mu_3)$ &$( -\gamma_2-\gamma_6,-\gamma_1-\gamma_3),(-\gamma_2,-\gamma_3)$  & $(-,-),(-,-)$ \ \\ \hline
           $F_4$ & $(\mu_7,\mu_2), (\mu_1,\mu_4)$ & $\small{(-\gamma_2-\gamma_4-\gamma_7,-\gamma_1-\gamma_3-\gamma_5-\gamma_6),(-\gamma_3-\gamma_5-\gamma_6, -\gamma_4-\gamma_2)}$  &  $(-,-),(-,-)$ \ \\ \hline
             $F_5$ & $(\mu_5,\mu_3), (\mu_6,\mu_7)$ &  $(-\gamma_4-\gamma_7, -\gamma_6-\gamma_5),(-\gamma_5, -\gamma_4)$ &  $(-,-),(-,-)$ \ \\ \hline
           \end{tabular}
    \caption{The mutation data for the five flips of the pentagon. }
    \label{5termtable}
\end{table}

There is no need to calculate the cluster coordinates following the detailed mutation sequences, which would be a really tedious calculation. 
The easy way is to use the extended quiver introduced for the purpose of green mutation:  we  do green mutations for the first two flips and then red mutation for the 
next three flips, and every node is green again after five flips.
There is a general theorem for the mutation sequence of an extended quiver: if the final quiver nodes are  all green, then the final cluster coordinate
is identical to the initial one up to a permutation. Using this theorem, we can easily verify that the cluster coordinates are back to themselves after five 
flips.`

\subsubsection{General $(A_{N-1}, A_{nN-1+j})$ theory}
Other punctures are needed for the general Argyres-Douglas theory considered in \cite{Xie:2012jd}. Let's now consider general $(A_{N-1}, A_{nN-1+j})$ theory with $ 0<j<N$, and
the Stokes matrices analysis suggests that there are $j$ more marked points which are labeled by simple Young Tableaux.  The number of marked points and their labels are
\begin{equation}
full: 2(n+1);~~~~~simple: j
\end{equation}
The cyclic distribution of the marked points on the boundary of the disc is the following: there are $2(n+1)$ full punctures followed by $j$ simple punctures.

We will try to find the finite chamber using the following idea: do the quadrilateral flip sequences on the green edge
and do the triangle flips at the end. Again, it is important to track the position of the quiver nodes, i.e.
whether it is on the edge or inside the triangle, etc.  The story is pretty the same as the full puncture cases, so we just 
give a simple example and the interested reader can do the similar exercises for the other BPS geometry.

\begin{figure}[htbp]
\small
\centering
\includegraphics[width=12cm]{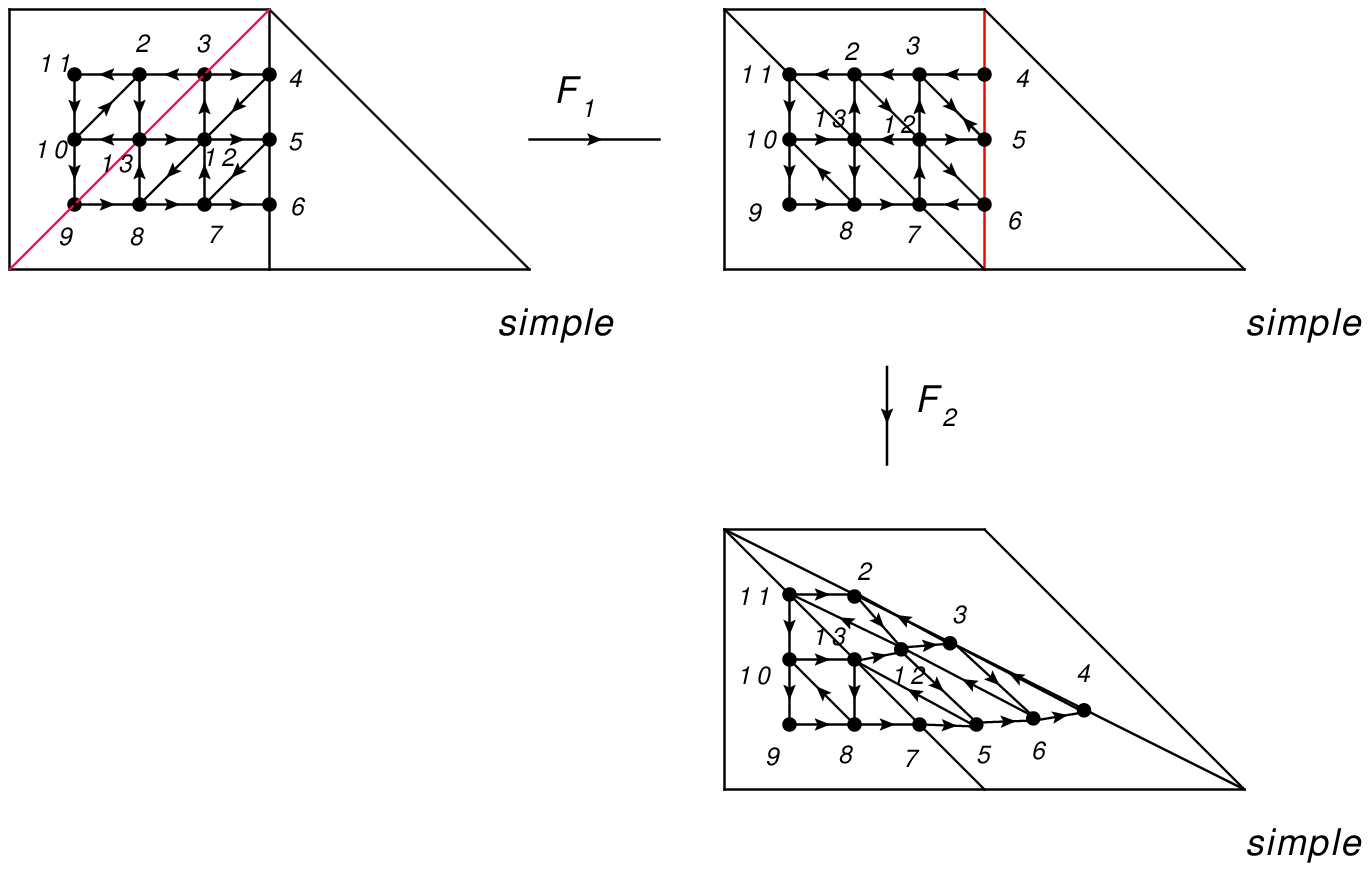}
\caption{The chamber involving two quadrilateral flips for the theory defined by a disc with four full punctures and one simple puncture. The 
quiver positions in each step are indicated. }
\label{nonful1}
\end{figure}

\begin{figure}[htbp]
\small
\centering
\includegraphics[width=12cm]{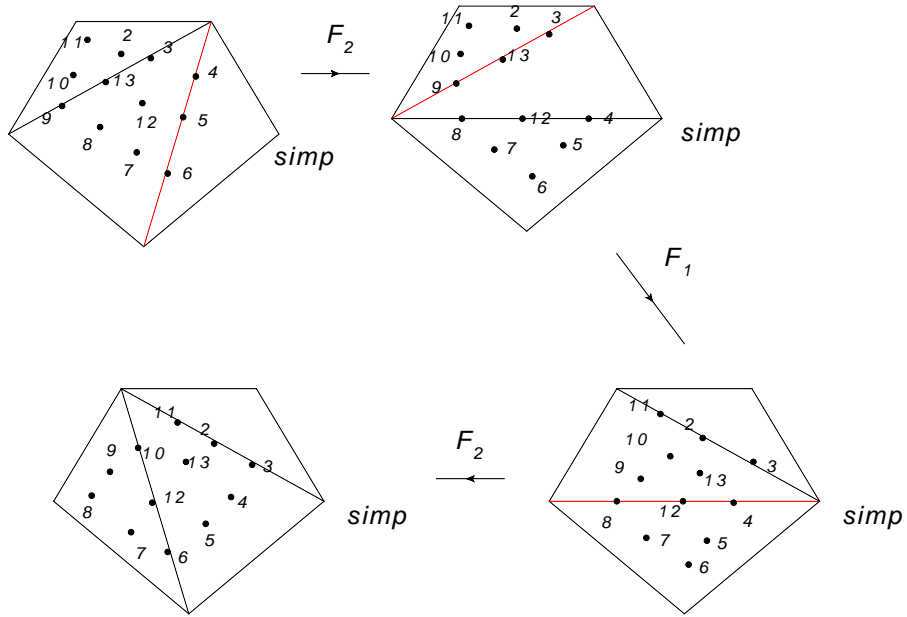}
\caption{The chamber involving three quadrilateral flips for the theory defined by a disc with four full punctures and one simple puncture. The 
quiver positions in each step are indicated. }
\label{nonful2}
\end{figure}

\begin{mydef}
Let's consider a five punctured disc with four full punctures and one simple puncture of $N=4$. The triangulations 
and the quiver are shown in figure. \ref{nonful1}. Using the two quadrilateral flips and triangle flips, one get the following maximal 
green mutation sequences:
\begin{align}
&F_1:~(\mu_{3},\mu_{13},\mu_9),(\mu_2,\mu_{10}, \mu_8, \mu_{12}), (\mu_{11}, \mu_{13}, \mu_{7}) \nonumber\\
&F_2:~(\mu_4,\mu_5,\mu_6), (\mu_3, \mu_{12}), \mu_2 \nonumber\\
&triangle~flip~1:~ \mu_9,( \mu_8,\mu_{10}), \mu_9  \nonumber\\
&triangle~ flip~2:~(\mu_6,(\mu_5, \mu_{12}), \mu_6  \nonumber\\
\end{align}
\end{mydef}
For another chamber involving three quadrilateral flips (see figure. \ref{nonful2}.), the mutations sequences are, 
\begin{align}
&F_2:~(\mu_4, \mu_5, \mu_6), (\mu_7, \mu_{12}), \mu_8, \nonumber\\
&F_1:~(\mu_3, \mu_9, \mu_{13}), (\mu_2, \mu_{10}), \mu_{11},\nonumber\\
&F_2:~(\mu_4,\mu_8,\mu_{12}), (\mu_5, \mu_7,\mu_9,\mu_{13}), (\mu_{6}, \mu_{12}, \mu_{10}), \nonumber\\
&triangle~flip~1: \mu_4, (\mu_5, \mu_{13}), \mu_4, \nonumber\\
&triangle~flip~2: \mu_8, (\mu_7,\mu_9), \mu_8,
\end{align}

\begin{figure}[htbp]
\small
\centering
\includegraphics[width=8cm]{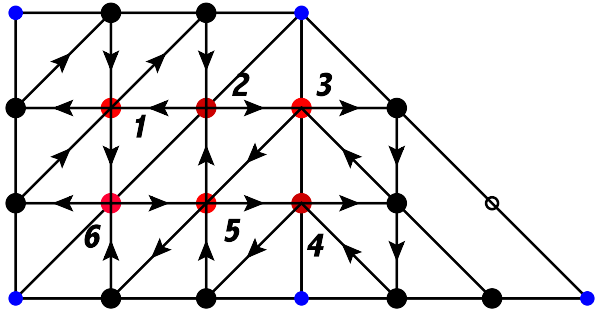}
\caption{ The quiver from pentagon with four full punctures and one simple puncture, here $N=3$.}
\label{pentagonII}
\end{figure}

\newpage
\paragraph{Generalized pentagon identity II}
The pentagon identity can be generalized to the case where there are several full punctures and simple punctures. Similarly, one 
chamber comes from doing two quadrilateral flips while the other one is derived using three flips. The triangulation and quiver for the pentagon with only one simple 
puncture are shown in figure. \ref{pentagonII} for $N=3$. 
The  mutation sequences and the charge vectors for two chambers is shown in table. \ref{onesimple}, and it
is straightforward to write the following pentagon identity (again the terms from the triangle flip can be 
dropped out):
\begin{align}
&E(y^{\gamma_2})E(y^{\gamma_6})E(y^{\gamma_1+\gamma_6})E(y^{\gamma_2+\gamma_5})E(y^{\gamma_3})E(y^{\gamma_4})E(y^{\gamma_4+\gamma_5})=\nonumber\\
&E(y^{\gamma_3})E(y^{\gamma_4})E(y^{\gamma_4+\gamma_5})E(y^{\gamma_2+\gamma_3})E(y^{\gamma_5+\gamma_6})E(y^{\gamma_1+\gamma_4+\gamma_5+\gamma_6})
E(y^{\gamma_2})E(y^{\gamma_5})E(y^{\gamma_2+\gamma_5})E(y^{\gamma_1+\gamma_6}).
\end{align}

\begin{table}[h]
\centering
    \begin{tabular}{|c|c|}
        \hline
                  ~&Chamber I  \\ \hline
          Mutations &$F_1: (\mu_2, \mu_6, \mu_1, \mu_5), F_2: (\mu_3, \mu_4, \mu_2), T_{flip}:(\mu_6,\mu_4)$  \\ \hline
           Charges&$(\gamma_2,\gamma_6, \gamma_1+\gamma_6, \gamma_2+\gamma_5), (\gamma_3,\gamma_4,\gamma_4+\gamma_5),(\gamma_1, \gamma_5)$\\ \hline
              ~&Chamber II \\ \hline
             Mutations &$F_2: (\mu_3, \mu_4, \mu_5), F_1: (\mu_2, \mu_6, \mu_1), F_2: (\mu_3, \mu_5, \mu_4, \mu_6), T_{flip}:(\mu_5,\mu_3)$  \\ \hline
   Charges&$(\gamma_3, \gamma_4, \gamma_4+\gamma_5), (\gamma_2+\gamma_3, \gamma_5+\gamma_6, \gamma_1+\gamma_4+\gamma_5+\gamma_6),
   (\gamma_2,\gamma_5, \gamma_2+\gamma_5, \gamma_1+\gamma_6), (\gamma_1, \gamma_5)$\\ \hline
             \end{tabular}
    \caption{The mutation data for two chambers of a pentagon with one simple puncture.}
    \label{onesimple}
\end{table}
The interested reader can work out the pentagon identity for other combinations of full punctures and  simple punctures.

\newpage
\paragraph{Sink-source sequences}
The BPS quiver for $(A_{N-1}, A_{k-1})$ from our network construction is mutation equivalent to the quiver formed by a product of $A_{N-1}$  and $A_{k-1}$ Dynkin diagram which 
gives the name for the theory in \cite{Cecotti:2010fi}.  The quiver mutation sequences for relating our quiver and the square $(A_{N-1}, A_{k-1})$ quivercan be readily found. Let's give an example for $N=5$ and $k=5$ whose
quiver is derived from a disc with four full punctures. The quiver mutations transforming the quiver from the triangulation to the $(A_4,A_4)$ form  are:
\begin{eqnarray}
\mu_1, (\mu_2, \mu_3), \mu_4, (\mu_5,\mu_6).
\end{eqnarray}
\begin{figure}[htbp]
\small
\centering
\includegraphics[width=12cm]{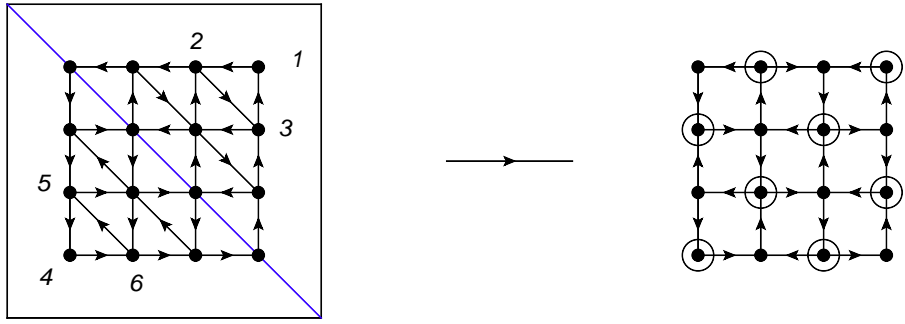}
\caption{The mutations transforming the quiver from the triangulation to the $(A_4,A_4)$ quiver.}
\label{A4A4}
\end{figure}

Let's now focus on the $(A_{N-1}, A_{k-1})$ quiver and choose a convention that the horizon direction is $(A_{k-1})$ quiver. 
The special feature of the subquiver in the horizontal direction and the vertical direction are that they are bipartite, and the quiver arrows form cyclic squares, see the left quiver in figure. \ref{A4A4}. 
Moreover, if a quiver node is a sink node in the horizontal direction, it would be a source node in the vertical direction.  
Let's denote the signature of a quiver node as $(-, +)$ if it is a sink in the horizontal direction and source in the 
vertical direction.  Define the quiver mutation sequences 
\begin{eqnarray}
\tau_{1}=\mu_{-+},\nonumber\\
\tau_{2}=\mu_{+-},
\end{eqnarray}
here $\mu_{-+}$ (resp. $\mu_{+-}$) is the quiver mutation on all the quiver nodes with signature $(-,+)$ (resp. $(+,-)$). It can be checked explicitly that
$(\tau_{1})^{h^{'}}$ and $(\tau_2)^{h}$ give two maximal green mutation sequences, where $h^{'}$ (resp. $h$) is the Coxeter number for group $A_{k-1}$ ($A_{N-1}$).  
Such sequences are found in \cite{Cecotti:2010fi}, here the charge vectors can be easily found using the maximal green mutations. When $N=2$, the quiver is just the bipartite quiver 
of $A_{k-1}$ Dynkin diagram. The Coxeter number of $A_1$ group and $A_{k-1}$ group are $two$ and $k$, and $\tau_2^2$ is the source green mutation sequence and 
gives a total of $k-1$ states which is actually the minimal chamber; $\tau_1^k$ is the sink green mutation sequence and gives a total of ${k(k-1)\over 2}$ states which is 
the maximal chamber.  In general, one of the green mutation sequence is the minimal chamber, but the other one might not be the maximal chamber.

\newpage

\subsubsection{Other AD theories from the disc}
The $(A,A)$ type AD theories involve only the full and simple punctures, and more general punctures appear for other 
type of AD theories considered in detail in \cite{Xie:2012jd}.  There are two more general classes whose BPS geometry 
involves a single disc. The type II AD theory which also has only full punctures and the simple punctures, but the  boundary
nodes of the simple puncture are gauged (we include the node on the edge of the simple puncture into our BPS quiver). 
 The strategy of finding the BPS spectrum is the following: 
 doing the quadrilateral flips and triangle flips as we did for the $(A, A)$ theory, and then do arbitrary green mutation sequences involving the boundary nodes! 

The BPS geometry of type III AD theory also involves a single disc, and the Young Tableaux for the marked points are more  fruitful.
The definition of type III AD theory includes a sequences of Young Tableaux which satisfies the following condition 
\begin{equation}
Y_n\subset Y_{n-1}\ldots \subset Y_{1},
\end{equation}
where  $Y_{1}$ is taken to be the full Young Tableaux so that the BPS quiver has the simple description \cite{Xie:2012jd}, 
and $Y_{i-1}$ is derived by further decomposing the columns of  $Y_i$.  The BPS geometry  is a disc with $2(n-1)$ Young Tableaux:
 $Y_n, Y_{n-1}, \ldots, Y_2, Y_n,\ldots, Y_2$ are arranged in cyclic order.
The strategy for finding the finite states chamber for type III AD theory is the following: do the mutation sequences corresponding to flips; and 
then do the green mutations on the edge nodes and internal nodes repeatedly.
This mutation sequences are very useful since many superconformal field theory engineered using Riemann surface and regular punctures has an realization 
as type III AD theory.
\begin{figure}[htbp]
\small
\centering
\includegraphics[width=10cm]{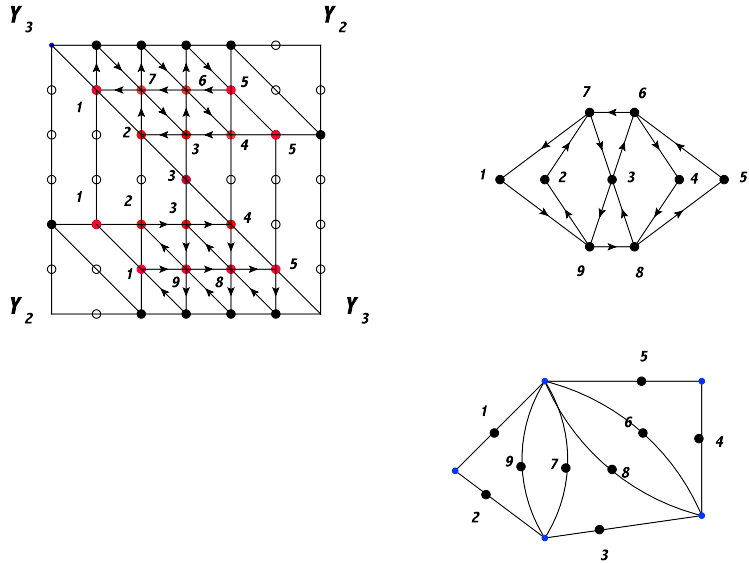}
\caption{The triangulation of the fourth punctured disc with ordered puncture $(Y_3, Y_2, Y_3, Y_2)$. The bottom part shows 
the triangulation of $A_1$ theory on a sphere with 5 punctures whose quiver is equivalent to the disc configuration. }
\label{typeIII}
\end{figure}

\begin{mydef}
Let's take $Y_3=[2,4]$, $Y_2=[2,1,1,1,1]$, and $Y_1=[1,1,1,1,1,1]$, and the dot diagram and quiver are shown in figure. \ref{typeIII}. 
This theory is in fact equivalent to the one engineered using  $A_1$ theory on 
a sphere with five punctures.  After doing the flip sequences, one need to do green mutations on the edge nodes and internal nodes repeatedly. 
One maximal green mutation sequences  are
\begin{equation}
(\mu_2,\mu_3,\mu_4), (\mu_6,\mu_7,\mu_8,\mu_9), (\mu_1, \mu_5, \mu_3),  (\mu_2, \mu_7, \mu_6, \mu_4, \mu_8, \mu_9, \mu_4, \mu_6, \mu_7, \mu_2)
\end{equation}
\end{mydef}

For the theory engineered using six dimensional $A_1$ theory compactified on a  sphere with $k$ punctures ,  there is a higher rank 
realization using type III AD theory with the following Young tableaux
\begin{equation}
Y_3=[2,k-1],~~Y_2=[2,1,1,\ldots,1],~~Y_1=[1,1,1,\ldots,1].
\end{equation}
The BPS geometry is a fourth punctured disc with  marked points $Y_3, Y_2, Y_3, Y_2$. It is not hard to find the finite chamber
using the mutation sequences representing the flips of this quadrilateral.

\newpage
\subsection{Riemann surface without punctures}
\subsubsection{Annulus with one marked point on boundary}
The next simplest BPS geometry is the annulus with one marked point on each boundary, which 
represents a gauge group coupled with two matter sectors. To find the finite spectrum,  one can not 
not do random quadrilateral flips on the green edges. This is the case where the flip sequences found in $A_1$ theory 
plays an important role here: we do the flip sequences found from the finite spectrum of the corresponding
$A_1$ theory with the same type of BPS geometry, and finally do the green mutations on the quiver nodes inside
each triangle.

\begin{figure}[htbp]
\small
\centering
\includegraphics[width=12cm]{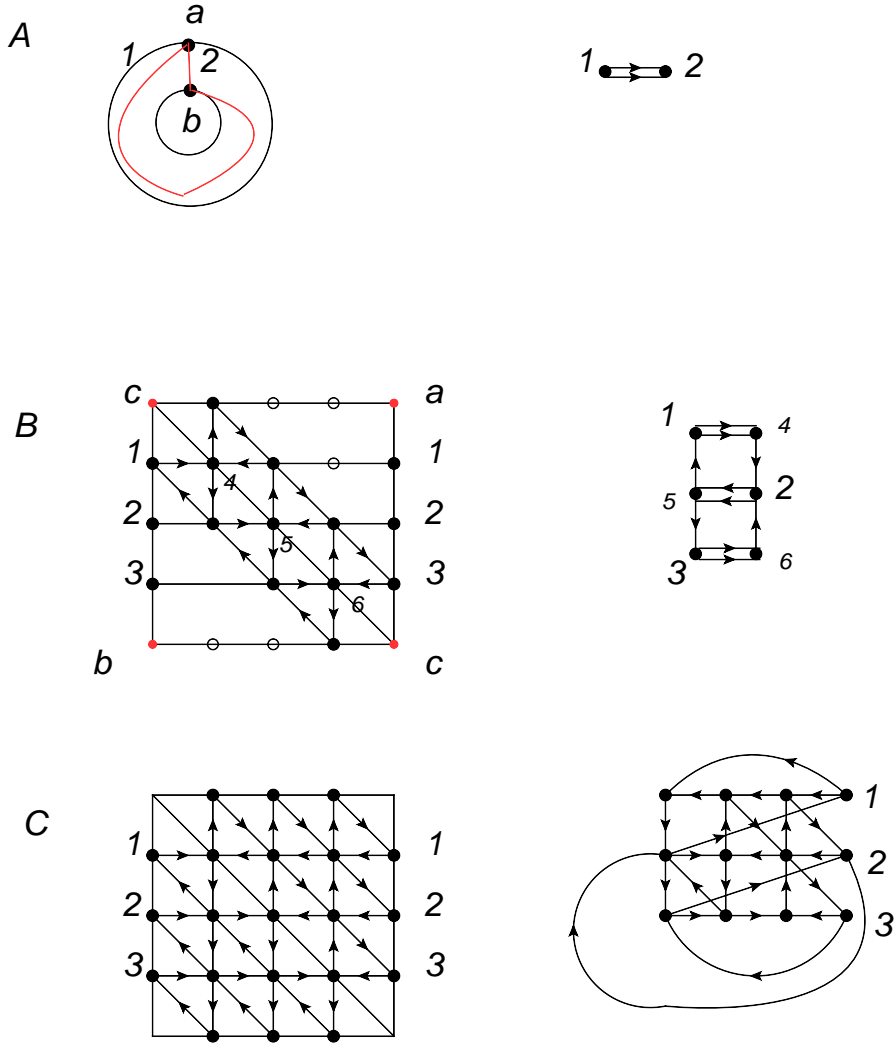}
\caption{A: The triangulation and quiver for the annulus with one marked point on each boundary, here $N=2$. B: The quiver for annulus  where each boundary has a simple puncture, here $N=4$.
C: The  quiver with two full punctures on each boundary of the annulus, here $N=4$.}
\label{SYM}
\end{figure}

\begin{mydef}
The BPS geometry is  an annulus with a simple puncture on each boundary from which a quiver can be found from the dot diagram on the triangulation, see figure. \ref{SYM}B.
We do the flip sequences found from the $A_1$ version: first mutate edge $1$ and then mutate edge $2$. Here some new features appear: one need to do more rounds of 
mutations.  The flip one edge $1$ is realized by the quiver mutations on node $1$, $2$, $3$, etc. The maximal green mutation 
sequences for the pure $SU(4)$ theory is
\begin{equation}
(\mu_1, \mu_2, \mu_3),~(\mu_4, \mu_5, \mu_6), ~(\mu_1, \mu_2, \mu_3),~(\mu_4, \mu_5, \mu_6).
\end{equation}
The order of mutations in each bracket is not important since the corresponding quiver nodes are disconnected.
There are 12 total BPS states, and in general this chamber for pure SU(N) theory has the following number of BPS states
\begin{equation}
N_{bps}=N(N-1),
\end{equation}
In fact, the BPS quiver is a sink-source product of affine $A_1$ diagram and $A_{N-1}$ Dynkin diagram, and the mutation 
sequence is just a generalization of the bipartite quiver. The result is the same as found in \cite{Lerche:2000uy, Alim:2011kw}.
\end{mydef}

\paragraph{Unique finite chamber for pure $SU(N)$ theory:}

\begin{figure}[htbp]
\small
\centering
\includegraphics[width=12cm]{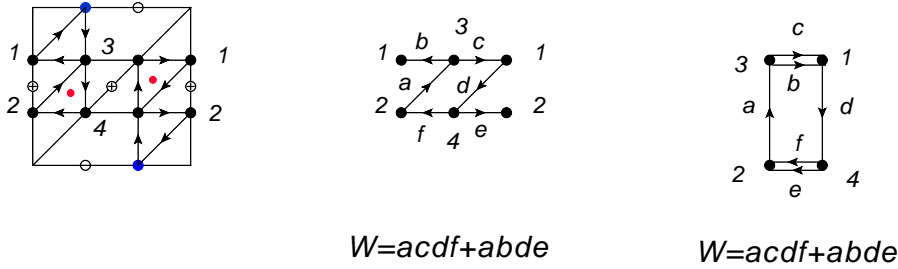}
\caption{The triangulation, BPS quiver and the superpotential of the pure $SU(3)$ theory. }
\label{puresu3}
\end{figure}
Let's now argue that the finite chamber for $SU(N)$ theory is unique: the above chamber is the only finite chamber. 
Let's do the analysis for the $SU(3)$ theory, and the BPS quiver and 
the superpotential can be easily found from the network construction, see figure. \ref{puresu3}.  The potential is shown in figure. \ref{clue}, and the $F$ term equations from the quiver are
\begin{equation}
cdf+bde=0, fac+eab=0~~dea=0,~~dfa=0,~~acd=0,~~abd=0
\end{equation}
The maximal green mutation sequences are
\begin{equation}
(\mu_3,\mu_4),(\mu_1, \mu_2),(\mu_3,\mu_4).
\end{equation}
Let's now give a proof that this is the unique sequences for finding a finite chamber using the quiver representation theory.
The dimension vector for two indecomposable representations representing vector bosons are
\begin{equation}
P_1=(1,0,1,0),~~P_2=(0,1,0,1),~
\end{equation}
 The two corresponding subquivers are the affine $A_1$ quiver and the sink nodes are $1$ and $2$, therefore we have to mutate
 node $3$ or node $4$ in the first step (the analysis of the quiver representation theory is 
 the same as the BPS quiver of the pure $SU(2)$ theory.). If we mutate node $3$ first, and the node $1$ and node $2$ would form an affine $A_1$ quiver (both of them are green), with
 node $1$  as the sink node, and they represent the vector boson too. So we can not mutate on node $2$ and node $1$ in this second step, and we have to mutate node $4$ in second step.
 Similar analysis can be done for the following quiver mutation sequences \footnote{The crucial point is the following:  if there are two \textbf{green} quiver nodes connected 
 by two arrows, the sink node can not be mutated.} , and the above mutation sequences are the unique one to  find a finite spectrum.
 The proof can be easily generalized to higher rank pure $SU(N)$ theory, and our conclusion is that there is only one finite chamber.
 
\begin{mydef}
The BPS quiver for $SU(N)$ with $N_f\leq 2N$ theory is shown in figure. \ref{flavor}.  The quiver has a main body formed by the quiver of pure $SU(N)$ theory, and each additional 
flavor adds a new vertex and new triangle to the quiver. The strategy of finding the finite spectrum is very simple:  doing the quiver mutation sequences for the 
pure SYM and then do random green mutations on the extra flavor nodes.
\begin{figure}[htbp]
\small
\centering
\includegraphics[width=10cm]{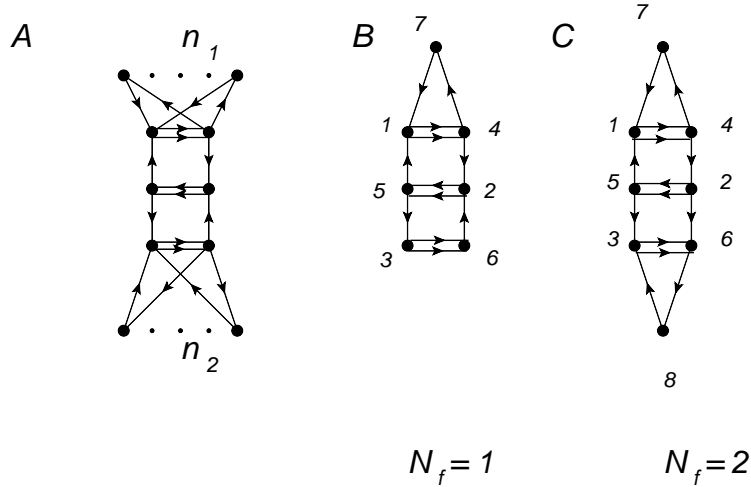}
\caption{The BPS quiver for $SU(4)$ theory with $n_1+n_2$ fundamental flavors.}
\label{flavor}
\end{figure}
\end{mydef}

Let's list one maximal green mutation sequences for the $SU(4)$ gauge theory with $N_f=1$ (the BPS quiver is shown in figure. \ref{flavor}B):
\begin{align}
& (\mu_1, \mu_2, \mu_3),~(\mu_4, \mu_5, \mu_6), ~(\mu_1, \mu_2, \mu_3),~(\mu_4, \mu_5, \mu_6), \nonumber \\
& \mu_7, \mu_4, \mu_1, \mu_5, \mu_2, \mu_6, \mu_3.
\end{align}
Similarly, for $SU(4)$ gauge theory with $N_f=2$ (the BPS quiver is shown in figure. \ref{flavor}C), a maximal green mutation sequence is 
\begin{align}
& (\mu_1, \mu_2, \mu_3),~(\mu_4, \mu_5, \mu_6), ~(\mu_1, \mu_2, \mu_3),~(\mu_4, \mu_5, \mu_6), \nonumber \\
& \mu_7, \mu_4, \mu_1, \mu_5, \mu_2, \mu_6, \mu_3, \nonumber \\
& \mu_8, \mu_2, \mu_6, \mu_1, \mu_5, \mu_7, \mu_4.
\end{align}

In general, each extra flavor would need $2N-1$ extra green mutations (one need to mutate the extra node once and all the other quiver nodes for the pure $SU(N)$ theory once), and the finite chamber has the following number of states
\begin{equation}
N(N-1)+(2N-1)N_f
\end{equation}
This is in agreement with the result presented in \cite{Alim:2011kw}. Notice that there are other finite chambers if we start mutating the extra quiver nodes first, and 
 the above number is not necessarily the minimal chamber for $SU(N)$ with $N_f$ flavors, since the number might be smaller
if we start with a quiver which is mutation equivalent to the above one.

\begin{mydef}
If there is one simple puncture on one boundary and one full puncture on another boundary. When $N$ is even, the underlying $\mathcal{N}=2$ theory is actually a linear quiver: 
\begin{equation}
SU(N)-SU(N-2)-\ldots-SU(2), 
\end{equation}
and it is the following linear quiver
\begin{equation}
SU(N)-SU(N-2)-\ldots-SU(3)-1,
\end{equation}
when $N$ is odd.
 \begin{figure}[htbp]
\small
\centering
\includegraphics[width=12cm]{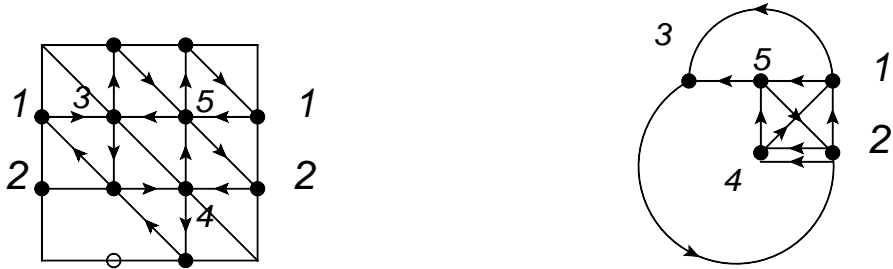}
\caption{The dot diagram for the annulus with one simple and one full puncture on each boundary, and the quiver is shown on the right. }
\label{SYM1}
\end{figure}

Let's look at the example shown in figure. \ref{SYM1} with $N=3$, which geometry actually represents $SU(3)$ gauge theory with $N_f=1$. The maximal green mutation sequence from our prescriptions are
\begin{align}
& F_1:~(\mu_1,\mu_2,\mu_5), \nonumber \\
& F_2:~(\mu_3,\mu_4,\mu_1),  \nonumber \\
& Triangle:~(\mu_3,\mu_5).
\end{align}
The first two steps are the mutation sequences for the quadrilateral flips, and the last step is some kind of generalized triangle flip.
\end{mydef}

In general, there are are two quadrilateral flips and each flip involves ${1\over 2}N(N-1)$ steps. Finally one need to do the triangle flip on one triangle with
three full punctures which gives ${1\over 6}N(N-1)(N-2)$ mutations, moreover, one need to do green mutations on the quiver nodes on the edges which involve $2(N-2)$ mutations.

\begin{mydef}
Let's now consider an annulus with two full punctures on each boundary.  The underlying $\mathcal{N}=2$ theory is actually a linear quiver\begin{equation}
SU(2)-SU(4)-\ldots-SU(N-2)-SU(N)-SU(N-2)-\ldots-SU(4)-SU(2), 
\end{equation}
 when $N$ is even, and it is the linear quiver 
\begin{equation}
1-SU(3)-\ldots SU(N-2)-SU(N)-SU(N-2)-\ldots-SU(3)-1.
\end{equation}
when $N$ is odd. The BPS geometry and the quiver is shown in figure. \ref{SYM}. There are quiver nodes inside each triangle, so we need to 
do triangle flip first, and then do two flips, so the total number of BPS states in this chamber is
\begin{equation}
N_{bps}=2N_s+2N_m={1\over 3}N(N-1)(2N-1).
\end{equation}
\end{mydef}

\paragraph{Some funny numerology}
We have found some sequences of numbers which seems to have the following pattern on dependence of rank $N$ if all the punctures are
 simple or full:

a. The number of BPS states is at most cubic in $N$.

b. The number has factor $N(N-1)$.

The first fact might be related to $N^3$ behavior of six dimensional $(2,0)$ theory; the latter fact is 
a manifestation that when $N=0$ and $N=1$, there is no BPS states. These two facts suggest that the dependence on $N$ has 
very fewer parameters and elegant form. For example, if the number of BPS states with a fixed BPS geometry has $N^2$ behavior, 
then the number of BPS states has only one free parameter and it must take the following form
\begin{equation}
f_1(N)=aN(N-1).
\end{equation}
If there is a $N^3$ behavior for the number of BPS states, then the formula has a maximal three parameters and take the following simple 
form
\begin{equation}
f_2(N)=N(N-1)(aN+b).
\end{equation}
So it is easy to determine those parameters using the result of the lower rank theory. For instance, when there are two simple punctures, the 
number of BPS states under large $N$ has only one parameter $a$ which can be fixed as $1$ by substituting the result $f_1(2)=2$. Similarly,  if there
are two full punctures, 
using the result $f_2(2)=2$ and $f_2(3)=10$, we find $a={2\over 3}$ and $b=-{1\over 3}$ which reproduce the results from 
explicit counting.

\subsubsection{More marked points and more boundaries}
The situation is quite similar for the Riemann surface with more marked points and more boundaries. One simply recalled the flip sequences from the maximal
green mutations of the corresponding $A_1$ theory which is described in detail in previous section, and then use the same flip sequences to the higher rank theory. 

\begin{figure}[htbp]
\small
\centering
\includegraphics[width=12cm]{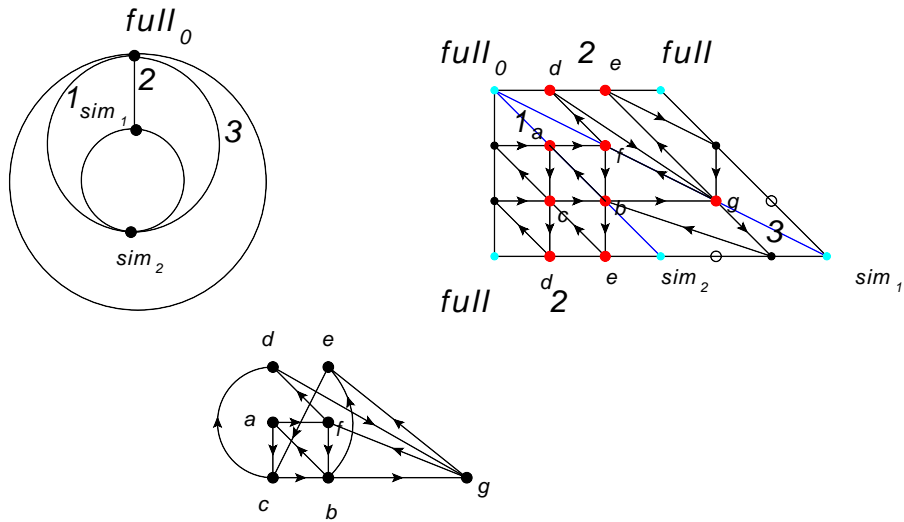}
\caption{The triangulation and dot diagram of annulus with two simple punctures on one boundary and one full puncture on another boundary. The quiver is also shown.}
\label{twoone}
\end{figure}

\begin{mydef}
Let's consider an annulus with one full puncture on one boundary, and two simple punctures on the other boundary. The triangulation and 
the quiver is shown in figure. \ref{twoone}. The minimal flip sequences for the $A_1$ theory  is  $F_1, F_2, F_3$,  and one maximal green mutation sequences for $N=3$ are
\begin{align}
& F_1: (\mu_a, \mu_b, \mu_c), \nonumber \\
& F_2: (\mu_f, \mu_g,\mu_a,\nonumber \\
& F_3:(\mu_d, \mu_e, \mu_f), \nonumber\\
& triangle~ flip: (\mu_d, \mu_c).
\end{align}

\end{mydef}

\newpage
\subsection{Once punctured disc }
This is the case where the geometric knowledge of the $A_1$ theory can not be generalized  since the 
quiver mutation will create self-folded triangle, and furthermore one need to do mutation on quiver nodes corresponding to 
an edge of the self-folded triangle. The mutation on this special edge does not have a good meaning in the higher rank case, i.e.
we do not know the mutation sequences  for such flip. So one need some new clues about the mutation sequences. The irregular 
realization is crucial in providing the clues about the mutation sequences.

Let's consider the simplest  BPS geometry of this type: a once punctured digon and all punctures are full. The experimental rule we find is the following:
first mutate the quiver nodes inside each triangle and use the mutation sequences corresponding to the triangle flip; then mutate the 
quiver nodes on two edges; and then do the quiver mutations on the inside nodes again, etc. This strategy is pretty successful in 
finding the maximal green mutation sequences.
 
\begin{mydef}
The BPS geometry is the once punctured digon which represents an AD theory, and we 
take all the puncture as full. This type of AD theory is actually isomorphic to the $A_{N-2}$ theory compactified on a sphere with 
$N$ simple punctures and a full puncture. The theory in one duality frame has the Lagrangian description and is given by 
\begin{equation}
1-SU(2)-SU(3)-\ldots-SU(N-1)-N
\end{equation}
 The theory is  isomorphic to $SU(2)$ with four flavors when $N=3$. We conjecture that the minimal chamber has the following number of BPS states:
\begin{equation}
N_{bps}=N(N-1)^2.
\end{equation}
\end{mydef}

The method for finding the spectrum is the following: first do the triangle flips for all the quiver nodes inside 
two triangles, and then do the quiver mutations on the quiver nodes on the edge. One need to mutate multiple times 
in this fashion, and the shape of the quivers inside the quiver nodes might change, however, the triangle flips 
can be done for each connected subquiver inside the triangle. The maximal green mutation sequences for $N=4$ is 
\begin{align}
&(\mu_7,\mu_8,\mu_9, \mu_7; \mu_{10}, \mu_{11},\mu_{12},\mu_{10}),  (\mu_1, \mu_2,\mu_3,\mu_4,\mu_5,\mu_6), \nonumber\\
&(\mu_{10}, \mu_{11},\mu_{12}, \mu_{7},\mu_8, \mu_{9}), ( \mu_2,\mu_3, \mu_5, \mu_6), \nonumber\\
&(\mu_{10},\mu_{11}, \mu_{12}, \mu_{7},\mu_8, \mu_{9}),( \mu_2,\mu_3, \mu_5, \mu_6). \nonumber\\
\end{align}

\begin{figure}[htbp]
\small
\centering
\includegraphics[width=12cm]{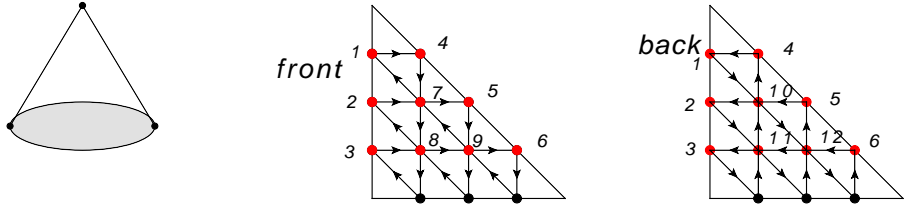}
\caption{The triangulation and quiver for the once punctured digon, all the punctures are full. There are two triangles and the quiver is 
derived by identifying the quiver nodes on the edges of the triangulation. }
\label{dtype}
\end{figure}

\begin{figure}[htbp]
\small
\centering
\includegraphics[width=8cm]{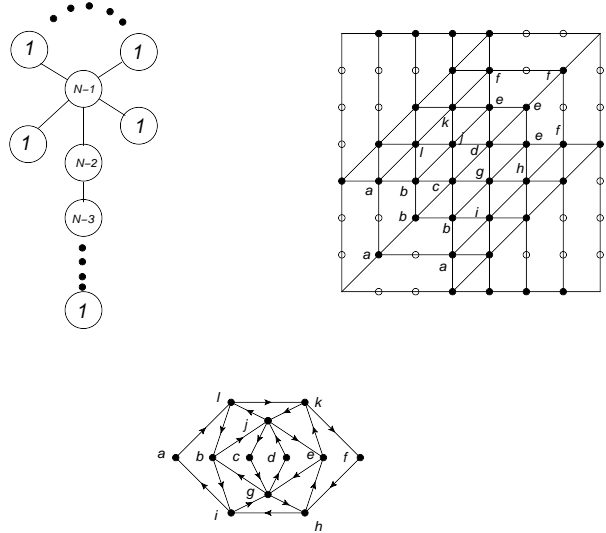}
\caption{Top: The irregular realization can be seen from the 3d mirror of this theory, and we draw the dot diagram. Bottom: The quiver from this realization, which is the same as in figure. (\protect \ref{dtype}). }
\label{dtypeirr}
\end{figure}

This type of sequences can be seen from the irregular realization of the same theory, i.e. it can be realized by 
a rank $2N-1$ theory compactified on a sphere with an order three irregular singularity with the type
\begin{equation}
Y_3=[N,N-1],~Y_2=[N-1, 1,1,\ldots,1],~Y_3=[1,1,\ldots, 1].
\end{equation}
The BPS geometry is a fourth punctured disc with cyclic ordered marked points $(Y_3, Y_2, Y_3, Y_2)$, and
we draw the dot diagram and the quiver for $N=4$ in figure. \ref{dtypeirr}, which is the same as the quiver using the $A_{N-1}$ representation.
In the irregular realization, all the mass nodes are put on the 
diagonal edge, and one can start doing triangle flip first, and then do the quadrilateral flip, which gives the above sequences.

For other type of theories,  one do random  quadrilateral flips and if we encounter the above once punctured digon , we do the above specific mutation sequences and then 
keep going until there is no green edge left, as usual, we would need to do triangle flip at the end. In this way, we can find many maximal mutation sequences.

\newpage
\subsection{Closed Riemann surface}
We consider four dimensional superconformal field theory derived by compactifying six dimensional $A_{N-1}$ theory on a Riemann surface 
with regular punctures. The scaling dimensions of Coulomb branch operators are integer and usually there are marginal coupling constants.
We are going to find the maximal mutation sequences for these class of theories.

\paragraph{$T_N$ theory}:
The $T_N$ theory is realized as the sphere with three full punctures. The BPS quiver is derived from the triangulation of the three
punctures sphere: there are two triangles with the same boundaries.  We conjecture that the minimal chamber has the following number of states:
\begin{equation}
N_{bps}=2N(N-1)^2.
\end{equation}
The reasoning is still based on our conjecture that the BPS states is of the order $N^3$, and the BPS states is zero for $N=0$ and $N=1$, so the 
BPS states has the following form
\begin{equation}
f(N)=N(N-1)(aN+b),
\end{equation}
Using the result for $N=2$ and $N=3$ found in \cite{Alim:2011kw}, we get the conjectured form. 
There are 72 states for $N=4$, 160 states for $N=5$ which we have checked explicitly. The charge vectors and the phase order can be found from the maximal green mutations. 
Interestingly, this formula gives the correct answer for $N=2$ even if the mutation method can not be applied to that quiver.

 \begin{figure}[htbp]
\small
\centering
\includegraphics[width=10cm]{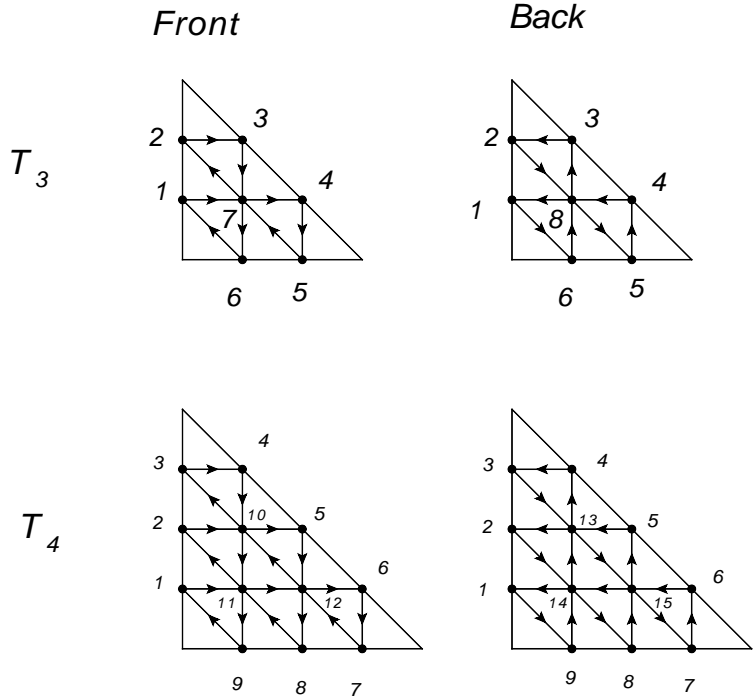}
\caption{The triangulation and the quivers for $T3$ and $T4$ theory. }
\label{TN}
\end{figure}

\begin{mydef}
The triangulations and dot diagram for $T3$ and $T4$ theory is shown in figure. \ref{TN}. 
The maximal green mutations for $T3$ theory is
\begin{align}
(\mu_7,\mu_8), (\mu_1,\mu_2,\mu_3,\mu_4,\mu_5,\mu_6) \\
(\mu_7,\mu_8), (\mu_1,\mu_2,\mu_3,\mu_4,\mu_5,\mu_6) \\
(\mu_7,\mu_8), (\mu_1,\mu_2,\mu_3,\mu_4,\mu_5,\mu_6) 
\end{align}

The  Green mutation sequences for the $T4$ theory is
\begin{eqnarray}
[\mu_{10},(\mu_{11},\mu_{12}),\mu_{10}],~[\mu_{13},(\mu_{14},\mu_{15}),\mu_{13}]\nonumber\\
\mu_1,\mu_2,\mu_3,\mu_4,\mu_5,\mu_6,\mu_7,\mu_8,\mu_9 \nonumber\\
\mu_{10},\mu_{11},\mu_{12}, \mu_{13},\mu_{14},\mu_{15} \nonumber\\
\mu_1,\mu_2,\mu_3,\mu_4,\mu_5,\mu_6,\mu_7,\mu_8,\mu_9 \nonumber\\
\mu_{10},(\mu_{11},\mu_{12}),\mu_{10},~\mu_{13},(\mu_{14},\mu_{15}),\mu_{13}\nonumber\\
\mu_1,\mu_2,\mu_3,\mu_4,\mu_5,\mu_6,\mu_7,\mu_8,\mu_9 \nonumber\\
\mu_{10},\mu_{11},\mu_{12}, \mu_{13},\mu_{14},\mu_{15} \nonumber\\
\mu_2,\mu_5,\mu_8 \nonumber\\
\mu_{10},(\mu_{11},\mu_{12}),\mu_{10},~\mu_{13},(\mu_{14},\mu_{15}),\mu_{13}\nonumber\\
\mu_1,\mu_3,\mu_4,\mu_6,\mu_7,\mu_9 \nonumber\\
\end{eqnarray}
\end{mydef}

\begin{figure}[htbp]
\small
\centering
\includegraphics[width=12cm]{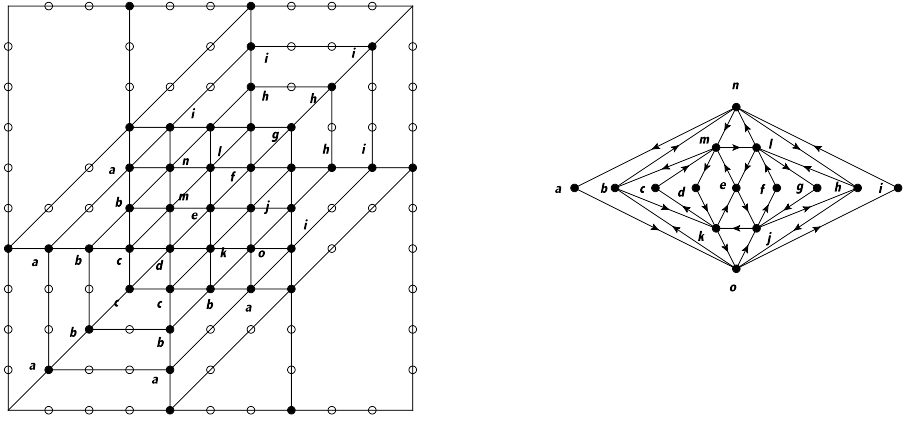}
\caption{The irregular realization of $T4$ theory. }
\label{T4irr}
\end{figure}

The basic rule is to mutate the internal nodes using the triangle flip such that no green nodes left, then mutate all the 
boundary nodes. One need to do such sequences in several  rounds. The motivation is coming from the irregular 
realization of the $T_N$ theory. For example, the $T4$ theory has another realization using rank $9$ theory compactified on
a sphere with the following order three type III irregular singularity 
\begin{equation}
Y_3=[6,4],~~Y_2=[4,3,3],~~Y_1=[1,1,1,\ldots,1]
\end{equation}
and the quiver for this irregular singularity is a fourth punctured sphere with cyclic ordered 
punctures $[Y_3, Y_2, Y_3, Y_2]$, the BPS quiver is shown in figure. \ref{T4irr}. It is easy to check that
the two quivers from different realization are the same.
Now all the mass nodes in $A_3$ realization are living on the single diagonal edges of the 
irregular realization, and the quiver nodes inside triangles are living inside the triangles in the irregular 
realization too. The mutation sequences for the flip is  mutating the quiver nodes on 
the boundary nodes first, and then mutating the quiver nodes inside two triangles. We need to do 
more than one rounds following our early study on this type of AD theories. That is how we 
find the above mutation sequences.

\begin{figure}[htbp]
\small
\centering
\includegraphics[width=10cm]{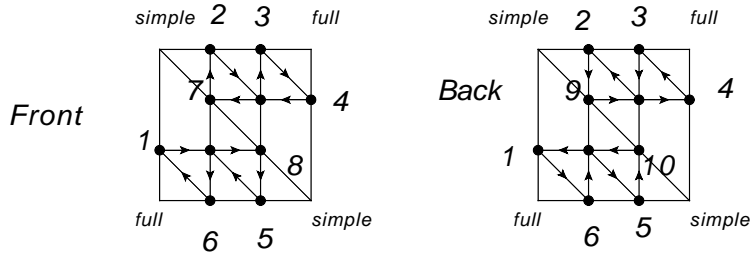}
\caption{ The triangulation and quivers for the theory defined by sphere with two full punctures and two simple punctures.}
\label{2Nf}
\end{figure}

\newpage
\begin{mydef}
$SU(N)$ with $N_f=2N$: This is represented by a sphere with two full punctures and two simple punctures. The triangulation and the quiver 
is shown in figure. \ref{2Nf}.  We find a finite chamber with following number of states:
\begin{equation}
N_{bps}=2N(2N-1).
\end{equation}
\end{mydef}

The mutation sequences for $SU(3)$ with $6$ flavor is the following
\begin{align}
&\mu_1, \mu_2,\mu_3,\mu_4,\mu_5,\mu_6 \\
&\mu_7,\mu_8,\mu_9,\mu_{10} \\
&\mu_1, \mu_2,\mu_3,\mu_4,\mu_5,\mu_6 \\
&\mu_7,\mu_8,\mu_9,\mu_{10} \\
&\mu_1, \mu_2,\mu_3,\mu_4,\mu_5,\mu_6\\
&\mu_7,\mu_8,\mu_9,\mu_{10} \\
\end{align}

In general, one first mutate the $2N$ quiver nodes on the boundary, and then $(2N-2)$ quiver nodes on the internal edges. The number  of
mutations for one cycle is $4N-2$, and one need to do $N$ cycles, so the total number of states are $N(4N-2)$.

For more general theory defined by a  sphere with punctures, it is possible to find a realization using the irregular singularity. Such realization
is always possible if all the Young Tableaux has the form $[n_1,1,1,1\ldots,1]$, in particular, it is possible if all the punctures are full.  
The irregular realization uses a even higher rank six dimensional group and a  
type III irregular singularity with an order $3$ pole,  see \cite{Xie:2012jd} for the exact map.  The BPS geometry is a disc with four punctures. 
One can use the mutation sequences for the flip to find the maximal green mutations.

\begin{figure}[htbp]
\small
\centering
\includegraphics[width=16cm]{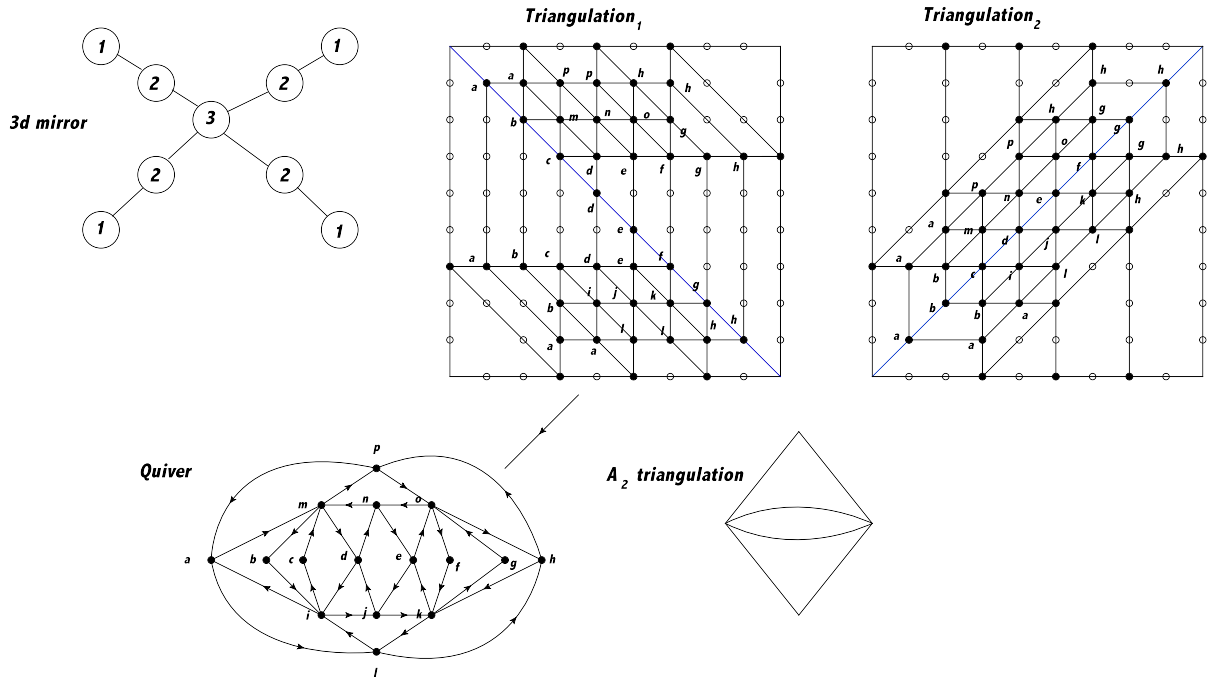}
\caption{The quiver for the theory defined by a $A_2$ theory compactified on a sphere with four full punctures.}
\label{fourfull}
\end{figure}

\begin{mydef}
Let's consider the theory defined by $A_2$ theory compactified on a sphere with four full punctures. This is a superconformal 
gauge theory with a $SU(3)$ group coupled to two $T_3$ theory. The three dimensional mirror for this theory is shown in 
figure. \ref{fourfull} from which we can read an irregular realizations, and we take the following rank $8$ realization. 
\begin{equation}
Y_3=[6,3],~~Y_2=[3,2,2,2],~~Y_1=[1,1,1,1,1,1,1,1,1].
\end{equation}
\end{mydef}
The BPS quiver is shown in figure. \ref{fourfull}. This quiver is the same as the quiver derived from the triangulations of the fourth puncture
sphere, we indicate the specific triangulations on figure. \ref{fourfull}. The interested reader can check that the two quivers 
are indeed the same. Basically, the flavor nodes of the theory lives on the diagonal edge of the triangulation of fourth
punctured disc. One of the maximal green mutation sequences which has 60 states is the following
\begin{align}
& (\mu_a,\mu_b,\mu_c,\mu_d, \mu_e, \mu_f, \mu_g, \mu_h), (\mu_m,\mu_n,\mu_o, \mu_i,\mu_j,\mu_k), (\mu_p,\mu_l), \nonumber \\
& (\mu_a,\mu_b,\mu_c,\mu_d, \mu_e, \mu_f, \mu_g, \mu_h), (\mu_p,\mu_n, \mu_m,\mu_o,\mu_p,\mu_n), (\mu_j,\mu_l,\mu_i,\mu_k,\mu_j,\mu_l), \nonumber \\
&  (\mu_a,\mu_b,\mu_c,\mu_d, \mu_e, \mu_f, \mu_g, \mu_h), (\mu_p,\mu_n,\mu_j,\mu_l), \nonumber \\
& (\mu_a, \mu_d, \mu_e, \mu_h),(\mu_m,\mu_o, \mu_i, \mu_k), (\mu_p,\mu_n,\mu_j,\mu_l). \nonumber \\
\end{align}

For general $N$, based on the assumption that the number of BPS states in minimal chamber is a polynomial function of $N$ and the 
scaling behavior is $N^3$, we conjecture that the function is 
\begin{equation}
f(N)=2N(N-1)(2N-1).
\end{equation} 
Using this method, one can find finite chambers for any theory defined on a sphere with regular punctures. 

\newpage
\section{Spectral generator}
\begin{figure}[htbp]
\small
\centering
\includegraphics[width=12cm]{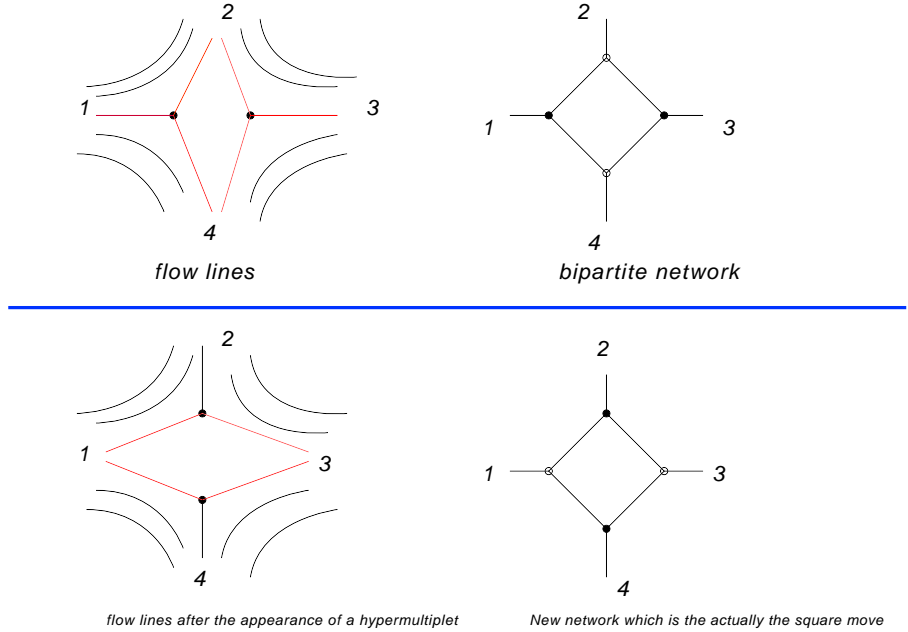}
\caption{ Top: The local picture of the flow lines and the equivalent bipartite network. Bottom: The new flow line 
after an appearance of the hypermulitplet and the new bipartite network, which is actually the square move 
of the original network.}
\label{flow}
\end{figure}

Let's first review the meaning of the spectral generator of $A_1$ case introduced in \cite{Gaiotto:2009hg}, and we will
rephrase it in terms of the language of cluster algebra. The Seiberg-Witten curve of the theory is 
\begin{equation}
x^2=\Phi(z),
\end{equation}
where $z$ is the coordinate on the Riemann surface, $x$ is the coordinate on the cotangent bundle, and
 $\phi(z)$ is a quadratic differential defined on the Riemann surface.  The Seiberg-Witten differential is $\lambda=xdz=\sqrt{\Phi(z)}dz$ and
 one can use it to  define a foliation on the Riemann surface from the following flow equation
 \begin{equation}
\lambda{ dz\over dt}=e^{i\theta},
 \end{equation}
here $\theta$ is a  fixed angle and in fact is the slop of the BPS hypermultiplet for some critical value $\theta$.
 For more details on the  structure of the foliation, please see \cite{Klemm:1996bj,Shapere:1999xr, Gaiotto:2009hg}. What we want to point out is that the topology of 
the foliation is exactly equivalent to the bipartite network introduced earlier and therefore also equivalent to the 
triangulation, see \cite{Heckman:2012jh}.  The branch points of the Seiberg-Witten
differential is the vertex for the foliation which is identified with the black vertices of our network, see figure. \ref{flow}.

When $\theta$ is changed, the foliation is also changed smoothly. However, when $\theta$ arrives at a critical value, then
the topology of the foliation is changed due to the appearance of a hypermultiplet which is represented by 
the flow lines connected by two branch points. Locally, this change of topology is just the square move for the network which 
then corresponds to the quiver mutations. The $\theta$ actually parameterizes one half plane of the central charge,
 therefore by rotating $\theta$ angle 180 degree, one can probe all the BPS particles by tracking the change of the topology of the network. 
 This is essentially the same as finding a maximal green mutation of the original quiver (for the chamber with finite number of states).
  
Now for each face of the network, one can associate a cluster $X$ coordinate which parameterizes the framed moduli space of 
flat connections defined on the Riemann surface. The appearance of the hypermultiplet acts like a square move on the network and therefore
acts like a cluster transformation on the coordinate.  When the $\theta$ angle is rotated by 180 degrees, the cluster 
coordinates for each face are changed to a certain value.  Moreover, the final cluster coordinates do not depend on 
the chamber one probe as long as long as initial foliation is given.

However, it is usually difficult to track the change of the network and therefore hard to find the BPS spectrum. 
There is an easy way of getting the final cluster coordinate without knowing the detailed spectrum information.
The method uses the definition of the cluster coordinates from the cross ratio of
the flags \cite{fock-2003} attached on each marked points. The rotation by 180 degree acts like a $Z_2$ action on the attached flag, 
and the change of the cluster coordinates can be worked out explicitly using the definition of the cross ratio. The final 
cluster coordinates is the spectral generator from which one could find sensible factorization and therefore 
the explicit spectrum information.

In this paper, we are not pursing a similar geometric derivation of the spectral generator for the higher rank theory, 
which we will discuss in another occasion. However, we will
 give a derivation using the information of the BPS spectrum though, in particular, the maximal green mutation 
 is very useful.
 
\begin{figure}[htbp]
\small
\centering
\includegraphics[width=12cm]{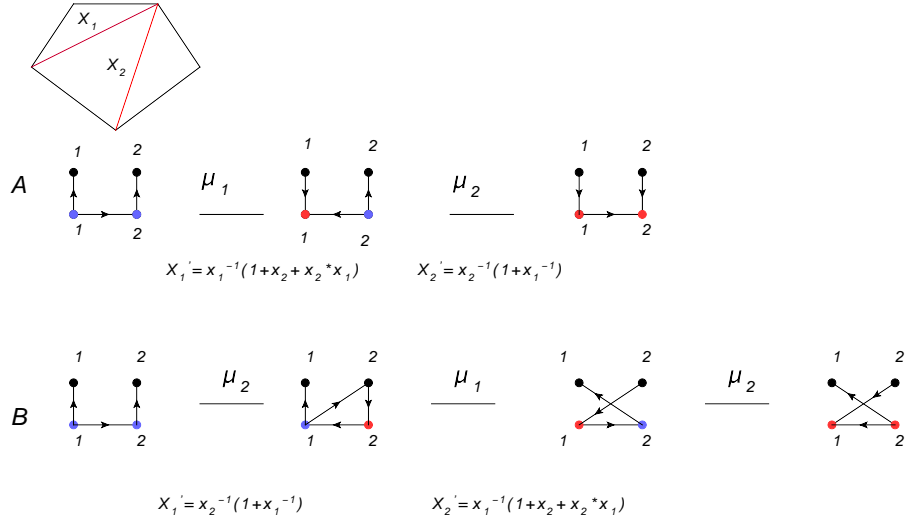}
\caption{ Top: The triangulation for the five punctured disc which gives the BPS quiver for $A_2$ AD theory. Bottom: Two BPS chambers from the maximal green mutations
and the final cluster coordinates.}
\label{cluster}
\end{figure}

One of the  remarkable feature of the maximal green mutation is that it keeps track of the position of the quiver nodes. Let's
label the frozen nodes as $(1,2,\ldots,n)$ which is the same labeling of the original quiver. Then after doing the maximal green mutations, 
 the cluster $X$ coordinate for $n$th quiver node is equal to the coordinate of the quiver nodes connected to the $n$th frozen node. 
 Let's check this explicitly using the simple pentagon geometry of the $A_1$ theory. The two chambers and the final cluster coordinates 
 are shown in figure. \ref{cluster}. The final coordinates from chamber 1 is 
 \begin{equation}
 X_{1}^{f}=X_{1}^{'}=x_1^{-1}(1+x_2+x_2x_1),~~~X_2^{f}=X_{2}^{'}=x_2^{-1}(1+x_1)^{-1}.
 \label{classical}
 \end{equation}
 and the results are the same from chamber 2 due to the permutation of the quiver nodes!  This expression is 
 clearly the same as the one given in \cite{Gaiotto:2009hg}, see formula $11.27$ and $11.28$ (with a slightly different convention for the definition of the coordinate.).
 
\begin{figure}[htbp]
\small
\centering
\includegraphics[width=12cm]{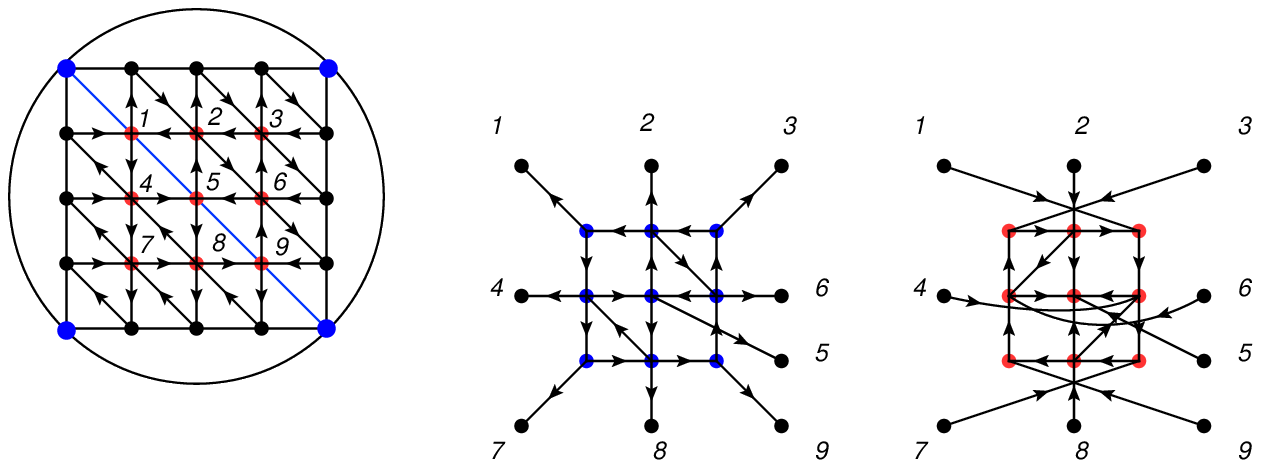}
\caption{The initial configuration and the final configuration of the maximal green mutation of the quiver from the disc with four full punctures.}
\label{spectral2}
\end{figure}

 Exact similar consideration can be generalized to higher rank theories: the spectral generator can be written down using 
 the explicit maximal green mutation sequences.  Let's consider a disc with four full punctures, and the initial configuration 
and the final configuration of the maximal green mutation is listed in figure. \ref{spectral2}. The mutation sequences are
\begin{align}
& (1,5,9), (2,4,6, 8), (3,5,7), \nonumber\\
& (1,2,4,1), (9,6,8,9), \nonumber\\
\end{align}
and the final coordinates or the spectral generator can be written down using the mutation formula for the $X$ coordinate and the permutation:
\begin{align}
&X_1^{f}=X_3^{'},~X_2^{f}=X_2^{'},~X_3^{f}=X_1^{'}, \nonumber\\
&X_4^{f}=X_6^{'},~X_5^{f}=X_5^{'},~X_6^{f}=X_4^{'}, \nonumber\\
&X_7^{f}=X_9^{'},~X_8^{f}=X_8^{'},~X_6^{f}=X_9^{'}. 
\end{align}

The expression is rather long and the interested reader can find the expression in the appendix A. 
One can write down the spectral generator for all the theories whose BPS spectrum
is discovered in this paper, many examples are given in the appendix A.
It would be interesting to find other factorizations of the spectral generator
 which will give the BPS spectrum in other chamber.

Since there is a quantum cluster algebra, then one could also define a refined version of the spectral generator, which would then 
tell us the spin information of the BPS particles. The refined 
spectral generator can be easily found using the explicit mutation sequences, let's consider the example shown in figure. \ref{cluster},  the noncommutative commutation relation is
\begin{equation}
X_{\alpha}X_{\beta}=q^{\epsilon_{\alpha\beta}}X_\beta X_{\alpha}
\end{equation}
and the quantum cluster transformation is 
\begin{align}
& X_k^{'}=X_k^{-1}, \nonumber \\
& X_i^{'}=X_i(\prod_{a=1}^{|\epsilon_{ik}|}(1+q^{a-1/2} X_k^{-sgn(\epsilon_{ik})}))^{-sgn(\epsilon_{ik})}.
\end{align}
Using the above quantum cluster transformation, the refined spectral generator can be written down using the mutation sequences from chamber 1 
\begin{equation}
X_1^{f}=X_1^{-1}+q^{1\over 2}X_1^{-1}X_2+X_2,~~~X_2^{f}={1\over X_2+q^{-{1\over2}}X_2X_1}.
\end{equation}
In the limit $q\rightarrow 1$, the refined spectral generator is the same as the classical one given in [\ref{classical}]. A more complicated example of  refined spectral generator 
is given in appendix B.

\section{Vector multiplets}
\subsection{Wall crossing between chamber with Infinite number of states}
Previous sections focused on the finite chambers and their wall crossing behavior.  In this section, we would like to say something about the 
chamber with higher spin states using the quiver representation theory. Our treatment is very elementary and we hope to 
do a more thorough analysis in the future.

\begin{figure}[htbp]
\small
\centering
\includegraphics[width=8cm]{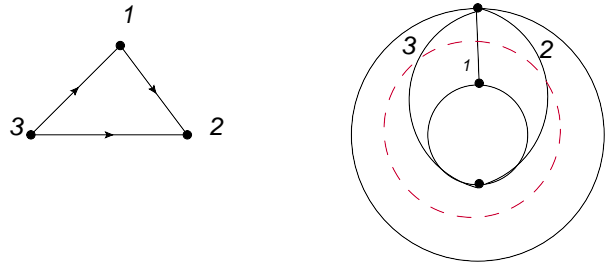}
\caption{The triangulation and BPS quiver for $SU(2)$ theory with one flavor, and the curve corresponding to the $W$ boson is drawn as the red curve. }
\label{Su2one}
\end{figure}

There are also wall crossing behavior between two chambers with infinite states, as discovered in original paper of Seiberg-Witten \cite{Seiberg:1994aj}.
We can actually see this type of wall crossing behavior using the quiver representation theory for some simple examples.
Consider  $SU(2)$  theory with one flavor whose BPS quiver is of the type $\tilde{A}(2,1)$ shown in figure. \ref{Su2one}, 
the possible BPS states are represented by the curves on the triangulated Riemann surface,  which are listed in table. \ref{Su2onerep}.

\begin{table}[h]
\centering
    \begin{tabular}{|c|c|}
        \hline
                  ~&charge \\ \hline
          Hypermultiplet &$(\gamma_i+\gamma_j, i\neq j)$, $(n\gamma_i+n\gamma_j+(n\pm1)\gamma_k,i\neq j \neq k)$ \\ \hline
           Vector multiplet &$\gamma_1+\gamma_2+\gamma_3$\\ \hline
           \end{tabular}
    \caption{The indecomposable representation of the quiver $\tilde{A}(2,1)$. }
    \label{Su2onerep}
\end{table}

\begin{table}[h]
\centering
    \begin{tabular}{|c|c|c|}
        \hline
               Name&   Indecomposable&subrepresentation\\ \hline
                  A&$\gamma_1+\gamma_2+\gamma_3$& $B, S_2$ \\ \hline
        B& $ \gamma_1+\gamma_2$ &$S_2$ \\ \hline
        C&   $\gamma_1+\gamma_3$&$S_1$\\ \hline
          D& $\gamma_2+\gamma_3$& $S_2$ \\ \hline
           $E_n$&$n\gamma_1+n\gamma_2+(n+1)\gamma_3$ & $A,B, S_2,D$\\ \hline
                  $F_n$ &   $n\gamma_1+n\gamma_2+(n-1)\gamma_3$ & $A,B, S_2$\\ \hline
           $G_n$&$n\gamma_1+n\gamma_3+(n+1)\gamma_2$ & $B, S_2$\\ \hline
                   $H_n$&   $n\gamma_1+n\gamma_3+(n-1)\gamma_2$ &  $A,B, S_2$\\ \hline
               $I_n$&$n\gamma_2+n\gamma_3+(n+1)\gamma_1$ & $A,B, S_2$\\ \hline
                        $J_n$&      $n\gamma_2+n\gamma_3+(n-1)\gamma_1$ & $A,B, S_2,D$\\ \hline

           \end{tabular}
    \caption{The subrepresentations of the indecomposable representation of the quiver $\tilde{A}(2,1)$. }
    \label{Su2sub}
\end{table}

\begin{figure}[htbp]
\small
\centering
\includegraphics[width=12cm]{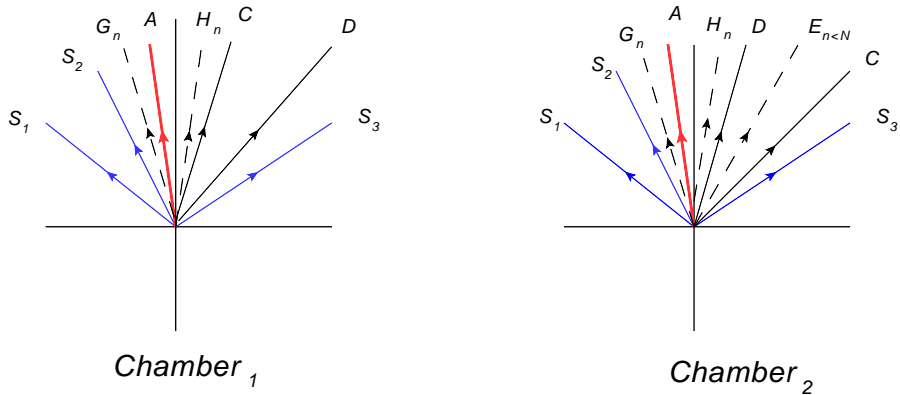}
\caption{Two choices of stability condition which will make $W$ boson stable. }
\label{Stabke}
\end{figure}

Representation $A$ is the $W$ boson whose subrepresentations are $\gamma_2, \gamma_1+\gamma_2$ \footnote{Let's give a little bit explanation of the sub-representation listed in table. For example, 
$D$ is a subrepresentation of series $E_1$ because one can select a one dimensional subspace of $V_3=2$ such that it maps to zero of $V_1=1$, etc, and therefore the representation $D=(0,1,1)$ is a subrepresentation of $E_1$.}, so it is stable if the slop of its subrepresentations are smaller than $\gamma_2$ and $\gamma_1+\gamma_2$.  There are two choices of the stability conditions on the simple representations which will make the 
$W$ boson stable,  one of them is  shown in figure. \ref{Stabke} (another is given by exchanging $S_1$ and $S_2$.).  There are more choices depending on the relative slop of  other 
representations. Let's explain this using the results shown in figure. \ref{Stabke}.
f
In chamber one,  representations $F_n$ and $I_n$ are  unstable since their subrepresentation $A$ is on their right. The reason is the following: consider  representations $F_n$  whose 
charge vector can be decomposed as:
\begin{equation}
n\gamma_1+n\gamma_2+(n-1)\gamma_3=(n-1)(\gamma_1+\gamma_2+\gamma_3)+\gamma_1+\gamma_2=(n-1)A+B,
\end{equation}
so $F_n$ is lying in between $A$ and $B$, and $A$ as its subrepresentation is on its right, so it is unstable!  Similarly $J_n$ is also unstable since its subrepresentation $D$ 
is always on its right. $C, D, G_n, H_n$ are always stable since all their proper sub representations have higher slops.  Finally the stability of $E_n$ depends on on the relative
 position of $C$ and $D$,  since  $E_n$ series lie in between the charge vector $C$ and $A$,    all $E_n$ are unstable if $D$ is on the right of $C$  as shown in chamber 1 of  figure. \ref{Stabke},
 On the other hand, if $D$ is on the left of $C$, then some of the representations of $E_n$ series are lying in between $C$ and $D$ and they 
 are stable, the truncation of the $E_n$ series depend on the relative position of $C$ and $D$,  so we have an infinite number of chambers! 
 
 If the relative order of $S_1$ and $S_2$ are interchanged, then $B=\gamma_1+\gamma_2$ are stable and the analysis of other stable particles are the same, the results 
 are shown in table. \ref{infinite}. 

\begin{table}[h]
\centering
    \begin{tabular}{|c|c|}
        \hline
                  ~& Stable particles \\ \hline
         Chamber 1 &$S_1, S_2, G_n, A, H_n, C, D, S_3$ \\ \hline
           Chamber 2&$S_1, S_2, G_n, A, H_n, D, E_{n<N}, C, S_3$\\ \hline
                    Chamber 3 &$S_2, B, S_1 G_n, A, H_n, C, D, S_3$ \\ \hline
           Chamber 2&$S_2, B S_1, G_n, A, H_n, D, E_{n<N}, C, S_3$\\ \hline
           \end{tabular}
    \caption{The BPS spectrum with infinite number of states of $SU(2)$ with one flavor.}
    \label{infinite}
\end{table}

\begin{figure}[htbp]
\small
\centering
\includegraphics[width=14cm]{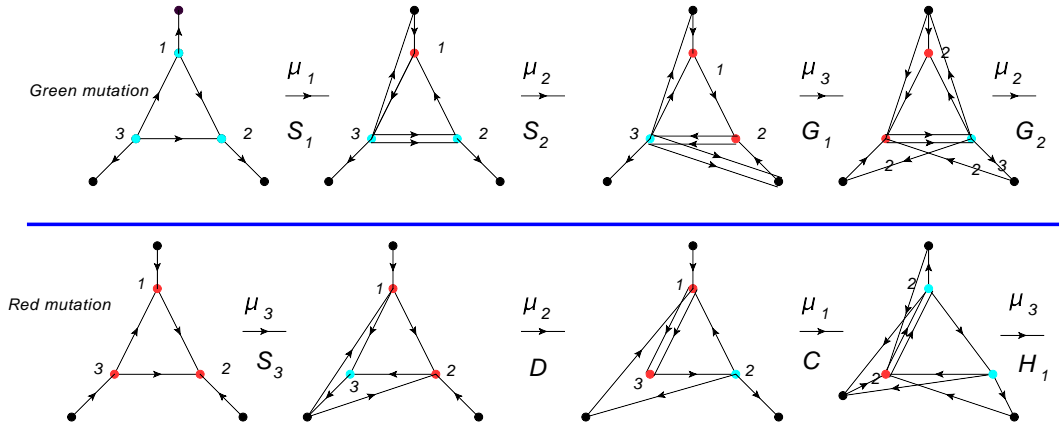}
\caption{Top: The green mutation with infinite number of steps. Bottom: The red mutation with infinite number of steps which approach to the same limit as the green mutation at the top.}
\label{grded}
\end{figure}

The above infinite chambers can be found using the green mutation and red mutation method. 
Let's consider chamber 1, we can do the green mutations
\begin{equation}
\mu_1, \mu_2, \mu_3,\mu_2, \mu_3, \mu_2 \ldots
\end{equation}
The charge vectors for the infinite mutation sequences are exactly the states on the left of the $W$ boson:
\begin{equation}
S_1,~~S_2,~~G_n,
\end{equation}
see the top of figure. \ref{grded}. The green mutation only probes this part of the spectrum. To probe the other part of the spectrum, one need to use 
the red mutations. Since $\gamma_1$ and $\gamma_2$ are already probed, one can only start mutating node $3$; After first step, one 
can mutate either node $2$ or node $1$, we choose to mutate node $2$. In third step, there are still two choices: either mutate node $3$ or node $1$, however,
 the charge vector for node $3$ is $\gamma_1$ which has already found by doing green mutation, therefore we can only mutate node $1$ in this step, there
 is no ambiguity in the later mutations, the red mutations are
 \begin{equation}
\mu_3, \mu_2, \mu_1,\mu_3, \mu_1, \mu_3, \mu_1, \ldots
\end{equation}
and the charge vectors for this part are
\begin{equation}
S_3, ~~D,~~C,~~H_n.
\end{equation}
Combining the green mutation and red mutation, we find $W$ boson as the common limit and recover the BPS states in chamber one. 
It is very easy to recover the other chambers using the mutation method.

\newpage

\subsection{Theory without finite chamber}
\begin{figure}[htbp]
\small
\centering
\includegraphics[width=4cm]{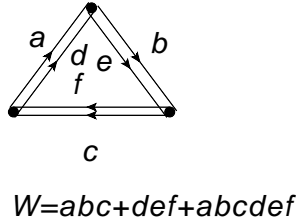}
\caption{The BPS quiver for $SU(2)$ with one massive adjoint, the superpotential is also given. }
\label{SU2ad}
\end{figure}

The  $SU(2)$ $N=2^{*}$ theory has no finite chamber \cite{Alim:2011kw}. Here let's give a very simple proof using the quiver representation theory. The 
BPS $(Q,W)$ for this theory is shown in figure. \ref{SU2ad}.  The three $W$ boson are represented by the subquiver of  pure $SU(2)$ theory, and
 each node is a sink for the $W$ boson, so if we start mutating any of the quiver nodes,  one of the $W$ boson would be stable, and there is 
no way we can find a finite chamber.

\begin{figure}[htbp]
\small
\centering
\includegraphics[width=10cm]{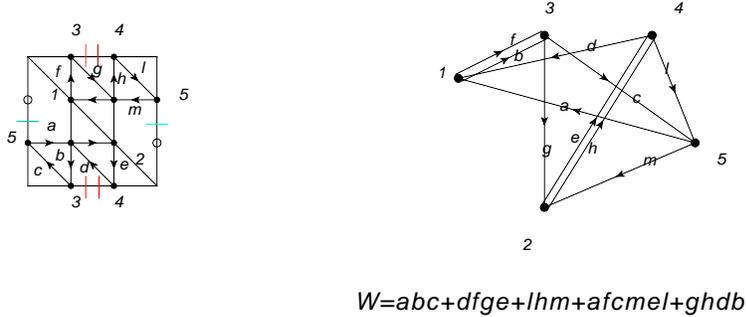}
\caption{The BPS quiver for $SU(3)$ with one massive adjoint, and the superpotential is also given. }
\label{SUNad}
\end{figure}

It is easy to generalize the above analysis to  $\mathcal{N}=2^{*}$ $SU(N)$ theory. The BPS quiver for $SU(3)$ with one massive adjoint is given in \cite{Xie:2012dw}. The triangulation, dot 
diagram and the quiver is shown in figure. \ref{SUNad} for $N=3$. 

The subquivers representing the $W$ bosons are  listed in table. \ref{suadjoint}. Let's now look at the green mutation sequences.  Because of the 
existence of the vector boson represented by $W1$ and $W2$, one could not mutate node $3$ and $4$ in first step. If we mutate
node $2$ and $1$ in first step, then in second step we can only mutate nodes $3$ and $4$, since mutating node $5$ would 
make $W3$ and $W4$ stable. Similarly, one can only mutate nodes $1$ and $2$ in third step to avoid activating $W3$ or $W4$, 
but this will finally make $W5$ stable! Therefore there is no finite chamber for $SU(3)$ with a massive adjoint. Similar analysis 
can be done to the general $SU(N)$ theory and the conclusion is that there is no finite chamber for $SU(N)$ theory with a massive adjoint!

\begin{table}[h]
\centering
    \begin{tabular}{|c|c|c|c|}
        \hline
                  Name&Quiver nodes&Source&Sink \\ \hline
          W1&(1,3)&1&3 \\ \hline
         W2& (2,4)&2&4\\ \hline
          W3&(2,3,5)&3&2\\ \hline
          W4&(1,4,5)&4&1 \\ \hline
          W5&(1,3,5,4)&4&5 \\  \hline
           \end{tabular}
    \caption{The subquiver of vector bosons  and the source-sink analysis for the BPS quiver of SU(3) with one adjoint.}
    \label{suadjoint}
\end{table}

\section{Conclusion}
Finite BPS chambers are found for a large class of 4d $\mathcal{N}=2$ theories engineered from  six dimensional  $A_{N-1}$ $(2,0)$ theory on a Riemann surface
with regular and irregular singularity.  Our results 
greatly extend the knowledge of the BPS spectrum of various kinds of higher rank $\mathcal{N}=2$ quantum field  theories, and 
should be a first step towards a full understanding of the BPS spectrum of these theories. 
There are many open questions which deserve further study:

a. It is  interesting to find  all the finite chambers for a given BPS  quiver of a given theory. Since an explicit combinatorial algorithm is given, 
 it may be possible to do it using computer scanning. It is also interesting to do the similar scanning for  all the BPS quivers of a given theory, and try to answer
 the following questions: what is the minimal chamber and what is the maximal chamber? whether any length between the 
 minimal and maximal length is realized as the length of a BPS chamber, etc. 

b. In fact, the spectrum we found should be called potential chamber of the theory, and it would be interesting to see if 
such spectrum is truly realized on the Coulomb branch. It seems to us that all the finite chamber can be realized on the moduli space, but this definitely 
needs further study. Although the detailed factorization might not be realized physically, the spectral generator is 
the correct one regardless of the chamber, and it is  interesting to explore how to find the sensible factorizations of the given spectral generator and 
therefore find new chambers.

c. We have not found any efficient way to deal with chamber with higher spin states.  It would be nice to find 
new methods to deal with this problem.

d. The BPS counting in the supergravity context is studied in \cite{Chuang:2008aw,Jafferis:2008uf, Aganagic:2009cg, Aganagic:2010qr,Yamazaki:2010fz,Andriyash:2010yf, Andriyash:2010qv, Manschot:2010qz, Alexandrov:2011ac,Manschot:2012rx}, can we  apply our combinatorial method to that context?

With the BPS spectrum on hand, there are many physical questions one could  ask:

a. Why the spectrum of a given theory has the specific structure? Can we learn about the UV theory from the BPS data? 
Recently, it is proposed that the BPS spectrum can be used to calculate the index of the superconformal field theory \cite{Lockhart:2012vp}, 
it would be interesting to carry out this explicitly. 
Furthermore, the finite spectrum usually happens in the strongly coupled region of the Lagrangian theory, and the 
massless BPS particles at the singularity of the Coulomb branch should be included into the stable BPS spectrum. Since the BPS particles 
at the singularity is very important in understanding the IR physics,  can we learn something deep
 about the quantum dynamics like the exact solutions and confinement from the explicit BPS spectrum?

b. The quantum dilogarithm identify has important implications for mirror symmetry of 3d $\mathcal{N}=2$ theory \cite{Dimofte:2011ju,Cecotti:2011iy}. We have 
found a huge number of new quantum dilogarithm identities, can we use them to find new 3d Mirror pairs? The quantum 
dilogarithm identity is closed related to the integrable system \cite{Keller:arXiv1001.1531}, it would be interesting to find the detailed connection.

c: The finite spectrum has an interesting $N^3$ scaling behavior in the large N limit,  which is in agreement with the degree of freedom of $N$ M5 branes. Since 
the BPS states can be thought of  as self-dual string wrapping on various one cycles on the punctured Riemann surface, it is natural to think this $N^3$ scaling 
should be related to M5 brane dynamics.

d. The BPS spectrum is important for finding the Coulomb branch metric of the corresponding 3d theory derived by compactifying 4d theory on a circle,
and  it would be interesting to find out  the Hyperkahler metric explicitly this using the BPS spectrum \cite{Xie:2012hh}.

\begin{flushleft}
\textbf{Acknowledgments}
\end{flushleft}

We thank M. Alim, M. Del Zotto, D. Gaiotto,  A. Neitzke,  C. Vafa and M. Yamazaki  for help discussions. We thank Yu-tin Huang for help on 
a mathematica code.
This research is supported in part by Zurich Financial services membership and by the U.S. Department of Energy, grant DE- FG02-90ER40542 (DX). 
We thank the 10th Simons Summer Workshop in Mathematics and Physics, and the Simons Center for Geometry and Physics for hospitality where
part of this work is finished. 

\appendix

\section{Explicit spectral generator}
\subsection{$A_3$ quiver}
\begin{figure}[htbp]
\small
\centering
\includegraphics[width=8cm]{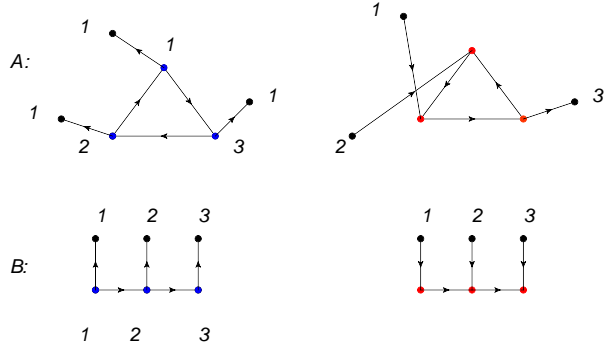}
\caption{Two quivers in $A_3$ mutation class, we show the initial configuration and final configuration of the maximal green mutation. }
\label{app2}
\end{figure}

In this section, we give some explicit expression for the spectral generator
using the cluster transformation rule for the $X$ coordinates under mutation on node $k$, 
\begin{align}
&X_k^{'}=X_k^{-1} \nonumber\\
&X_i^{'}=X_i(1+X_k^{-sgn(\epsilon_{ik})})^{-\epsilon_{ik}}.
\end{align}
 For the $A_3$ quiver  shown in figure. \ref{app2}A, the maximal green mutation sequences are
\begin{equation}
\mu_1, \mu_2, \mu_3, \mu_1.
\end{equation}
The final cluster coordinates are:  \footnote{We only show the cluster coordinates due to the mutation, the permutation can be found using the green mutation as 
we show in figure. \ref{app2}A, the spectral generator is derived by combining the permutation and the cluster transformation.}

\begin{doublespace}
\noindent\(\pmb{\hat{x}[1]=\frac{1+x_1+x_1 x_2}{x_2 \left(1+x_3+x_1 x_3\right)},}\\
\pmb{\hat{x}[2]=\frac{1+x_3+x_1 x_3}{x_1 \left(1+x_2+x_2 x_3\right)},}\\
\pmb{\hat{x}[3]=\frac{1+x_2+x_2 x_3}{\left(1+x_1+x_1 x_2\right) x_3};}\)
\end{doublespace}

If we use another quiver in the same mutation class as the above quiver, say the quiver in figure. \ref{app2}B,  then one maximal green mutation sequences are 
\begin{equation}
\mu_1, \mu_2, \mu_3,
\end{equation}
and the final cluster coordinates would be:

\begin{doublespace}
\noindent\(\pmb{\hat{X }[1]=\frac{1+x_2+x_1 x_2}{x_1},}\\
\pmb{\hat{X }[2]=\frac{1+x_3+x_2 x_3+x_1 x_2 x_3}{\left(1+x_1\right) x_2},}\\
\pmb{\hat{X }[3]=\frac{1}{\left(1+x_2+x_1 x_2\right) x_3}.}\)
\end{doublespace}

Notice that the final coordinates depend on the initial quiver, but the quantum dilogarithm identity
is independent of the quiver, they are all equal by changing the basis of the quantum torus from 
one quiver to another quiver.

\subsection{Disc with three $A_{N-1}$ full punctures}
\begin{figure}[htbp]
\small
\centering
\includegraphics[width=8cm]{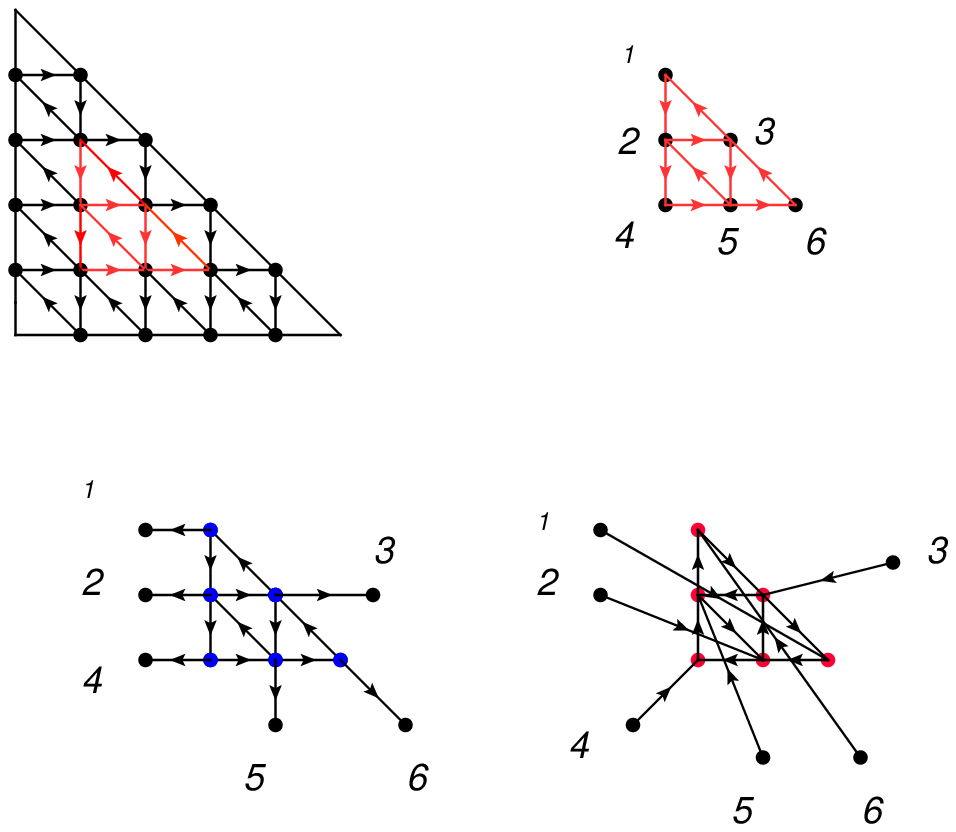}
\caption{The quiver from disc with three full punctures and $N=5$. The initial and final configuration of the maximal green mutation are indicated. }
\label{app1}
\end{figure}

Let's consider a disc with three full punctures and $N=5$ , the BPS quiver is shown in figure. \ref{app1}, and the maximal mutation sequence is 
\begin{equation}
\mu_1, \mu_2,\mu_3, \mu_4, \mu_5, \mu_6, \mu_1, \mu_2, \mu_3, \mu_1
\end{equation}
We find the spectral generator (final cluster coordinates) as following

\begin{doublespace}
\noindent\(\pmb{\hat{X }[1]=\frac{1+x_3+x_2 x_3+x_3 x_6+x_2 x_3 x_6+x_2 x_3 x_5 x_6}{\left(1+x_5+x_3 x_5+x_4 x_5+x_3 x_4 x_5+x_2 x_3 x_4 x_5\right)
x_6},}\\
\pmb{\hat{X }[2]=\frac{\left(1+x_2+x_1 x_2+x_2 x_5+x_1 x_2 x_5+x_1 x_2 x_3 x_5\right) \left(1+x_6+x_5 x_6+x_4 x_5 x_6\right)}{\left(1+x_4+x_2 x_4+x_1
x_2 x_4\right) x_5 \left(1+x_3+x_2 x_3+x_3 x_6+x_2 x_3 x_6+x_2 x_3 x_5 x_6\right)},}\\
\pmb{\hat{X }[3]=\frac{\left(1+x_5+x_3 x_5+x_4 x_5+x_3 x_4 x_5+x_2 x_3 x_4 x_5\right) \left(1+x_1+x_1 x_3+x_1 x_3 x_6\right)}{x_3 \left(1+x_2+x_1
x_2+x_2 x_5+x_1 x_2 x_5+x_1 x_2 x_3 x_5\right) \left(1+x_6+x_5 x_6+x_4 x_5 x_6\right)},}\\
\pmb{\hat{X }[4]=\frac{1+x_5+x_3 x_5+x_4 x_5+x_3 x_4 x_5+x_2 x_3 x_4 x_5}{x_4 \left(1+x_2+x_1 x_2+x_2 x_5+x_1 x_2 x_5+x_1 x_2 x_3 x_5\right)},}\\
\pmb{\hat{X }[5]=\frac{\left(1+x_4+x_2 x_4+x_1 x_2 x_4\right) \left(1+x_3+x_2 x_3+x_3 x_6+x_2 x_3 x_6+x_2 x_3 x_5 x_6\right)}{x_2 \left(1+x_5+x_3
x_5+x_4 x_5+x_3 x_4 x_5+x_2 x_3 x_4 x_5\right) \left(1+x_1+x_1 x_3+x_1 x_3 x_6\right)},}\\
\pmb{\hat{X }[6]=\frac{1+x_2+x_1 x_2+x_2 x_5+x_1 x_2 x_5+x_1 x_2 x_3 x_5}{x_1 \left(1+x_3+x_2 x_3+x_3 x_6+x_2 x_3 x_6+x_2 x_3 x_5 x_6\right)}.}\)
\end{doublespace}
Here $x_i$  is the initial cluster variable. This result is the same as the formula  $[7.13-7.15]$ in \cite{Gaiotto:2012db} if we identify the initial coordinates in the following way
\begin{equation}
x_1=r_{200},~~x_2=r_{110}~~x_3=r_{011}  ,~~ x_4=r_{020},~~x_5=r_{101},~~x_6=r_{002}.
\end{equation}

\newpage
\subsection{Disc with four $A_{N-1}$ full punctures}
\begin{figure}[htbp]
\small
\centering
\includegraphics[width=12cm]{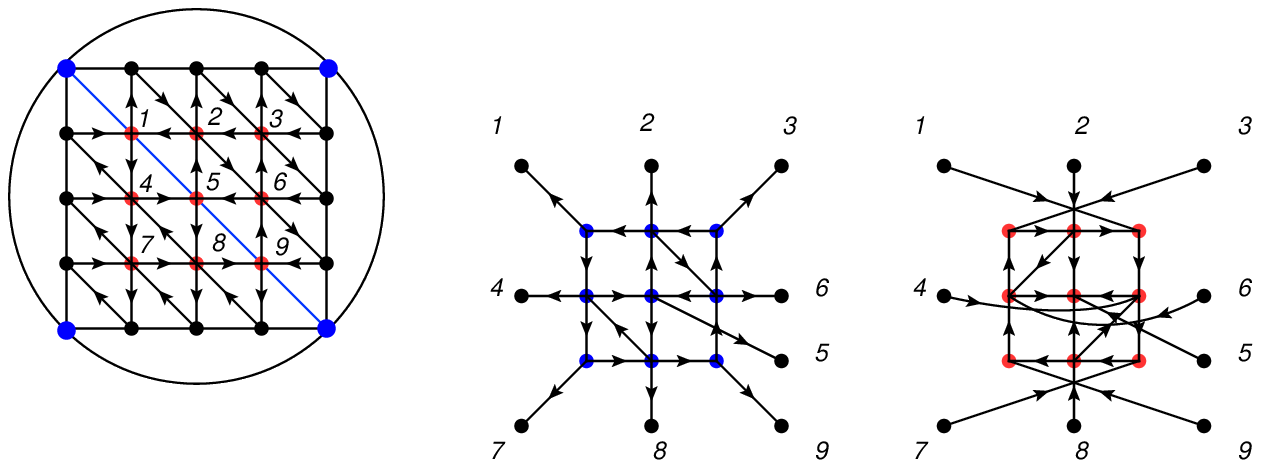}
\caption{The initial configuration and the final configuration of the maximal green mutation of the quiver from the disc with four full punctures.}
\label{app4}
\end{figure}

The quiver is shown in figure. \ref{app4}, and the green mutation sequence is 
\begin{align}
& (1,5,9), (2,4,6, 8), (3,5,7), \nonumber\\
& (1,2,4,1), (9,6,8,9), \nonumber\\
\end{align}

we take $N=4$ and the spectral generator is

\begin{doublespace}
\noindent\(\pmb{\hat{X }[1]=\frac{1+x_2+x_2 x_3+x_2 x_5+x_2 x_3 x_5+x_2 x_4 x_5+x_2 x_3 x_4 x_5+x_2 x_3 x_5 x_6+x_2 x_3 x_4 x_5 x_6+x_2 x_3 x_5 x_6
x_9+x_2 x_3 x_4 x_5 x_6 x_9+x_2 x_3 x_4 x_5 x_6 x_8 x_9}{x_3 \left(1+x_6+x_2 x_6+x_6 x_9+x_2 x_6 x_9+x_6 x_8 x_9+x_2 x_6 x_8 x_9+x_2 x_5 x_6 x_8
x_9+x_6 x_7 x_8 x_9+x_2 x_6 x_7 x_8 x_9+x_2 x_5 x_6 x_7 x_8 x_9+x_2 x_4 x_5 x_6 x_7 x_8 x_9\right)},}\\
\pmb{\hat{X }[3]=\frac{1+x_4+x_1 x_4+x_1 x_2 x_4+x_1 x_2 x_3 x_4+x_4 x_8+x_1 x_4 x_8+x_1 x_2 x_4 x_8+x_1 x_2 x_3 x_4 x_8+x_1 x_2 x_4 x_5 x_8+x_1
x_2 x_3 x_4 x_5 x_8+x_1 x_2 x_3 x_4 x_5 x_6 x_8}{x_1 \left(1+x_2+x_2 x_3+x_2 x_5+x_2 x_3 x_5+x_2 x_4 x_5+x_2 x_3 x_4 x_5+x_2 x_3 x_5 x_6+x_2 x_3
x_4 x_5 x_6+x_2 x_3 x_5 x_6 x_9+x_2 x_3 x_4 x_5 x_6 x_9+x_2 x_3 x_4 x_5 x_6 x_8 x_9\right)},}\\
\pmb{\hat{X }[7]=\frac{1+x_6+x_2 x_6+x_6 x_9+x_2 x_6 x_9+x_6 x_8 x_9+x_2 x_6 x_8 x_9+x_2 x_5 x_6 x_8 x_9+x_6 x_7 x_8 x_9+x_2 x_6 x_7 x_8 x_9+x_2
x_5 x_6 x_7 x_8 x_9+x_2 x_4 x_5 x_6 x_7 x_8 x_9}{\left(1+x_8+x_5 x_8+x_5 x_6 x_8+x_7 x_8+x_5 x_7 x_8+x_4 x_5 x_7 x_8+x_1 x_4 x_5 x_7 x_8+x_5 x_6
x_7 x_8+x_4 x_5 x_6 x_7 x_8+x_1 x_4 x_5 x_6 x_7 x_8+x_1 x_2 x_4 x_5 x_6 x_7 x_8\right) x_9},}\\
\pmb{\hat{X }[9]=\frac{1+x_8+x_5 x_8+x_5 x_6 x_8+x_7 x_8+x_5 x_7 x_8+x_4 x_5 x_7 x_8+x_1 x_4 x_5 x_7 x_8+x_5 x_6 x_7 x_8+x_4 x_5 x_6 x_7 x_8+x_1
x_4 x_5 x_6 x_7 x_8+x_1 x_2 x_4 x_5 x_6 x_7 x_8}{x_7 \left(1+x_4+x_1 x_4+x_1 x_2 x_4+x_1 x_2 x_3 x_4+x_4 x_8+x_1 x_4 x_8+x_1 x_2 x_4 x_8+x_1 x_2
x_3 x_4 x_8+x_1 x_2 x_4 x_5 x_8+x_1 x_2 x_3 x_4 x_5 x_8+x_1 x_2 x_3 x_4 x_5 x_6 x_8\right)}.}\)
\end{doublespace}

\begin{doublespace}
\noindent\(\pmb{\hat{X }[2]= \frac{\text{A2}*\text{B2}}{\text{C2}*\text{D2}}}\\
\pmb{\text{A2}=\left(1+x_1+x_1 x_2+x_1 x_2 x_3+x_1 x_2 x_5+x_1 x_2 x_3 x_5+x_1 x_2 x_3 x_5 x_6+x_1 x_2 x_3 x_5 x_6 x_9\right), }\\
\pmb{ \text{B2}=\left(1+x_6+x_2 x_6+x_6 x_9+x_2 x_6 x_9+x_6 x_8 x_9+x_2 x_6 x_8 x_9+x_2 x_5 x_6 x_8 x_9+x_6 x_7 x_8 x_9\right.}\\
\pmb{\left.+x_2 x_6 x_7 x_8 x_9+x_2 x_5 x_6 x_7 x_8 x_9+x_2 x_4 x_5 x_6 x_7 x_8 x_9\right),}\\
\pmb{\text{C2}=\left(1+x_5+x_4 x_5+x_1 x_4 x_5+x_5 x_6+x_4 x_5 x_6+x_1 x_4 x_5 x_6+x_1 x_2 x_4 x_5 x_6+x_5 x_6 x_9+x_4 x_5 x_6 x_9\right.}\\
\pmb{\left.+x_1 x_4 x_5 x_6 x_9+x_1 x_2 x_4 x_5 x_6 x_9+x_4 x_5 x_6 x_8 x_9+x_1 x_4 x_5 x_6 x_8 x_9+x_1 x_2 x_4 x_5 x_6 x_8 x_9+x_1 x_2 x_4 x_5^2
x_6 x_8 x_9\right) ,}\\
\pmb{\text{D2}=x_2 \left(1+x_3+x_3 x_6+x_3 x_6 x_9+x_3 x_6 x_8 x_9+x_3 x_6 x_7 x_8 x_9\right),}\)
\end{doublespace}

\begin{doublespace}
\noindent\(\pmb{\hat{X }[4]=\frac{\text{A4}*\text{B4}}{\text{C4}*\text{D4}}}\\
\pmb{\text{A4}=\left(1+x_5+x_4 x_5+x_1 x_4 x_5+x_5 x_6+x_4 x_5 x_6+x_1 x_4 x_5 x_6+x_1 x_2 x_4 x_5 x_6+x_5 x_6 x_9+x_4 x_5 x_6 x_9\right.}\\
\pmb{+x_1 x_4 x_5 x_6 x_9+x_1 x_2 x_4 x_5 x_6 x_9+x_4 x_5 x_6 x_8 x_9+x_1 x_4 x_5 x_6 x_8 x_9+x_1 x_2 x_4 x_5 x_6 x_8 x_9}\\
\pmb{\left.+x_1 x_2 x_4 x_5^2 x_6 x_8 x_9\right), }\\
\pmb{\text{B4}=\left(1+x_3+x_3 x_6+x_3 x_6 x_9+x_3 x_6 x_8 x_9+x_3 x_6 x_7 x_8 x_9\right),}\\
\pmb{\text{C4}=x_6 \left(1+x_2+x_2 x_3+x_2 x_5+x_2 x_3 x_5+x_2 x_4 x_5+x_2 x_3 x_4 x_5+x_2 x_3 x_5 x_6+x_2 x_3 x_4 x_5 x_6\right.}\\
\pmb{\left.+x_2 x_3 x_5 x_6 x_9+x_2 x_3 x_4 x_5 x_6 x_9+x_2 x_3 x_4 x_5 x_6 x_8 x_9\right),}\\
\pmb{ \text{D4}=\left(1+x_9+x_8 x_9+x_5 x_8 x_9+x_7 x_8 x_9+x_5 x_7 x_8 x_9+x_4 x_5 x_7 x_8 x_9+x_1 x_4 x_5 x_7 x_8 x_9\right),}\\
\pmb{}\\
\pmb{\hat{X }[5]=\frac{\text{A5}*\text{B5}}{\text{C5}*\text{D5}}}\\
\pmb{}\\
\pmb{\text{A5}=\left(1+x_8+x_5 x_8+x_5 x_6 x_8+x_7 x_8+x_5 x_7 x_8+x_4 x_5 x_7 x_8+x_1 x_4 x_5 x_7 x_8+x_5 x_6 x_7 x_8\right.}\\
\pmb{\left.+x_4 x_5 x_6 x_7 x_8+x_1 x_4 x_5 x_6 x_7 x_8+x_1 x_2 x_4 x_5 x_6 x_7 x_8\right), }\\
\pmb{\text{B5}=\left(1+x_2+x_2 x_3+x_2 x_5+x_2 x_3 x_5+x_2 x_4 x_5+x_2 x_3 x_4 x_5+x_2 x_3 x_5 x_6+x_2 x_3 x_4 x_5 x_6\right.}\\
\pmb{\left.+x_2 x_3 x_5 x_6 x_9+x_2 x_3 x_4 x_5 x_6 x_9+x_2 x_3 x_4 x_5 x_6 x_8 x_9\right),}\\
\pmb{\text{C5}=x_5 \left(1+x_4+x_1 x_4+x_1 x_2 x_4+x_1 x_2 x_3 x_4+x_4 x_8+x_1 x_4 x_8+x_1 x_2 x_4 x_8+x_1 x_2 x_3 x_4 x_8\right.}\\
\pmb{\left.+x_1 x_2 x_4 x_5 x_8+x_1 x_2 x_3 x_4 x_5 x_8+x_1 x_2 x_3 x_4 x_5 x_6 x_8\right),}\\
\pmb{\text{D5}= \left(1+x_6+x_2 x_6+x_6 x_9+x_2 x_6 x_9+x_6 x_8 x_9+x_2 x_6 x_8 x_9+x_2 x_5 x_6 x_8 x_9+x_6 x_7 x_8 x_9\right.}\\
\pmb{\left.+x_2 x_6 x_7 x_8 x_9+x_2 x_5 x_6 x_7 x_8 x_9+x_2 x_4 x_5 x_6 x_7 x_8 x_9\right),}\\
\pmb{}\\
\pmb{\hat{X }[6]=\frac{\text{A6}*\text{B6}}{\text{C6}*\text{D6}}}\\
\pmb{}\\
\pmb{\text{A6}=\left(1+x_7+x_4 x_7+x_1 x_4 x_7+x_1 x_2 x_4 x_7+x_1 x_2 x_3 x_4 x_7\right),}\\
\pmb{ \text{B6}=\left(1+x_5+x_4 x_5+x_1 x_4 x_5+x_5 x_6+x_4 x_5 x_6+x_1 x_4 x_5 x_6+x_1 x_2 x_4 x_5 x_6\right.}\\
\pmb{+x_5 x_6 x_9+x_4 x_5 x_6 x_9+x_1 x_4 x_5 x_6 x_9+x_1 x_2 x_4 x_5 x_6 x_9+x_4 x_5 x_6 x_8 x_9}\\
\pmb{\left.+x_1 x_4 x_5 x_6 x_8 x_9+x_1 x_2 x_4 x_5 x_6 x_8 x_9+x_1 x_2 x_4 x_5^2 x_6 x_8 x_9\right),}\\
\pmb{\text{C6}=x_4 \left(1+x_8+x_5 x_8+x_5 x_6 x_8+x_7 x_8+x_5 x_7 x_8+x_4 x_5 x_7 x_8+x_1 x_4 x_5 x_7 x_8\right.}\\
\pmb{\left.+x_5 x_6 x_7 x_8+x_4 x_5 x_6 x_7 x_8+x_1 x_4 x_5 x_6 x_7 x_8+x_1 x_2 x_4 x_5 x_6 x_7 x_8\right) ,}\\
\pmb{\text{D6}=\left(1+x_1+x_1 x_2+x_1 x_2 x_3+x_1 x_2 x_5+x_1 x_2 x_3 x_5+x_1 x_2 x_3 x_5 x_6+x_1 x_2 x_3 x_5 x_6 x_9\right),}\)
\end{doublespace}

\begin{doublespace}
\noindent\(\pmb{\hat{X }[8]=\frac{\text{A8}*\text{B8}}{\text{C8}*\text{D8}}}\\
\pmb{\text{A8}=\left(1+x_4+x_1 x_4+x_1 x_2 x_4+x_1 x_2 x_3 x_4+x_4 x_8+x_1 x_4 x_8+x_1 x_2 x_4 x_8\right.}\\
\pmb{\left.+x_1 x_2 x_3 x_4 x_8+x_1 x_2 x_4 x_5 x_8+x_1 x_2 x_3 x_4 x_5 x_8+x_1 x_2 x_3 x_4 x_5 x_6 x_8\right), }\\
\pmb{\text{B8}=\left(1+x_9+x_8 x_9+x_5 x_8 x_9+x_7 x_8 x_9+x_5 x_7 x_8 x_9+x_4 x_5 x_7 x_8 x_9+x_1 x_4 x_5 x_7 x_8 x_9\right),}\\
\pmb{\text{C8}=\left(1+x_7+x_4 x_7+x_1 x_4 x_7+x_1 x_2 x_4 x_7+x_1 x_2 x_3 x_4 x_7\right) x_8 ,}\\
\pmb{\text{D8}=\left(1+x_5+x_4 x_5+x_1 x_4 x_5+x_5 x_6+x_4 x_5 x_6+x_1 x_4 x_5 x_6+x_1 x_2 x_4 x_5 x_6\right.}\\
\pmb{+x_5 x_6 x_9+x_4 x_5 x_6 x_9+x_1 x_4 x_5 x_6 x_9+x_1 x_2 x_4 x_5 x_6 x_9+x_4 x_5 x_6 x_8 x_9}\\
\pmb{\left.+x_1 x_4 x_5 x_6 x_8 x_9+x_1 x_2 x_4 x_5 x_6 x_8 x_9+x_1 x_2 x_4 x_5^2 x_6 x_8 x_9\right),}\)
\end{doublespace}

\subsection{$A_2$ pentagon}
\begin{figure}[htbp]
\small
\centering
\includegraphics[width=4cm]{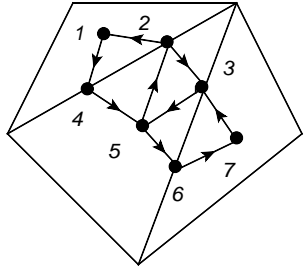}
\caption{The quiver for $A_2$ pentagon with all punctures full. }
\label{app3}
\end{figure}

The quiver is shown in figure. \ref{app3}, and the maximal mutation sequence is 
\begin{equation}
(\mu_2, \mu_4, \mu_1, \mu_5),~~(\mu_3, \mu_6, \mu_2, \mu_7),~(\mu_3, \mu_4, \mu_6).
\end{equation} 

So the final cluster coordinates are

\begin{doublespace}
\noindent\(\pmb{\hat{x}[1]=\frac{1+x_5+x_3 x_5+x_4 x_5+x_1 x_4 x_5+x_3 x_4 x_5+x_1 x_3 x_4 x_5+x_1 x_2 x_3 x_4 x_5+x_3 x_5 x_7+x_3 x_4 x_5 x_7+x_1
x_3 x_4 x_5 x_7+x_1 x_2 x_3 x_4 x_5 x_7}{x_4 \left(1+x_1+x_1 x_2+x_1 x_2 x_5\right)},}\\
\pmb{\hat{x}[2]=\frac{1+x_7+x_6 x_7+x_5 x_6 x_7+x_4 x_5 x_6 x_7+x_1 x_4 x_5 x_6 x_7}{x_6 \left(1+x_5+x_3 x_5+x_4 x_5+x_1 x_4 x_5+x_3 x_4 x_5+x_1
x_3 x_4 x_5+x_1 x_2 x_3 x_4 x_5+x_3 x_5 x_7+x_3 x_4 x_5 x_7+x_1 x_3 x_4 x_5 x_7+x_1 x_2 x_3 x_4 x_5 x_7\right)},}\\
\pmb{\hat{x}[3]=\frac{1+x_3+x_2 x_3+x_3 x_7+x_2 x_3 x_7+x_3 x_6 x_7+x_2 x_3 x_6 x_7+x_2 x_3 x_5 x_6 x_7+x_2 x_3 x_4 x_5 x_6 x_7}{\left(1+x_6+x_5
x_6+x_3 x_5 x_6+x_4 x_5 x_6+x_1 x_4 x_5 x_6+x_3 x_4 x_5 x_6+x_1 x_3 x_4 x_5 x_6+x_1 x_2 x_3 x_4 x_5 x_6\right) x_7},}\\
\pmb{\hat{x}[4]=\frac{1+x_4+x_1 x_4+x_1 x_2 x_4}{x_1 \left(1+x_2+x_2 x_5+x_2 x_4 x_5\right)},}\\
\pmb{\hat{x}[5]=\frac{\left(1+x_1+x_1 x_2+x_1 x_2 x_5\right) \left(1+x_3+x_2 x_3+x_3 x_7+x_2 x_3 x_7+x_3 x_6 x_7+x_2 x_3 x_6 x_7+x_2 x_3 x_5 x_6
x_7+x_2 x_3 x_4 x_5 x_6 x_7\right)}{x_2 \left(1+x_5+x_3 x_5+x_4 x_5+x_1 x_4 x_5+x_3 x_4 x_5+x_1 x_3 x_4 x_5+x_1 x_2 x_3 x_4 x_5+x_3 x_5 x_7+x_3 x_4
x_5 x_7+x_1 x_3 x_4 x_5 x_7+x_1 x_2 x_3 x_4 x_5 x_7\right)},}\\
\pmb{\hat{x}[6]=\frac{\left(1+x_2+x_2 x_5+x_2 x_4 x_5\right) \left(1+x_6+x_5 x_6+x_3 x_5 x_6+x_4 x_5 x_6+x_1 x_4 x_5 x_6+x_3 x_4 x_5 x_6+x_1 x_3
x_4 x_5 x_6+x_1 x_2 x_3 x_4 x_5 x_6\right)}{\left(1+x_4+x_1 x_4+x_1 x_2 x_4\right) x_5 \left(1+x_3+x_2 x_3+x_3 x_7+x_2 x_3 x_7+x_3 x_6 x_7+x_2 x_3
x_6 x_7+x_2 x_3 x_5 x_6 x_7+x_2 x_3 x_4 x_5 x_6 x_7\right)},}\\
\pmb{\hat{x}[7]=\frac{1+x_5+x_3 x_5+x_4 x_5+x_1 x_4 x_5+x_3 x_4 x_5+x_1 x_3 x_4 x_5+x_1 x_2 x_3 x_4 x_5+x_3 x_5 x_7+x_3 x_4 x_5 x_7+x_1 x_3 x_4 x_5
x_7+x_1 x_2 x_3 x_4 x_5 x_7}{x_3 \left(1+x_2+x_2 x_5+x_2 x_4 x_5\right) \left(1+x_7+x_6 x_7+x_5 x_6 x_7+x_4 x_5 x_6 x_7+x_1 x_4 x_5 x_6 x_7\right)}.}\)
\end{doublespace}

\subsection{$SU(2)$ with four flavors}
\begin{figure}[htbp]
\small
\centering
\includegraphics[width=4cm]{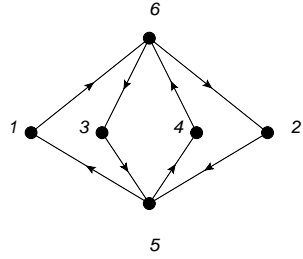}
\caption{The quiver for $SU(2)$ with four flavors. }
\label{app5}
\end{figure}

The quiver is shown in figure. \ref{app5}. The maximal green mutation sequences are
\begin{equation}
\mu_1, \mu_2, \mu_3, \mu_4, \mu_5, \mu_6,\mu_1, \mu_2, \mu_3, \mu_4, \mu_5, \mu_6.
\end{equation}
So the spectral generator is 

\begin{doublespace}
\noindent\(\pmb{\hat{x}[1]=\frac{\text{A1}}{\text{B1}}}\\
\pmb{\text{A1}=\left(1+x_6+x_1 x_6+x_4 x_6+x_1 x_4 x_6+x_1 x_5 x_6+x_4 x_5 x_6+2 x_1 x_4 x_5 x_6+x_1 x_2 x_4 x_5 x_6+\right.}\\
\pmb{\left.x_1 x_3 x_4 x_5 x_6+x_1 x_4 x_5^2 x_6+x_1 x_2 x_4 x_5^2 x_6+x_1 x_3 x_4 x_5^2 x_6+x_1 x_2 x_3 x_4 x_5^2 x_6+x_1 x_2 x_3 x_4 x_5^2 x_6^2\right),}\\
\pmb{\text{B1}=x_4 \left(1+x_5+x_2 x_5+x_3 x_5+x_2 x_3 x_5+x_2 x_5 x_6+x_3 x_5 x_6+2 x_2 x_3 x_5 x_6+x_1 x_2 x_3 x_5 x_6+\right.}\\
\pmb{\left.x_2 x_3 x_4 x_5 x_6+x_2 x_3 x_5 x_6^2+x_1 x_2 x_3 x_5 x_6^2+x_2 x_3 x_4 x_5 x_6^2+x_1 x_2 x_3 x_4 x_5 x_6^2+x_1 x_2 x_3 x_4 x_5^2 x_6^2\right),}\\
\pmb{}\\
\pmb{\hat{x}[2]=\frac{\text{A2}}{\text{B2}}}\\
\pmb{\text{A2}=\left(1+x_5+x_2 x_5+x_3 x_5+x_2 x_3 x_5+x_2 x_5 x_6+x_3 x_5 x_6+2 x_2 x_3 x_5 x_6+x_1 x_2 x_3 x_5 x_6+\right.}\\
\pmb{\left.x_2 x_3 x_4 x_5 x_6+x_2 x_3 x_5 x_6^2+x_1 x_2 x_3 x_5 x_6^2+x_2 x_3 x_4 x_5 x_6^2+x_1 x_2 x_3 x_4 x_5 x_6^2+x_1 x_2 x_3 x_4 x_5^2 x_6^2\right),}\\
\pmb{\text{B2}=x_3 \left(1+x_6+x_1 x_6+x_4 x_6+x_1 x_4 x_6+x_1 x_5 x_6+x_4 x_5 x_6+2 x_1 x_4 x_5 x_6+x_1 x_2 x_4 x_5 x_6+\right.}\\
\pmb{\left.x_1 x_3 x_4 x_5 x_6+x_1 x_4 x_5^2 x_6+x_1 x_2 x_4 x_5^2 x_6+x_1 x_3 x_4 x_5^2 x_6+x_1 x_2 x_3 x_4 x_5^2 x_6+x_1 x_2 x_3 x_4 x_5^2 x_6^2\right),}\\
\pmb{}\\
\pmb{\hat{x}[3]=\frac{\text{A3}}{\text{B3}}}\\
\pmb{\text{A3}=\left(1+x_5+x_2 x_5+x_3 x_5+x_2 x_3 x_5+x_2 x_5 x_6+x_3 x_5 x_6+2 x_2 x_3 x_5 x_6+x_1 x_2 x_3 x_5 x_6+\right.}\\
\pmb{\left.x_2 x_3 x_4 x_5 x_6+x_2 x_3 x_5 x_6^2+x_1 x_2 x_3 x_5 x_6^2+x_2 x_3 x_4 x_5 x_6^2+x_1 x_2 x_3 x_4 x_5 x_6^2+x_1 x_2 x_3 x_4 x_5^2 x_6^2\right),}\\
\pmb{\text{B3}=x_2 \left(1+x_6+x_1 x_6+x_4 x_6+x_1 x_4 x_6+x_1 x_5 x_6+x_4 x_5 x_6+2 x_1 x_4 x_5 x_6+x_1 x_2 x_4 x_5 x_6+\right.}\\
\pmb{\left.x_1 x_3 x_4 x_5 x_6+x_1 x_4 x_5^2 x_6+x_1 x_2 x_4 x_5^2 x_6+x_1 x_3 x_4 x_5^2 x_6+x_1 x_2 x_3 x_4 x_5^2 x_6+x_1 x_2 x_3 x_4 x_5^2 x_6^2\right),}\\
\pmb{}\\
\pmb{\hat{x}[4]=\frac{\text{A4}}{\text{B4}}}\\
\pmb{\text{A4}=\left(1+x_6+x_1 x_6+x_4 x_6+x_1 x_4 x_6+x_1 x_5 x_6+x_4 x_5 x_6+2 x_1 x_4 x_5 x_6+x_1 x_2 x_4 x_5 x_6+\right.}\\
\pmb{\left.x_1 x_3 x_4 x_5 x_6+x_1 x_4 x_5^2 x_6+x_1 x_2 x_4 x_5^2 x_6+x_1 x_3 x_4 x_5^2 x_6+x_1 x_2 x_3 x_4 x_5^2 x_6+x_1 x_2 x_3 x_4 x_5^2 x_6^2\right),}\\
\pmb{\text{B4}=x_1 \left(1+x_5+x_2 x_5+x_3 x_5+x_2 x_3 x_5+x_2 x_5 x_6+x_3 x_5 x_6+2 x_2 x_3 x_5 x_6+x_1 x_2 x_3 x_5 x_6+\right.}\\
\pmb{\left.x_2 x_3 x_4 x_5 x_6+x_2 x_3 x_5 x_6^2+x_1 x_2 x_3 x_5 x_6^2+x_2 x_3 x_4 x_5 x_6^2+x_1 x_2 x_3 x_4 x_5 x_6^2+x_1 x_2 x_3 x_4 x_5^2 x_6^2\right),}\\
\pmb{}\\
\pmb{\hat{x}[5]=\frac{\text{A5}*\text{C5}}{\text{B5}*\text{D5}}}\\
\pmb{\text{A5}=\left(1+x_1+x_1 x_5+x_1 x_2 x_5+x_1 x_3 x_5+x_1 x_2 x_3 x_5+x_1 x_2 x_3 x_5 x_6+x_1^2 x_2 x_3 x_5 x_6\right) ,}\\
\pmb{\text{C5}=\left(1+x_4+x_4 x_5+x_2 x_4 x_5+x_3 x_4 x_5+x_2 x_3 x_4 x_5+x_2 x_3 x_4 x_5 x_6+x_2 x_3 x_4^2 x_5 x_6\right),}\\
\pmb{\text{B5}=x_5 \left(1+x_2+x_2 x_6+x_1 x_2 x_6+x_2 x_4 x_6+x_1 x_2 x_4 x_6+x_1 x_2 x_4 x_5 x_6+x_1 x_2^2 x_4 x_5 x_6\right),}\\
\pmb{ \text{D5}=\left(1+x_3+x_3 x_6+x_1 x_3 x_6+x_3 x_4 x_6+x_1 x_3 x_4 x_6+x_1 x_3 x_4 x_5 x_6+x_1 x_3^2 x_4 x_5 x_6\right),}\\
\pmb{}\\
\pmb{\hat{x}[5]=\frac{\text{A6}*\text{C6}}{\text{B6}*\text{D6}}}\\
\pmb{\text{A6}=\left(1+x_2+x_2 x_6+x_1 x_2 x_6+x_2 x_4 x_6+x_1 x_2 x_4 x_6+x_1 x_2 x_4 x_5 x_6+x_1 x_2^2 x_4 x_5 x_6\right) ,}\\
\pmb{\text{C6}=\left(1+x_3+x_3 x_6+x_1 x_3 x_6+x_3 x_4 x_6+x_1 x_3 x_4 x_6+x_1 x_3 x_4 x_5 x_6+x_1 x_3^2 x_4 x_5 x_6\right),}\\
\pmb{\text{B6}=x_6 \left(1+x_1+x_1 x_5+x_1 x_2 x_5+x_1 x_3 x_5+x_1 x_2 x_3 x_5+x_1 x_2 x_3 x_5 x_6+x_1^2 x_2 x_3 x_5 x_6\right) ,}\\
\pmb{\text{D6}=\left(1+x_4+x_4 x_5+x_2 x_4 x_5+x_3 x_4 x_5+x_2 x_3 x_4 x_5+x_2 x_3 x_4 x_5 x_6+x_2 x_3 x_4^2 x_5 x_6\right).}\)
\end{doublespace}

\section{Refined spectral generator}
Let's describe a simple way of finding the refined spectral generator from the classical spectral generator. The $q$ deformation of the cluster algebra is 
\begin{equation}
X_{i}X_{j}=q^{\epsilon_{ij}}X_{i}X_{j},
\end{equation}
where $\epsilon_{ij}$ is the antisymmetric tensor read from the quiver, and the cluster transformation on quantum cluster algebra is 
\begin{align}
& X_k^{'}=X_k^{-1}, \nonumber\\
& X_i^{'}=X_i(\prod_{a=1}^{|\epsilon_{ik}|}(1+q^{a-1/2}X_k^{-sgn(\epsilon_{ik})}))^{-sgn(\epsilon_{ik})}.
\end{align}
Given the mutation sequences, one could find the final quantum cluster coordinates. 

In practice, there is a way of reading the quantum cluster algebra from the classical one using the $*$ invariance.
One could define a $*$ action on 
the quantum cluster algebra
\begin{equation}
*(q)=q^{-1},~~*(X_i)=X_i,~~*(X_iX_j)=*(X_j)*(X_i)=X_jX_i,
\end{equation}
and any monomial appearing in the quantum cluster algebra should be invariant under this $*$ action, this would uniquely fix 
the $q$ factor before each monomial. So we will first find the classical spectral generator and then find the quantum 
version by adding the $q$ factor before each monomial to make it $*$ invariant.
\begin{figure}[htbp]
\small
\centering
\includegraphics[width=10cm]{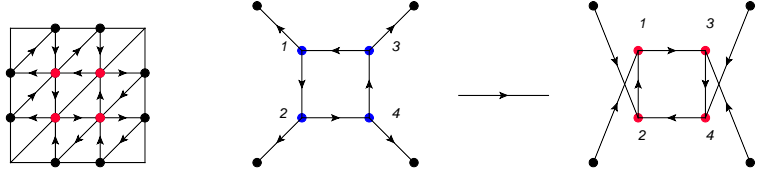}
\caption{The quiver and maximal green mutation for a quiver from a disc with four $A_2$ full punctures. }
\label{A2A2}
\end{figure}

Let's look at an example which is represented by a $A_2$ theory on a disc with four full punctures. The quiver and the initial and 
final configuration of the maximal green mutation sequences is shown in figure. \ref{A2A2}. The maximal green mutation sequences are
\begin{equation}
\mu_2,\mu_3,\mu_1,\mu_4,\mu_2,\mu_3.
\end{equation}
and the final cluster coordinates are

\begin{doublespace}
\noindent\(\pmb{\hat{X}[1]=\frac{1+x_4+x_2 x_4+x_1 x_2 x_4}{x_2 \left(1+x_1+x_1 x_3+x_1 x_3 x_4\right)},}\\
\pmb{\hat{X}[2]=\frac{1+x_2+x_1 x_2+x_1 x_2 x_3}{x_1 \left(1+x_3+x_3 x_4+x_2 x_3 x_4\right)},}\\
\pmb{\hat{X}[3]=\frac{1+x_3+x_3 x_4+x_2 x_3 x_4}{\left(1+x_2+x_1 x_2+x_1 x_2 x_3\right) x_4},}\\
\pmb{\hat{X}[4]=\frac{1+x_1+x_1 x_3+x_1 x_3 x_4}{x_3 \left(1+x_4+x_2 x_4+x_1 x_2 x_4\right)}.}\)
\end{doublespace}

Using the $*$ invariance, we can easily find the refined spectral generator:

\begin{doublespace}
\noindent\(\pmb{\hat{X}[1]=\frac{1}{x_2+q^{-1/2}x_1x_2+x_1x_2 x_3+x_1 x_2x_3 x_4}+\frac{1}{q^{1/2}x_2x_4{}^{-1}+x_1x_2x_4{}^{-1}+x_1x_2 x_3x_4{}^{-1}+x_1
x_2x_3 }+}\\
\pmb{\frac{1}{x_4{}^{-1}+x_1x_4{}^{-1}+x_1 x_3x_4{}^{-1}+q^{1/2}x_1 x_3 }+\frac{1}{x_1{}^{-1}x_4{}^{-1}+x_4{}^{-1}+ q^{-1/2}x_3x_4{}^{-1}+x_3},}\\
\pmb{\hat{X}[2]=\frac{1}{x_1+q^{1/2} x_1x_3+q x_1x_3 x_4+x_1 x_2x_3 x_4}+\frac{1}{q^{1/2}x_1x_2{}^{-1}+q x_1x_2{}^{-1}x_3+q^2x_1x_2{}^{-1} x_3x_4+q
x_1 x_3x_4 }+}\\
\pmb{\frac{1}{x_2{}^{-1}+x_2{}^{-1}x_3+q x_2 {}^{-1}x_3x_4+q^{1/2}x_3 x_4 }+\frac{1}{x_2{}^{-1}x_3{}^{-1}+x_2{}^{-1}+ q^{1/2}x_2x_4{}^{-1}+x_4},}\\
\pmb{}\\
\pmb{\hat{X}[3]=\frac{1}{x_4+q^{-1/2}x_2x_4+q^{-1}x_1x_2 x_4+x_1 x_2x_3 x_4}+}\\
\pmb{\frac{1}{q^{-1/2}x_3{}^{-1}x_4+q^{-1}x_2x_3{}^{-1}x_4+q^{-2}x_1x_2x_3{}^{-1} x_4+q^{-1}x_1 x_2x_4}+\frac{1}{x_3{}^{-1}+x_2x_3{}^{-1}+q^{-1}x_1x_2x_3{}^{-1}
+x_1 x_2 }+}\\
\pmb{\frac{1}{x_2{}^{-1}x_3{}^{-1}+x_3{}^{-1}+q^{-1/2}x_1x_3{}^{-1} +x_1 },}\\
\pmb{}\\
\pmb{\hat{X}[4]=\frac{1}{x_3+q^{1/2}x_3x_4+x_2 x_3x_4+x_1 x_2x_3 x_4}+\frac{1}{q^{-1/2}x_1{}^{-1}x_3+x_1{}^{-1}x_3x_4+x_1{}^{-1}x_2 x_3x_4+ x_2x_3
x_4}+}\\
\pmb{\frac{1}{x_1{}^{-1}+x_1{}^{-1}x_4+x_1{}^{-1}x_2 x_4+q^{-1/2} x_2 x_4}+\frac{1}{x_1{}^{-1}x_4{}^{-1}+x_1{}^{-1}+q^{1/2}x_1{}^{-1}x_2 + x_2 }.}\\
\pmb{}\\
\pmb{}\)
\end{doublespace}

\bibliographystyle{utphys} 
 \bibliography{PLforRS}    
\end{document}